\documentclass[%
reprint,
 prd,
superscriptaddress,
nofootinbib,
aps,
dvipsnames
]{revtex4-2}
\usepackage[utf8]{inputenc}

\usepackage{amsmath, amsthm, amsfonts,  amssymb, latexsym}
\usepackage{graphicx}
\usepackage{dcolumn}
\usepackage{bm}
\usepackage{hyperref}
\usepackage[mathlines]{lineno}
\usepackage{orcidlink}
\usepackage{breakurl}

\usepackage{xcolor}

\usepackage{enumerate}
\usepackage{mathrsfs}
\usepackage{graphicx}

\usepackage{subcaption}
\usepackage{xfrac}
\usepackage{lmodern}
\usepackage{multirow}

\usepackage{tikz}
\usetikzlibrary{angles,
                quotes}

\newcommand{\bphi}{\bar \phi}
\newcommand{\bx}{\bar x}
\newcommand{\bt}{\bar t}

\newcommand{\ud}{\mathrm{d}}

\newcommand{\llangle}{\langle\langle}
\newcommand{\rrangle}{\rangle\rangle}

\renewcommand{\S}{{\mathcal S}}
\newcommand{\E}{{\mathcal E}}
\newcommand{\T}{{\mathcal T}}
\renewcommand{\L}{{\mathcal L}}
\renewcommand{\O}{{\mathcal O}}

\newcommand{\sM}{{\mathscr M}}

\def\SLg{\ensuremath{\mathrm{SL}_2(\mathbb R)}}

\let\edth\relax

\font\ec=ecrm0800 at 12pt
\def\th{\hbox{\ec\char'336}}
\def\edth{\hbox{\ec\char'360}}

\DeclareMathOperator{\thorn}{\text{\rm \th}}

\newcommand{\GHPw}[2]{\left\{ #1, #2 \right\}}

\def\Sh{\ensuremath{{}_s S_{\ell m \omega}}}
\def\Rh{\ensuremath{ {}_s R_{\ell m \omega }}}

\renewcommand{\Re}{\operatorname{Re}}
\renewcommand{\Im}{\operatorname{Im}}

\newcommand{\dd}{{\rm d}}

\def\no{\nonumber \\}

\begin{document}
	
	\title{Extremal Black Hole Weather}
	
	\author{Claudio Iuliano \orcidlink{0009-0009-3498-0485}}%
	\email{iuliano@mis.mpg.de}
	\affiliation{%
		Institute of Theoretical Physics, Leipzig University, Br{\"u}derstra{\ss}e 16, 04103 Leipzig, Germany
	}%
	\affiliation{%
	Max Planck Institute for Mathematics in Sciences (MiS), Inselstra{\ss}e 22, 04103
	Leipzig, Germany
}
	
		\author{Stefan Hollands \orcidlink{0000-0001-6627-2808}}
	\email[Contact author: ]{stefan.hollands@uni-leipzig.de}
	\affiliation{%
	Institute of Theoretical Physics, Leipzig University, Br{\"u}derstra{\ss}e 16, 04103 Leipzig, Germany
}%
\affiliation{%
	Max Planck Institute for Mathematics in Sciences (MiS), Inselstra{\ss}e 22, 04103
	Leipzig, Germany
}

\author{Stephen R. Green  \orcidlink{0000-0002-6987-6313}}
	\email{stephen.green2@nottingham.ac.uk}
	\affiliation{%
	School of Mathematical Sciences, University of Nottingham, University Park, Nottingham NG7 2RD, United Kingdom}

\author{Peter Zimmerman}
\email{zimmerator@protonmail.com}

	\begin{abstract}
We consider weakly non-linear gravitational perturbations of a near-extremal Kerr black hole governed by the second order vacuum Einstein equation. Using the GHZ formalism [Green et al., Class. Quant. Grav. 7(7):075001,
2020], these are parameterized by a Hertz potential. We make an ansatz for the Hertz potential as a series of zero-damped quasinormal modes with time-dependent amplitudes, and derive a non-linear dynamical system for them. We find that our dynamical system has a time-independent solution within the near horizon scaling limit. This equilibrium solution is supported on axisymmetric modes, with amplitudes scaling as $c_\ell \sim C^{\rm low} 2^{-\ell/2} \ell^{-\frac{7}{2}}$ for large polar angular momentum mode number $\ell$, where $C^{\rm low}$ is a cumulative amplitude of the low $\ell$ modes. We interpret our result as evidence that the dynamical evolution will approach, for a parametrically long time as extremality is approached, a distribution of mode amplitudes dyadically exponentially suppressed in $\ell$, hence as the endpoint of an inverse cascade. It is reminiscent of weather-like phenomena in certain models of atmospheric dynamics of rotating bodies. During the timescale considered, the decay of the QNMs themselves plays no role given their parametrically long half-life. Hence, our result is due entirely to weakly non-linear effects.
	\end{abstract}

	\maketitle

\section{Introduction}\label{sec:intro}
It is intriguing to ask whether there are dynamical regimes for excited Kerr black holes governed by a weakly non-linear self-interaction of gravitational waves over a physically relevant time scale; for recent controversial discussions see e.g., \cite{mitman2023nonlinearities, cheung2023nonlinear, ma2024excitation, lagos2023generation, sberna2022nonlinear, bourg2024quadratic,Khera:2024yrk}. A particularly natural regime to look for such effects is the highly spinning Kerr black hole, characterized by a small but non-zero extremality parameter 
\begin{equation}
\label{extrempar}
\varepsilon = \frac{r_+-r_-}{2r_+}>0, 
\end{equation}
where $r_+, r_-$ are the radii of the outer and inner horizons.

The near horizon region of such black holes may be imagined as a leaky cavity \cite{andersson2000superradiance}, supporting  a tower of exponentially decaying but parametrically long-lived, standing gravitational waves called zero-damped quasi normal modes (QNMs). The corresponding QNM frequencies scale as \cite{Hod:2012bw, yang2013branching, Yang2013}
\begin{equation}
\label{QNMf}
\omega \approx \frac{m -i\varepsilon(N+h_{\ell m})}{2M} \quad 
\text{for $\varepsilon \ll 1$.}
\end{equation}
Here $\ell,m$ are standard angular momentum type labels of spin-weighted spheroidal harmonics \cite{TorresdelCastillo2003,Breuer1977,flammer2014spheroidal}, $h_{\ell m}$ is a parameter called the conformal weight\footnote{
Near horizon QNMs form lowest weight $h_{\ell m}$ modules of the near horizon isometry group $\SLg$, which is also the conformal group of a lightray, see App. \ref{SL2R}.
} [see Eq.~\eqref{eq:hpm}], $N = 0,1,2, \dots$ is the so-called the overtone number, and $M \approx r_+$ is the mass of the nearly extremal ($\varepsilon \ll 1$) Kerr black hole. The zero-damped QNMs pile up at the superradiant 
bound $m\Omega_H$ [$\approx m/(2M)$ for a nearly extremal black hole] and are co-rotating i.e., $m/\Re(\omega) \ge 0$. They are long-lived 
because $e^{-i\omega t}$ is very slowly decaying due 
to the smallness of $\Im(\omega) = -O(\varepsilon)$.

One reason for suspecting that near extremal Kerr black holes may support significant non-linear effects is that the long-lived nature of the QNMs \eqref{QNMf} affords ample time for self-interaction before dissipative effects are expected to take over\footnote{For Schwarzschild- and slowly rotating Kerr or Kerr-deSitter black holes, non-linear stability with quantitative decay has been mathematically proven by \cite{Hafner:2019kov, Klainerman:2021qzy, Dafermos:2021cbw}.}. In fact, the QNMs \eqref{QNMf} have been linked  \cite{casals2016horizon,Gralla:2016sxp} at the linear level to the Aretakis phenomenon \cite{Aretakis:2011ha} for exactly extremal black holes, see e.g., \cite{Bantilan:2017kok} for a refined numerical analysis, or \cite{angelopoulos2023late} for a recent mathematical analysis. 

Furthermore, \cite{yang2015turbulent} observed that, since the real part of the QNM frequencies $m/(2M)$ is basically an integer, there arises the possibility of a (near) resonant interaction between long-lived QNMs once non-linear effects in perturbation theory are taken into account, see 
\cite[Sec. I]{green2020teukolsky} for a rough estimation of various competing effects. In particular, \cite{yang2015turbulent}
suggested that certain energy transfer processes between modes of different energy might occur, possibly leading to a sort of inverse cascade. 

Yet another reason for suspecting effects of this kind for nearly extremal Kerr black holes is that the near horizon region is geometrically a fibration over an AdS$_2$ space \cite{bardeen1972rotating,bardeen1999extreme,amsel2009no}.\footnote{This fact, and the correspondingly enhanced conformal symmetry group $\SLg$ is the basis for the ``Kerr-CFT correspondence, see e.g., \cite{bredberg2011lectures} for a review.} It is known that, due to non-linear interactions, the Einstein equations (EEs) in AdS$_4$ have weakly turbulent solutions \cite{Bizon:2015pfa} recently confirmed mathematically by \cite{Moschidis:2018ruk}.\footnote{In the Einstein-Vlasov system on AdS$_4$.} Such effects can be understood from the point of view of resonant non-linear interaction between ``normal''\footnote{There is no dissipation in AdS$_4$, hence these are ordinary ``Fourier''-type modes with real frequency.} modes \cite{Balasubramanian:2014cja, Craps:2014vaa}.

However, there are also good reasons for thinking that this intuition about non-linear interactions between long-lived QNMs might {\it not} be qualitatively correct. Indeed, AdS$_4$ is qualitatively different from a warped product of AdS$_2$. 
In AdS$_4$, there is a direct cascade, not an {\it inverse} cascade. As opposed to normal modes, QNMs are very far from a complete basis of perturbations even in the linear regime.

Furthermore, while the proposal by \cite{yang2015turbulent} is intriguing and suggestive, they do not provide---nor claim to do so---a self-consistent framework of the non-linear effects. In fact, they do not consider the EEs but instead a linear scalar model equation on a background perturbed by a single given QNM. By design of their setup,  \cite{yang2015turbulent} are restricted to an analysis of how the background ``parent'' QNM sources a doublet of the dynamical scalar field ``daughter'' QNMs. Such an analysis might give a reliable indication of the nature of energy transfer in a situation with only a single triplet of resonant modes, as happens e.g. for certain triplets of quasi bound state modes of Schwarzschild-AdS$_4$ \cite{Kehle}. But its status is in our view not totally clear in the present situation, given that {\it all} the zero-damped QNMs are resonant, which is more suggestive of a coherent superposition of many QNMs.
Such a superposition simply cannot be captured by the setup of \cite{yang2015turbulent}.
 
In this work, we develop the weak turbulence idea in a new framework of infinite dimensional dynamical systems for QNM amplitudes.

\subsection{\texorpdfstring{$1+1$}{}D toy model}
In order to give the reader a basic idea of our approach, in this Introduction we consider first a toy model of a real-valued scalar field $\Phi(t,\phi)$, $2\pi$-periodic in the angular coordinate $\phi$, obeying the non-linear Klein-Gordon (KG) equation
\begin{equation}
\label{KGnonlin}
    (\partial_t^2 - \partial_\phi^2) \Phi = \alpha \Phi^2.
\end{equation}
Here, $\alpha$ is a constant characterizing the strength of the interaction. The corresponding linear KG equation, where $\alpha=0$, has ordinary Fourier modes with real frequency $\omega$ (``normal'' modes, NMs)
\begin{equation}
    u_m(t,\phi) = \frac{1}{\sqrt{2\pi} \sqrt{2\omega}}e^{-i\omega t + im\phi},
\end{equation}
where\footnote{For the sake of simplicity of this discussion we ignore the zero mode, $m=0$. In the black hole context, this mode corresponds to axisymmetric perturbations and will be treated with appropriate care.} $m= \pm 1, \pm 2, \dots$ and $\omega = |m|$. These modes are normalized with respect to the KG inner product
\begin{equation}
    (\Phi_1, \Phi_2)_t = -i\int\limits_0^{2\pi} \dd \phi (
    \Phi_1 \partial_t \Phi_2^* - \Phi_2^* \partial_t \Phi_1)\Bigg|_{t},
\end{equation}
i.e. $(u_{m_1},u_{m_2})_t = \delta_{m_1,m_2}$. While the KG inner product does not depend on $t$ for solutions to the linear KG equation, it does for solutions $\Phi$ of the {\it non}-linear KG equation \eqref{KGnonlin}. In fact, we may define a $t$-dependent NM amplitude $a_m(t)$ by
\begin{equation}
\label{KGmodeproj}
    a_m(t) := (u_m,\Phi)_t \ . 
\end{equation}
$a_m$ is a complex valued function, analogous to the time-dependent annihilation operator in the LSZ approach to scattering in quantum field theory, see e.g., \cite[I.5]{srednicki2007quantum}. In fact, we may write
\begin{equation}
\label{KGmodesum}
    \Phi = \sum_m a_m(t) u_m + \text{c.c.}
\end{equation}
It is easy to show that the non-linear KG equation \eqref{KGnonlin} is equivalent to an infinite-dimensional Hamiltonian dynamical system for the $a_m$'s, of the schematic form 
\begin{equation}
\label{dynsys2}
    \frac{\dd}{\dd t} a_1 = \alpha \sum_{2,3} U_{123} a_2 a_3 + \dots,
\end{equation}
with ``overlap coefficients''
\begin{equation}
    U_{123} = e^{i(\omega_1-\omega_2-\omega_3) t} \delta_{m_1,m_2+m_3} \frac{1}{4\sqrt{\pi}\sqrt{\omega_1\omega_2\omega_3}}.
\end{equation}
The dots in Eq.~\eqref{dynsys2} stand for other terms with $a_2^* a_3, a_2 a_3^*, a_2^* a_3^*$ and correspondingly different overlap coefficients involving other combinations 
$e^{i(\omega_1\pm_2\omega_2\pm_3\omega_3) t} \delta_{m_1, \pm_2 m_2 \pm_3 m_3}$, where the signs $\pm_2, \pm_3$ depend on the particular combination.

Based on Eq. \eqref{dynsys2}, we may expect that the 
$a_m$'s change significantly over a timescale of order $\alpha^{-1}$, which, for small $\alpha$, is long compared to the  timescale on which the terms $e^{i(\omega_1\pm_2\omega_2\pm_3\omega_3) t}$ oscillate. These oscillations may be expected to cancel {\it unless} we have a resonant combination, meaning that 
\begin{equation}
\label{resonance}
 \omega_1\pm_2\omega_2\pm_3 \omega_3 = 0.   
\end{equation}
Since $\omega_i = |m_i|$ and the $m_i$ are integers, there many ways to satisfy these conditions. Terms in Eq. \eqref{dynsys2} meeting the resonance condition \eqref{resonance} ought to govern the long time evolution of the dynamical system. This effect may be captured e.g., by the ``two-timescale formalism'' \cite{Balasubramanian:2014cja} which has, in fact, been used to argue for a turbulent instability of the Einstein-scalar field equations in spherical symmetry in AdS$_4$ spacetime by analyzing the corresponding dynamical system of NM amplitudes in this case (see also \cite{Craps:2014vaa,Bizon:2015pfa}).

\subsection{Dynamical system for long-lived QNM amplitudes}

Up to a correction of order $O(\varepsilon)$, the real part of the long-lived QNM spectrum \eqref{QNMf} is resonant in the sense that Eq. \eqref{resonance} nearly holds whenever $m_1 \pm_2 m_2 \pm_3 m_3=0$.
Since QNMs of spin $s=\pm 2$ describe linear gravitational perturbations of Kerr, one may hope that an analysis analogous to that sketched for the KG field may be carried out for the EE in the leading non-linear approximation,
\begin{equation}
\label{EEquad}
    \mathcal{E}_{ab}[h] = 8\pi \alpha \mathcal{T}_{ab}[h,h], 
\end{equation}
where 
\begin{equation}
g_{ab} = \bar g_{ab} + \alpha h_{ab}
\end{equation}
is a metric describing a Kerr metric $\bar g_{ab}$ perturbed by 
a small field, $h_{ab}$. $\mathcal{E}_{ab}$ represents the linearized Einstein operator on $\bar g_{ab}$ [see Eq. \eqref{eq:linearE}], and the second order Einstein tensor $\mathcal{T}_{ab}$ [see Eq. \eqref{G2E}] 
represents the leading quadratic non-linearity on $\bar g_{ab}$ in the full EE $G_{ab}[\bar g+\alpha h]=0$.

By analogy with the treatment of the KG equation \eqref{KGnonlin},
one might consider decomposing $h_{ab}$ into QNMs (rather than NMs, which do not exist in Kerr), and derive a dynamical system analogous to \eqref{dynsys2}. Potential objections to such a scheme might be
\begin{enumerate}
    \item black holes are dissipative systems: energy and angular momentum may be absorbed or radiated away to infinity, potentially driving the black hole away from extremality,
    \item QNMs as normally considered in general relativity refer to the linear, scalar Teukolsky equations and not directly to the non-linear, tensorial EE \eqref{EEquad},
    \item unlike NMs, QNMs are very far from a complete set of functions,
    \item there is a priori no obvious analogue of the KG inner product, because complex conjugation is not a symmetry of the Teukolsky equation.
\end{enumerate}
Regarding 1), spin down rates due to dissipative effects were estimated in \cite[Sec. I]{green2020teukolsky}, where it was shown that sustained non-linear interactions of long-lived QNMs are possible within a suitable parameter range sufficiently near extremality. Objection 2) is a technical nuisance but no longer a fundamental 
obstruction thanks to a recent generalization \cite{green2020teukolsky,hollands2024metric}
of the metric reconstruction technique \cite{kegeles1979constructive,chrzanowski1975vector}
to the non-linear EE. In fact \cite{green2020teukolsky,hollands2024metric} showed that up to a so-called ``corrector tensor’’, $x_{ab}$, which can be dealt with straightforwardly, non-linear metric perturbations of Kerr can be written in so-called ``reconstructed form’’, i.e. in terms of a Hertz potential, $\Phi$, solving a sourced Teukolsky equation \cite{Teukolsky1973,TeukolskyII},
\begin{equation}
\label{eq:recost}
    h_{ab} = \Re \S^\dagger_{ab} \Phi + x_{ab}.
\end{equation}
Here, $\S^\dagger_{ab}$ [see Eq.~\eqref{eq:Sdag}] is the so-called reconstruction operator. We propose that objection 3) is not an issue in that we restrict attention to a dynamical regime during which the non-linear evolution of $h_{ab}$ is driven predominantly by the QNM part of $\Phi$ via Eq.~\eqref{eq:recost} up to and including second order. 

In accordance with this hypothesis, we informally write 
\begin{equation}
\label{Phisum3}
\Phi = \sum_q c_q(t) \Upsilon_q,
\end{equation}
[compare Eq. \eqref{KGmodesum}], where we call $c_q(t), q=(N,\ell,m)$ the ``QNM amplitudes'', and 
where the $\Upsilon_q \equiv \Upsilon_q(x^\mu) \propto e^{-i\omega_q t}$ are the separated QNM mode functions at linear order. Thus, we propose that the metric \eqref{eq:recost} can be accurately described, for a parametrically long Boyer Lindquist time diverging as $\varepsilon \to 0$,
by \eqref{Phisum3} and the corrector $x_{ab}$ which is obtained by the technique of \cite{casals2024spin} as a quadratic expression in the $c_q$'s.

At the level of the linearized EE, we do not require $x_{ab}$, and the $c_q$'s would be simply constant. To derive a dynamical system for them capturing the leading non-linearities of the EE, and in order to address objection 4), we use a ``scalar product'' for Teukolsky-like scalars recently introduced by \cite{Green2022a}. It will turn out that, in terms of this product and up to normalization factors,
we have, 
\begin{equation}
    c_q(t) = \langle\langle \Upsilon_q, \Phi \rangle\rangle_t
\end{equation}
[compare Eq. \eqref{KGmodeproj}]. This expression will enable us 
to derive a dynamical system for the $c_q$ analogous to Eq. \eqref{dynsys2}. Our system takes the general form\footnote{
There are certain selection rules implicit in the double summation e.g., 
$U_{123} \propto \delta_{m_1,m_2+m_3}$
or $V_{123} \propto \delta_{m_1,m_2-m_3}$. Hence the second term can be associated with ``mirror modes'' under a parity flip.
} [see Eq.~\eqref{G4}] 
\begin{equation}
\label{DYN}
\frac{\dd}{\dd t} c_1
= \alpha \sum_{2,3} \left( U_{123} c_2 c_3 + V_{123} c_2 c_3^* \right),
\end{equation}
and is valid, in principle, without the near extremal assumption. 

Eq. \eqref{DYN} describes the leading non-linear (quadratic) interaction between QNMs under the ansatz \eqref{Phisum3}. As such, it is well-suited to describe the resonant dynamics of QNMs of near-extreme black holes. Our ansatz, however, does not include what are commonly referred to as quadratic QNMs (QQNMs)~\cite{bourg2024quadratic, Khera:2024yrk, lagos2023generation, ma2024excitation}, which are generally driven off resonance. Indeed, QQNMs are particular solutions of the second-order Einstein equation for generic spin, which are sourced by quadratic products of QNMs. Since for generic spin, the driving frequency is off-resonance, QQNM mode functions are not typically close to those of QNMs. See Sec.~\ref{sec:qqnm} for further discussion.

In a linearization of the dynamical system \eqref{DYN}, 
one would pick some fixed ``background'' distribution 
$\{c_q\}$ and consider its linear perturbation $\{ \delta c_q \}$.  The linearization would read
\begin{equation}
\label{DYNlin}
\frac{\dd}{\dd t} \delta c_1
= \alpha \sum_{2,3} \Big[ 2U_{123} c_2 \delta c_3 + V_{123} (c_2 \delta c_3^* + \delta c_2 c_3^*) \Big].
\end{equation}
In particular, though this would almost certainly not be a solution to \eqref{DYN}, one could consider a background distribution with only a {\it single} QNM amplitude $c_{N\ell m}$ excited. 
This ``parent mode'' gives a linear system for the ``daughter modes'' described by the linear perturbation $\delta c_q$. There are certain selection rules implicit in the coupling coefficients such as $U_{123} \propto \delta_{m_1,m_2+m_3}$
or $V_{123} \propto \delta_{m_1,m_2-m_3}$, thus a parent mode can induce a coupling only between daughter modes satisfying these rules. 

These rules can be satisfied in infinitely many ways, so the resulting system still is in general infinite-dimensional. A possible coupling is e.g., for the parent mode to have magnetic mode number $2m$ and the daughter modes to have $\pm m$ i.e., half that of the parent mode. Ref. \cite{yang2015turbulent} basically follows this approach (in a simplified setting where the daughter modes have spin 0) and considers an ad-hoc truncation of the infinite-dimensional system for the daughter modes to $\delta c_{N\ell (\pm m)}$.\footnote{Since their daughter modes are furthermore associated with a real scalar field as opposed to a spin-2 gravitation perturbation, further less important simplifications arise.} The resulting $2 \times 2$ dimensional truncated system may be cast into a single ODE possessing exponentially growing solutions under certain conditions specified by \cite{yang2015turbulent}.
These growing solutions are interpreted by \cite{yang2015turbulent} as evidence for an inverse cascade, 
$2m \to m$, since they propose this process would replicate itself once the daughter mode sources further granddaughter modes, etc.

Our framework goes beyond the analysis by \cite{yang2015turbulent} in several ways. Firstly, we will not consider a linearization of the dynamical system, but instead self consistently consider the non-linear EE at the ``three wave interactions’’ non-linear level, which we will show, leads to \eqref{DYN}. Consequently, we will not make a distinction between parent and daughter QNMs but treat all QNM amplitudes on the same footing. As we have already indicated, we think that this is more appropriate if one has in mind a cascade which will excite many QNMs at roughly the same amplitude. Thirdly, for the same reason, we will not truncate the system ad-hoc to a $2 \times 2$ system. Lastly, in contrast to \cite{yang2015turbulent}, it will be important for us to take into account the finer detail of the long-lived QNM spectrum at order $O(\varepsilon/M)$ hidden in the conformal weight $h_{\ell m}$. 

Our analysis shows that our dynamical system of wave interactions governed by the EE displays unexpected and significant simplifications which we ascribe to the special structure of the EE off of Kerr. These simplifications will allow us to propose explicit turbulence spectra for near equilibrium solutions.

\subsection{Simplifications in the near extremal approximation}

While our dynamical system \eqref{DYN} has the same general form as in the KG toy model [see Eq. \eqref{dynsys2}] with 
$a_m$ suitably replaced by $c_q$ and the frequencies of the NMs replaced by those \eqref{QNMf} of the QNMs, it would be illusory to expect that one might be able to find simple, explicit formulas for the overlap coefficients $U_{123}, V_{123}$ in Kerr. 

We therefore proceed to consider them in the so-called near-near horizon extremal Kerr (nNHEK) formalism \cite{Baarden:1973ula,bardeen1999extreme,amsel2009no,TeukolskyII} in the regime $\varepsilon \ll 1$, with the assumption that significant interactions between the QNMs should take place only in the ``near zone'', 
\begin{equation}
\frac{r-r_+}{r_+} = O(\varepsilon) \quad \text{(near zone).}
\end{equation}
Still, it turns out that the nNHEK approximation is by itself not sufficient neither for obtaining explicit expressions for the overlap coefficients $U_{123}, V_{123}$ in the dynamical system, nor to ensure the near resonance condition \eqref{resonance}.
We therefore further restrict our attention to a subset of QNMs which either have a low, $\ell = O(1)$, or high, $\ell \gg 1$, angular momentum. The latter regime enables the so-called ``eikonal approximation'', where the long-lived QNM spectrum becomes {\it doubly} resonant, 
\begin{equation}
\omega \approx \frac{m - i\varepsilon(N+\ell+1)}{2M}, 
\end{equation}
so long as $\varepsilon(N+\ell) \ll 1$. QNMs with intermediate values of $\ell$, or ones with $\varepsilon(N+\ell) \gtrsim 1$ are discarded. Here, the idea is that the low $\ell$ modes provide a, possibly transient, pumping effect, and that the very high $\ell$ modes dissipate away due to their exponential decay in time. The hypothesis is thereby that they effectively decouple from the truncated system, in much the same way as one proceeds in weak wave turbulence \cite{zakharov2012kolmogorov}. 

The calculations required in order to find $U_{123}^{\rm near}, V_{123}^{\rm near}$ in the nNHEK and low/high $\ell$ approximation are still substantial even in this range, but, in the end, there appear several crucial simplifications in the structure of $U_{123}^{\rm near}, V_{123}^{\rm near}$:
\begin{itemize}
    \item The time derivative is now with respect to a ``slow time’’ $\bar t$, given by $\varepsilon t/(2M)$ in terms Boyer-Lindquist time $t$.
    \item Dividing the QNM amplitudes up into
    \begin{equation}
    c_q \to c_q^{\rm high} \quad \text{or} \quad c_q^{\rm low}
    \end{equation}
    according to whether $\ell$ is high or low and adopting a chemical reaction
    notation for the interactions inducing a temporal change in $c_q^{\rm high}, c_q^{\rm low}$ according to the dynamical system \eqref{DYN}, the dominant channels turn out to be
    \begin{equation}
    \begin{split}
        & \text{(high),(high)} \to \text{(high)}\\
        & \text{(high),(low)} \to \text{(high)}\\
        & \text{(high),(low)} \to \text{(low)}\\
        & \text{(high),(high)} \to \text{(low)}.
    \end{split}
    \end{equation}
    In each case, the corresponding ``transition amplitudes''
    $U_{123}^{\rm near}, V_{123}^{\rm near}$ can be found explicitly, see Sec. \ref{highlowelloverlap}, App. \ref{sec:exploverlap}.
    \item These transition amplitudes imply ``selection rules'' for the ``$m$'' QNM frequency labels, which are a reflection of the fact that the background spacetime is axisymmetric.
    \item Additionally, there are selection rules for the $\ell$ QNM frequency labels corresponding to those for the addition of angular momentum in quantum mechanics.
\end{itemize}
 
\subsection{Equilibrium}\label{sec:eqlbrm}
A distribution $\{ c_q \}$ of QNMs such that 
\begin{equation}
    \frac{\dd}{\dd t} c_q^{\rm eq} = 0
\end{equation}
for all high and low $\ell$
QNMs $q$ in the nNHEK approximation is considered an ``equilibrium distribution'', because for such a distribution, the corresponding metric, $g_{ab}^{\rm eq}$, 
\begin{equation}
\label{eq:recosteq}
    g_{ab}^{\rm eq} = \bar g_{ab} +\Re \S^\dagger_{ab} \Phi^{\rm eq} + x_{ab}^{\rm eq}.
\end{equation}
will not change over a parametrically large Boyer-Lindquist time scaling like some inverse power of the extremality parameter $\varepsilon$.

Finding an equilibrium distribution amounts to solving the quadratic system of equations 
\begin{equation}
    0 = \sum_{2,3} \Big[U_{123}^{\rm near} c_2^{\rm eq} c_3^{\rm eq} + V_{123}^{\rm near} c_2^{\rm eq} (c_3^{\rm eq})^*\Big].
\end{equation}
Using non-trivial simplifications of our dynamical system in the nNHEK limit and high/low $\ell$ approximation, we are able to find such a solution based on scaling considerations. It has the form 
\begin{equation}
    c^{\rm eq}_{N\ell m}
    \propto C^{\rm low} \cdot \delta_{N,0} \delta_{m,0} \cdot 
    2^{-\frac{\ell}{2}}
    \ell^{-\frac{7}{2}},
\end{equation}
where $C^{\rm low}$ is a certain weighted sum of the amplitudes of the low $\ell$ amplitudes, which may be viewed as a manifestation of the aforementioned pumping effect. We observe that the distribution is decreasing like a dyadic exponential in $\ell$, and we interpret this as saying that equilibrium is achieved when the QNM amplitudes are non-zero only for low $\ell$ (and zero $m$). We view this as a manifestation of an inverse cascade, hence a kind of ``weather'' see e.g., \cite{serazin2018inverse} in the context of oceanography.  

Note that our equilibrium solution only involves an infinite tower of axisymmetric modes undergoing 3 wave interactions. On the other hand, as we have described, \cite{yang2015turbulent} consider the coupling between $3$ waves with non-vanishing magnetic mode numbers $\pm m$ and $2m$. In order to understand more fully the validity of their truncation and the relationship to ours, one would have to study our dynamical system outside the axisymmetric sector. While we give the prerequisite formulas for the overlap coefficients, we leave such an analysis for future work.

\subsection{Connection to quadratic QNMs}
\label{sec:qqnm}

Finally, we wish to clarify the relationship between our perturbative solutions and quadratic QNMs~\cite{bourg2024quadratic, Khera:2024yrk, lagos2023generation, ma2024excitation}.  For instance, the starting point of \cite{ma2024excitation}, like ours, is \eqref{EE4}, where each `$\phi$' on the right side is a QNM at linear order. The approach by \cite{ma2024excitation} is to view `$\phi$' on the left side as the second-order correction. They solve Eq. \eqref{EE4} essentially by the usual method of separation of variables. However, given that the spectrum of QNMs is generally not resonant for slowly rotating Kerr black holes, the resulting second order `$\phi$' is in general not expected to be close to a QNM nor to be particularly long-lived. 

By contrast, we consider near extremal Kerr black holes for which the zero-damped modes are very nearly resonant and long-lived. In our approach, we project Eq. \eqref{EE4} onto such QNMs, thereby effectively assuming that
\eqref{EE4} may be solved using only their contribution to the retarded Green's function for the Teukolsky equation for a near-extremal Kerr black hole. Thus, we assume that the metric perturbation can, in effect, be considered to be dominated by the zero-damped QNMs. As described, this assumption leads to a dynamical system for the QNM amplitudes, $c_q$, not obtained by \cite{ma2024excitation}. Unlike \cite{ma2024excitation}, we do not divide the metric perturbation, nor the $c_q$'s, into first- and second-order contributions. Instead, we think of Eq.~\eqref{DYN} as a self-consistent approximation which takes into account any gradual drift of $c_q$ amplitudes away from their constant values in linear order.

Our analysis is thus complementary to typical approaches to QQNMs, and applies to physical regimes dominated by resonances. We refer to our modes as resonant quadratic QNMs (rQQNMs). As triplets of mode frequencies are not in general commensurate, rQQNMs are restricted to special regimes such as near-extreme Kerr. One other example where such resonant dynamics become important is within the gravity-fluid correspondence for large Schwarzschild-AdS~\cite{Bhattacharyya:2007vjd, Baier:2007ix}, which leads to highly turbulent dynamics~\cite{Adams:2013vsa, Green:2013zba}. The situation changes, however, if one considers third order interactions, since resonant quartets of modes do generically occur, thus making resonant cubic QNMs (rCQNMs) rather common~\cite{sberna2022nonlinear}. We leave the investigation of such higher order interactions for future work.

\medskip

This paper is organized as follows. In Sec. \ref{nhekgeo}, we introduce our notation related to the Kerr spacetime and recall its near-near horizon extreme Kerr (nNHEK) scaling limit as $\varepsilon \to 0$. In Sec. \ref{sec:Teukform} we recall some of relevant portions of the Teukolsky- \cite{Teukolsky1973,teukolsky1972rotating} and GHP (for Geroch, Held and Penrose \cite{Geroch1973}) formalisms, as well as our definition of the bilinear form \cite{Green2022a} (called a ``scalar product'' in this work) between solutions of Teukolsky's equation. In Sec. \ref{sec:mode}, we derive the general form of our dynamical system for the QNM amplitudes, however without taking as yet the near extremal condition $\varepsilon \ll 1$ into account. In Sec. \ref{nNHEK_QNM}, we analyze QNMs in the nNHEK limit. In particular, we compute the scalar products of QNMs in the nNHEK approximation. Building in part on this analysis, we then consider in Sec. \ref{sec:dynsysnNHEK} the simplifications in the detailed form of our dynamical system arising from $\varepsilon \ll 1$. In Sec. \ref{sec:randomphase}, we derive equilibrium distributions for the QNM amplitudes. A considerable amount of technical detail is deferred to various appendices. 

\medskip
\noindent
{\bf Notations and conventions} We generally follow the conventions of
\cite{wald2010general}, with the exception of the signature of $g_{ab}$
which is $(+, -, -, -)$ in this paper. When acting on weighted GHP scalars
\cite{Geroch1973}, 
$\mathcal{L}_\xi$ denotes the intrinsically defined GHP-invariant Lie-derivative introduced in \cite{edgar2000integration}.
An overbar over a quantity as in $\bar X$ is associated with a scaling limit (explained in detail below), and is not to be confused with the complex conjugate. The complex conjugate of a number $z \in \mathbb C$ is denoted by $z^*$. $O(x)$ denotes a function $\le {\rm const.} x$ for all $x\ge x_0$.

\section{Kerr and its extremal scaling limits} \label{nhekgeo}

The limit of a family spacetimes $(\mathscr{M}(\varepsilon), g_{ab}(\varepsilon))$ 
as a parameter $\varepsilon$ tends to some value depends on which coordinates one holds fixed as that limit is taken --- or, more mathematically, 
on the family of diffeomorphisms used to identify the spacetimes $\mathscr{M}(\varepsilon)$ with a reference spacetime manifold \cite{geroch1969limits}.   

In this paper, we consider the limit of the subextremal Kerr solutions in which the extremality parameter \eqref{extrempar}
$\varepsilon$ tends to zero, where $r_\pm = M \pm \sqrt{M^2 -a^2}$ are the inner and outer horizon radii in Boyer-Lindquist (BL) coordinates \eqref{eq:BL Kerr met}. 
The so-called ``far limits" and ``near limits" correspond to specific identifications. 
In the far limit, the extremality parameter $\varepsilon$ is taken to zero fixing BL 
coordinates $(t,r,\theta,\phi)$, in which the Kerr metric is
\begin{equation}
\label{eq:BL Kerr met}
\begin{split}
  \dd s^2 =& \left(1-\frac{2Mr}{\Sigma}\right) \dd t^2 + \frac{4Mar\sin^2\theta}{\Sigma}\dd t \dd\phi - \frac{\Sigma}{\Delta}\dd r^2\\
  & - \Sigma \dd\theta^2 - \frac{
  (r^2 + a^2)^2 - \Delta a^2 \sin^2\theta
  }{\Sigma} \sin^2\theta \dd\phi^2,
  \end{split}
\end{equation} 
where 
\begin{subequations}\label{eq:BL stuff} 
\begin{align}
  \Delta=& \ r^2+a^2-2Mr, \\
  \Sigma=& \ r^2+a^2\cos^2\theta.
 \end{align}
\end{subequations}
In the far limit the metric is simply given by \eqref{eq:BL Kerr met} with $a=M$.  That limit is naturally adapted to observers that dwell far away from the black hole, such as those measuring gravitational radiation at infinity. The far limit, however, fails to accurately describe the physics as experienced by near-horizon observers.\footnote{For instance,  in the far limit the coordinate location of the innermost stable circular orbit (ISCO) tends to the event horizon,  while the proper distance between the horizon and the ISCO diverges as $\log \varepsilon$ \cite{TeukolskyII}, see also 
\cite{Gralla:2018xzo,Compere2018}.} 
To portray the experience of these near-horizon observers which co-rotate with the horizon in the far limit, one can adopt a new set of coordinates \cite{TeukolskyII,Baarden:1973ula,bardeen1999extreme}  
which ``stretch out'' the near-horizon throat region and co-rotate the black hole. We consider the so-called ``near near horizon extremal Kerr (nNHEK) scaling'' where $\varepsilon$ is taken to zero at the same rate as the coordinates are scaled\footnote{One may also consider the so-called ``NHEK scaling'' which zooms in on length scales intermediate between the near NHEK zone and the far zone \cite{TeukolskyII,amsel2009no}.}, or 
\begin{equation}
\label{scaledcts}
\bar t= \frac{t \varepsilon}{2M} , \quad \bx=\frac{x}{\varepsilon},\quad \bar\theta=\theta, \quad \bphi=\phi-\frac{ t}{2M},
\end{equation}
where 
\begin{equation}
\label{xdef}
x=\frac{r-r_+}{r_+}.
\end{equation}
In other words, in the nNHEK limit we identify the spacetimes $\sM(\varepsilon)$
by identifying points with the same barred coordinates $\bar x^\mu$ \eqref{scaledcts}.

In nNHEK limit, the metric can be thought of as a fibration over a 2-dimensional AdS-spacetime $\dd \bar s^2$ with a constant\footnote{We mean that $\bar \star \dd \bar A$ is constant.} electromagnetic field $\bar A$ \cite{bardeen1999extreme},
\begin{equation}
\begin{split}
\dd s^2_{\rm nNHEK} =&- M^2 \bigg[(1+\cos^2\bar\theta)  \left( \dd \bar s^2 +  \dd \bar\theta^2 \right) \\
&+\frac{4\sin^2\bar\theta}{1+\cos^2\bar\theta}\left (\dd \bar\phi + \bar A \right)^2\bigg],
\end{split}
\end{equation}
where 
\begin{subequations}
\begin{align}
\dd \bar s^2 =& -f \dd\bar t^2 +  \frac{\dd\bar x^2}{f}, \\
f=&\bx(\bx+2), \\
\bar A=&\frac{f'}{2}\dd\bar t
\end{align}
\end{subequations}
are geometric data on the base AdS$_2$-spacetime. The nNEHK spacetime is known \cite{bardeen1999extreme} to possess a continuous\footnote{Discrete isometries will be discussed in Sec. \ref{discreteisometries}.} isometry group $\SLg \times \mathrm{U}(1)$ which enhances the continuous 
isometry group $\mathbb{R} \times \mathrm{U}(1)$ of Kerr comprised of time-translations and rotations. 


\section{Teukolsky formalism}
\label{sec:Teukform}

\subsection{NP tetrads}

Perturbative calculations in Kerr are much simplified using a complex null (Newman-Penrose, NP) tetrad 
aligned with its principal null directions such as e.g., the Kinnersley frame,
\begin{subequations}\label{eq:Kintet}
  \begin{align}
      l^a &= \frac{1}{\Delta} \big[ \left(r^2+a^2\right)\partial_t+ \Delta\partial_r + a\partial_\phi \big]^{a}, \\
    n^a &= \frac{1}{2 \Sigma} \big[ \left(r^2+a^2\right)\partial_t -\Delta\partial_r+a\partial_\phi \big]^a, \\
    m^a &= \frac{1}{\sqrt{2} (r+ia\cos\theta)} \big[ ia(\sin\theta)\partial_t +\partial_\theta+i(\csc\theta) \partial_\phi \big]^a, 
  \end{align}
\end{subequations}
where $x^\mu = (t,r,\theta,\phi)$ are the BL coordinates. The Kerr metric takes the form
\begin{equation}\label{eq:NP met}
g_{ab} = 2l_{(a}n_{b)}-2m_{(a}\bar m_{b)}
\end{equation}
in terms of the tetrad. Its form remains unchanged under
a local rotation preserving the real null pair and transforms the NP frame to
$(l^a, n^a, e^{i\Gamma} m^a, e^{-i\Gamma} \bar m^a)$,
and also under a local boost preserving the directions of the real
null pair, transforming the NP frame to
$(\Lambda l^a, \Lambda^{-1}n^a, m^a, \bar m^a)$. Here, $\Lambda$,
$\Gamma$ are smooth functions of real value which may be combined
into the complex function $\lambda^2 = \Lambda e^{i\Gamma}$.

The Kinnersley frame is singular in the nNHEK scaling when $\varepsilon \to 0$, which can be counteracted by choosing $\lambda = \varepsilon$ \cite{Compere2018}. 
With this, the limit of the above NP frame in nNHEK is \cite{Compere2018}
\begin{subequations}
\label{nNHEK_NP}
	\begin{align}
		l^a & =\left[\frac{1}{f}\partial_{\bar t}+ \partial_{\bar x}-\frac{(\bx+1)}{f}\partial_{\bar\phi}\right]^a \\
	n^a & =\frac{1}{2 M^2\left(1+\cos ^2 \bar \theta\right)}\big[\partial_{\bar t}-f\partial_{\bar x} -(\bx+1)\partial_{\bar \phi}\big]^{a}, \\
	m^a & =\frac{1}{\sqrt{2} M(1+i \cos \bar \theta)}\left[\partial_{\bar\theta}+ \frac{i\left(1+\cos ^2 \bar \theta\right)}{2 \sin \bar \theta}\partial_{\bar\phi}\right]^{a},
	\end{align}
	\end{subequations}
	in the coordinates $\bar x^\mu=(\bt, \bx, \bar \theta, \bphi)$ of \eqref{scaledcts}. 

\subsection{GHP formalism and Hertz potentials}

The GHP formalism \cite{Geroch1973} is a refinement of the NP formalism as described e.g., in \cite{wald2010general, chandrasekhar1998mathematical}. The main difference between the two apart from notations is that in GHP, only properly weighted scalars under a combined frame boost$+$rotation $(l^a, n^a, m^a) \to (\lambda \bar \lambda l^a, (\lambda \bar \lambda)^{-1}n^a, \lambda \bar \lambda^{-1} m^a)$ are considered, whereas all non-weighted scalars become part of a GHP covariant derivative. A properly weighted scalar, aka ``GHP scalar'', $\eta$, of 
weights $(p,q)$ by definition transforms as 
\begin{equation}
    \eta \to \lambda^p \lambda^q \eta,
    \label{GHPtrafo}
\end{equation} 
written as $\eta \circeq \GHPw{p}{q}$. Examples of GHP scalars in any spacetime are 
\begin{subequations}\label{eq:Weylcomp}
\begin{align}
\Psi_0  =& -C_{abcd} \; l^a m^b l^c m^d \circeq \GHPw{4}{0}, \\
\Psi_1  =& - C_{abcd} l^a n^b l^c m^d \circeq \GHPw{2}{0}, \\
\Psi_2  =& -\tfrac{1}{2} C_{abcd}( l^a n^b l^c n^d + l^a n^bm^c \bar{m}^d) \circeq \GHPw{0}{0},\\
\Psi_3  =& -C_{abcd} l^a n^b \bar{m}^c n^d \circeq \GHPw{-2}{0}, \\
\Psi_4  =& -C_{abcd} \; n^a \bar{m}^b n^c \bar{m}^d \circeq \GHPw{-4}{0} ,
\end{align} 
\end{subequations}
as well as the ``optical scalars'',
\begin{subequations}\label{GHPscalars}
\begin{align}
\kappa &=m^a l^b \nabla_b l_a \circeq \GHPw{3}{1} ,\\
\tau &= m^a n^b \nabla_b l_a \circeq \GHPw{1}{-1},\\
\sigma &= m^a m^b \nabla_b l_a \circeq \GHPw{3}{-1}, \\
\rho &= m^a  \bar{m}^b \nabla_b l_a  \circeq \GHPw{1}{1}.
\end{align}
\end{subequations}
In Kerr, we have $\kappa = \kappa' = \sigma = \sigma' = 0, \Psi_i = 0$ for $i \neq 2$ in an NP frame aligned with the principal null directions as considered in this paper, and in nNHEK, we additionally have $\rho = \rho' = 0$. A prime generally means the GHP operation $l^a \leftrightarrow n^a$, $m^a \leftrightarrow \bar m^a$. 
The values of all the non-zero GHP scalars for the frame \eqref{nNHEK_NP} in nNHEK are recalled in App. \ref{appA}. 

The GHP derivative is defined by 
\begin{equation}\label{nabladef}
\begin{split}
\Theta_a \eta &= \left[\nabla_a - \tfrac{1}{2}(p+q) n^b \nabla_a l_b + \tfrac{1}{2}(p-q) \bar{m}^b \nabla_a m_b\right]\eta \nonumber \\
 & \equiv [\nabla_a + l_a(p\epsilon'+q\bar{\epsilon}') + n_a(-p\epsilon-q\bar{\epsilon}) \\
 & \ \ \ \ - m_a(p \beta' -q \bar{\beta}) - \bar{m}_a(-p \beta +q \bar{\beta}')]\eta. 
 \end{split}
\end{equation}
It is covariant in the sense that it maps properly weighted GHP quantities to such quantities. The second line features the 
non-properly weighted remaining spin coefficients $\epsilon, \epsilon', \beta, \beta'$ in the GHP formalism. Their values in the frame \eqref{nNHEK_NP} 
in nNHEK are recalled in App. \ref{appA}. 

In the GHP formalism, Teukolsky's operator \cite{Teukolsky1973,teukolsky1972rotating} acting on 
a GHP scalar $\eta \circeq \GHPw{2s}{0}$ (i.e, spin $s$) reads \cite{araneda2016symmetry}
\begin{equation}
\label{Odef}
    {}_s \O\eta  := \left[g^{ab}(\Theta_a + 2s B_a)(\Theta_b + 2s B_b) - 4s^2 \Psi_2 \right] \eta,
\end{equation}
for $s \ge 0$, where $B^a := -(\rho n^a - \tau \bar m^a) \circeq \GHPw{0}{0}$. For $s \le 0$, we have ${}_s \O := ({}_{-s} \O)'$ 
in terms of the 
GHP priming operation. The GHP covariant directional derivatives along the NP tetrad legs are often easier to deal with computationally, and denoted traditionally by
\begin{subequations}
\label{GHPops}
	\begin{align}
	\thorn &= l^a \Theta_a = l^a\nabla_a-p\epsilon-q\bar\epsilon  \\
	\thorn'&= n^a \Theta_a = n^a\nabla_a+p\epsilon'+q\bar\epsilon' \\
	\edth &= m^a \Theta_a =m^a\nabla_a-p\beta +q\bar \beta ' \\
	\edth' &= \bar m^a \Theta_a = \bar m^a\nabla_a+p\beta' -q\bar \beta .
	\end{align}
	\end{subequations}
For a linearized metric perturbation $\delta g_{ab}$ satisfying $\delta G_{ab} = 8\pi T_{ab}$, the spin\footnote{In the GHP formalism, the spin is 
$s=(p-q)/2$.} $s=\pm 2$ perturbed 
Weyl scalars 
\begin{subequations}\label{eq:Weylcomppm2}
\begin{align}
{}_{+2} \psi :=& -\delta C_{abcd} \; l^a m^b l^c m^d \circeq \GHPw{4}{0}, \\
{}_{-2} \psi  :=& -\delta C_{abcd} \; n^a \bar{m}^b n^c \bar{m}^d \circeq \GHPw{-4}{0} ,
\end{align} 
\end{subequations}
satisfy the sourced Teukolsky equations \cite{Teukolsky1973,teukolsky1972rotating}
\begin{equation}
{}_s \O{}_s \psi = {}_s \S^{ab} T_{ab}.
\end{equation}
Here, ${}_s \O$ are the spin $s$ Teukolsky operators, see Eq. \eqref{Odef}. 
The operators ${}_s \S^{ab}$ prepare the Teukolsky source. For $s=+2$, we have ${}_{+2} \S^{ab} \equiv \S^{ab}$, where
\begin{widetext}
\begin{equation}
  \label{eq:S}
    \S^{ab} T_{ab} = (\edth - \bar \tau' - 4\tau)\Big[(\thorn - 2\bar \rho) T_{lm} - (\edth - \bar \tau') T_{ll} \Big] 
            + (\thorn - \bar \rho - 4\rho)\Big[(\edth - 2\bar \tau') T_{lm} - (\thorn - \bar \rho) T_{mm} \Big],   
\end{equation}
\end{widetext}
and for $s=-2$, we set ${}_{-2}\S^{ab} := (\S^{ab})'$. Eq. \eqref{eq:S} is valid in Kerr and simplifies in nNHEK because 
$\rho=\rho'=0$ in that case. 

So-called Hertz potentials are solutions to the formal adjoints of the Teukolsky equations ${}_s \O^\dagger {}_{s} \Phi = 0$, see e.g., \cite{hollands2024metric} for the GHP forms
of the operators ${}_s \O^\dagger$. Given a Hertz potential, the metric perturbation 
\begin{equation}
\delta g_{ab} = {\rm Re} \left( {}_{s}\S^\dagger {}_s \Phi \right)_{ab}
\end{equation}
is a solution to the linearized Einstein equation $\delta G_{ab}=0$ (for either $s=\pm 2$) \cite{kegeles1979constructive,chrzanowski1975vector}. The well-known concrete expression for the operator 
${}_{s}\S^\dagger$ in GHP form is  ${}_{+2} \S^\dagger \equiv \S^\dagger$ for $s=+2$, where
\begin{widetext}
\begin{equation}
\label{eq:Sdag}
 \left(\S^\dagger \Phi \right)_{ab} =
   - l_a l_b(\edth-\tau)(\edth+3\tau) \Phi - m_a m_b (\thorn - \rho)(\thorn + 3\rho) \Phi
   + l_{(a} m_{b)}\Big[(\thorn - \rho + \bar \rho)(\edth + 3\tau) + (\edth-\tau+\bar \tau')(\thorn +3\rho) \Big] \Phi . 
\end{equation}
\end{widetext}
For $s=-2$, we set ${}_{-2}\S^\dagger := (\S^\dagger)'$. Eq. \eqref{eq:Sdag} is valid in Kerr and simplifies in nNHEK because 
$\rho=\rho'=0$ in that case. 

\subsection{Bilinear form}
\label{sec:bilinear}
In this paper, we will prominently use an invariant, conserved bilinear form between spin $s$ solutions of the homogeneous Teukolsky equation introduced in \cite{Green2022a}. 
For the convenience of the reader, we briefly recall the definition of this object.
The bilinear form \cite{Green2022a} is based on the $t-\phi$ reflection isometry $\mathcal J$ and a ``symplectic current'', $\pi^a$ \cite{Prabhu2018}. This current depends on a pair of GHP 
scalars ${}_{-s} \phi \circeq \GHPw{-2s}{0}$ and ${}_{s} \psi \circeq \GHPw{2s}{0}$ and is defined as \cite{Prabhu2018}
 \begin{equation}
 \label{pidef}
   \pi^a =  {}_s \psi (\Theta^a - 2sB^a) {}_{-s} \phi - {}_{-s} \phi (\Theta^a + 2s B^a)  {}_s \psi , 
 \end{equation}
 where $B^a \equiv -(\rho n^a - \tau \bar m^a) \circeq \GHPw{0}{0}$.
The corresponding bilinear form --- associated with a time slice $\mathscr C$ and 
formally similar to the Klein-Gordon inner product of a charged scalar field --- is
\begin{equation}
\label{Pidef}
\Pi_{\mathscr C}[ {}_s \psi, {}_{-s} \phi ] = \int_{\mathscr{C}} \pi^a [{}_s \psi, {}_{-s} \phi] \dd S_a \ . 
\end{equation}
The current $\pi^a$ is conserved whenever ${}_s \O {}_s \psi = 0 = {}_s \O^\dagger {}_{-s} \phi$ \cite{Prabhu2018}. 
In fact, its construction is precisely such that 
\begin{equation}
 \label{pidef1}
   \nabla_a \pi^a = {}_{-s} \phi  ({}_s \O {}_s \psi) -({}_s \O^\dagger {}_{-s} \phi) {}_s \psi.
\end{equation}
By Gauss' theorem, 
the bilinear form is therefore unchanged if we deform $\mathscr C$ locally. Furthermore, 
it follows from the intertwining relation \eqref{intertwJ}
that if ${}_s \O {}_s \psi=0$, then ${}_{-s} \phi:= \Psi_2^{-2s/3} \mathcal{J} {}_s \psi$, where $\mathcal J$ is the action of the $t$-$\phi$ reflection isometry of Kerr on GHP scalars (see App. \ref{discreteisometries}), solves ${}_s \O^\dagger {}_{-s} \phi = 0$. Consequently, the ``scalar product'',
\begin{equation}
\label{sprod}
\llangle {}_s \psi_1, {}_s \psi_2 \rrangle_{\mathscr{C}} := \Pi_{\mathscr C} \left[ {}_s \psi_1,  \Psi_2^{-\frac{2s}{3}} \mathcal{J} {}_s \psi_2 \right]
\end{equation}
is not only invariantly defined for any two GHP scalars of weights $\circeq \GHPw{2s}{0}$ satisfying the spin $s$ Teukolsky equation
${}_s \O {}_s \psi_i=0, i=1,2$, but is also unchanged under local changes of the time-slice $\mathscr C$ \cite{Green2022a}. 

We will write the scalar product as $\llangle {}_s \psi_1, {}_s \psi_2 \rrangle_{t}$ when referring to a constant $t$ slice in BL coordinates in Kerr, 
and as $\llangle {}_s \psi_1, {}_s \psi_2 \rrangle_{\bar t}$ when referring to a constant $\bt$ slice in nNHEK, see Eq. \eqref{scaledcts}. For solutions 
${}_s \O {}_s \psi_i=0, i=1,2$ with a sufficient decay towards the horizon and spatial infinity, the scalar product does not depend on either $t$ respectively 
$\bt$. 

In the case $\vert{s}\vert=2$, the bilinear form $\Pi$ has a
relation with the symplectic form $W$
of vacuum general relativity \cite{Iyer}
first pointed out in \cite{Prabhu2018}. Letting $\phi \circeq \GHPw{-4}{0}$ be any solution of $\O^\dagger \phi = 0$ and $h_{ab}$ be any solution to the linearized EE, the relation is 
\begin{equation}
    W[h,\S^\dagger \phi] = \Pi[\mathcal{W}h, \phi] + B[h,\phi]
\end{equation}
where $B$ is a  term associated with 
the boundary $\partial \mathscr C$ whose explicit form is given in \cite[App. A]{Green2022a}. For a pair $h_{ab},\phi$ such that this boundary term does not contribute, we may combine the above relationship between $\Pi$ and $W$, the relationship between $\Pi$ and the scalar product, and Eq. \eqref{P4} to obtain a useful interplay between $\thorn^4$ and the scalar product:
\begin{equation}
\label{p4T}
\begin{split}
    \llangle \psi, \thorn^4 \phi^* \rrangle &= \Pi[\zeta^4 \mathcal{J}  \psi, \thorn^4  \phi^*]\\
    &= 4\Pi[\zeta^4 \mathcal{J}  \psi, \mathcal{W} \S^{\dagger *}\phi^*]\\
    &=4 W[ \S^{\dagger} \zeta^4 \mathcal{J}  \psi,  \S^{\dagger *}\phi^*]\\
    &=4 W[ \S^{\dagger *} \zeta^{* 4} \mathcal{J}  \psi^*,  \S^{\dagger }\phi]^*\\
    &=4 \Pi[ \mathcal{W} \S^{\dagger *} \zeta^{* 4} \mathcal{J}  \psi^*,  \phi]^*\\
    &= \Pi[ \thorn^4 \zeta^{* 4} \mathcal{J}  \psi^*,  \phi]^*\\
        &= \Pi[\zeta^{-4} \mathcal{J} \zeta^4 \mathcal{J} \thorn^4  \mathcal{J} \zeta^{* 4} \psi^*,  \phi]^*\\
      &= \llangle \zeta^4 \thorn^{\prime 4} \zeta^{* 4} \psi^*,  \phi \rrangle^*,
        \end{split}
\end{equation}
where we assumed that $\O \psi = 0, \psi \circeq \GHPw{4}{0}$, and where we used $\mathcal{J}\thorn\mathcal{J}=\thorn', \mathcal{J}\zeta\mathcal{J}=\zeta$, as follows from the identities provided in App. \ref{discreteisometries}

Eq. \eqref{pidef1} and the intertwining relation \eqref{intertwJ} imply that, if $_{+2} \psi,\, _{+2} \eta \circeq \GHPw{4}{0}$ are such that $\O {}_{+2} \eta=0$ but 
{\it not} necessarily $\O {}_{+2} \psi =0$ , then
\begin{equation}
\label{dtid}
\frac{\dd}{\dd t} \llangle {}_2 \eta , {}_{2} \psi \rrangle_t = \int\limits_{ \{ x^{0} = t\} } (\zeta^4 \mathcal{J}  {}_2 \eta) ( \O {}_{2} \psi) \, T^a \dd S_a,
\end{equation}
where here and in the following, we use the shorthand $\zeta:=\Psi_2^{-\frac{1}{3}}$, and where $T^a$ is the normalized asymptotically timelike Killing field of Kerr 
given in BL coordinates by $T^a = (\partial_t)^a$.

Further properties of the scalar product \eqref{sprod} are \cite{Green2022a}: It is symmetric, real linear in each entry, and QNMs are orthogonal with respect to it. 
QNMs are exponentially growing near the horizon and infinity on a constant BL time $t$ slice, so the definition of the scalar product as 
an integral over such a slice needs to be done with care by introducing a regulator, see \cite{Green2022a, cannizzaro2024relativistic} and below in Sec. \ref{sec:mode}. The scalar product is {\it not} positive definite nor even real valued, reflecting in a sense the dissipative nature of the dynamics of Teukolsky's equations in Kerr. 

\section{The dynamical system for QNM amplitudes}

\subsection{Mode solutions}
\label{sec:mode}
Mode solutions to the Teukolsky equations are denoted by ${}_s \Upsilon_q \circeq \GHPw{2s}{0}$ in this paper , where $q$ stands for the collection of mode labels, usually $q=(\omega,\ell,m)$. 
They are by definition solutions to the homogeneous spin $s$ Teukolsky equation  ${}_s \O {}_s \Upsilon = 0$ for $s \ge 0$ and to the {\it adjoint} homogeneous spin $s$ Teukolsky equation  ${}_{-s} \O^\dagger {}_s \Upsilon = 0$ for $s \le 0$. The mode solutions may be given in separated form,
\begin{equation}
\label{eq:modes}
 {}_s\Upsilon_{\ell m\omega}(x^\mu) = e^{-i\omega t + i m \phi} \Rh(r) \Sh(\theta),
\end{equation}
where $m \in \mathbb Z$ and $\omega \in \mathbb C$, and where $x^\mu = (t,r,\theta,\phi)$ are the BL coordinates.
In the Kinnersley frame, the spin $s$ Teukolsky equation 
can be separated into an angular equation, 
\begin{widetext}
\begin{align}\label{eq:Sph eq}
  \left[\frac{1}{\sin \theta} \frac{ \dd}{\dd \theta}\left(\sin \theta \frac{\dd \,}{\dd \theta} \right) \right. \left. + \left( {}_s E_{\ell m}(a\omega) -  \frac{m^2+s^2+2 m s \cos \theta}{\sin^2 \theta} + a^2 \omega^2 \cos^2 \theta-2 a \omega s \cos \theta \right) \right] \Sh(\theta) = 0,
\end{align}
and a radial equation,
\begin{align}
\label{eq:radial}
   \left[ \Delta^{-s} \frac{\dd}{\dd r} \left( \Delta^{s+1} \frac{\dd}{\dd r} \right) \right. \left. + \left( \frac{H^2 - 2 i s (r-M)H}{\Delta} + 4 i s \omega r+2 a m \omega  -a^2\omega^2- {}_s E_{\ell m}(a\omega) +s(s+1) \right) \right] \Rh(r) = 0,
\end{align}
\end{widetext}
where $H := (r^2+a^2)\omega - a m$, and where ${}_s E_{\ell m}(a\omega)$ is a separation
constant \cite{Teukolsky1973,teukolsky1972rotating}. In our convention, it only depends on $|s|$.

In order for Eq.~\eqref{eq:modes} to represent a smooth GHP scalar, one has to impose that 
$\Sh(\theta)$ remain finite at the poles $\theta=0,\pi$. This leads to a discrete set of modes $\Sh$ and separation
constants ${}_s E_{\ell m}(a\omega)$ labeled by $\ell$ for each fixed 
real $\omega$ and $m$. Traditionally, the indexing is chosen such that
$\ell \in \mathbb Z^{\ge \max(|m|, |s|)}$. 
$\Sh(\theta)$ are referred to as spin-weighted spheroidal
harmonics~\cite{flammer2014spheroidal, Breuer1977,TorresdelCastillo2003}. We choose them such that, for $\omega \in \mathbb R$, we have ${}_s S_{\ell m \omega}(\theta)^* = {}_s S_{\ell m \omega}(\theta)$ and
\begin{equation}\label{eq:theta-orthogonality}
  \int_0^\pi \dd\theta\, \sin\theta \, {}_sS_{\ell m\omega}(\theta) {}_sS_{\ell'm\omega}(\theta) = \delta_{\ell\ell'}.
\end{equation}
The forms of the radial and angular equations in nNHEK will be recalled below in Sec. \ref{nNHEK_QNM}.

We now discuss boundary conditions of the radial equation in terms of 
the Kerr tortoise coordinate $\dd r_*= (r^2+a^2)/\Delta \dd r$. 
For fixed $s,\ell,m,\omega$ one considers the  solutions
${}_s R_{\rm in}$ and ${}_s R_{\rm up}$ fixed by the asymptotic conditions ($s \le 0$)
\begin{subequations}\label{eq:R bcs}
  \begin{align}
    &{}_s R_{\rm in}  \sim  \frac{e^{-ikr_*}}{\Delta^{s}}, \qquad  r_*\to-\infty,\\
    &{}_s R_{\rm up}  \sim  \frac{e^{i\omega r_*}}{r^{2s+1}}, \qquad  r_*\to\infty,
  \end{align}
\end{subequations}
where 
$k \equiv \omega-m\Omega_H$, where $\Omega_H=a/(2Mr_+)$ is the angular frequency of the outer horizon, and where we recall that the radii of the inner- and outer horizons (roots of $\Delta$) are denoted by $r_\pm$, respectively. The asymptotic conditions for $s>0$ are analogous to Eq. \eqref{eq:R bcs}, except that we chose a prefactor 
in such a way that \eqref{TS} holds:
 \begin{equation}
 \label{TS}
 \tfrac{1}{4} \th^4  \, {}_{-2} \Upsilon_{(- \omega^*) \ell (-m)}^{*} =  {}_{+2} \Upsilon_{\omega \ell m},
 \end{equation}
for both in and up modes.\footnote{See \cite{Ori} for the precise value of this prefactor. It is given below in Eq. \eqref{Ncoeff} for the scaling limit(s) that we require in this paper.} 
Indeed, proportionality already follows from 
Eq. \eqref{intertw3} because both sides are in the kernel of Teukolsky's operator $\O$ for spin $s=+2$, and both sides satisfy the same boundary conditions up to a constant. The choice of that constant fixes the precise prefactor in the boundary conditions in the case $s =-2$ in Eq. \eqref{eq:R bcs}.\footnote{The other $s$ values $0,1$ can be dealt with similarly but this is not relevant for this paper.}

The intertwining relations \eqref{intertwJ}, \eqref{intertwI} imply that the $t-\phi$ and $\theta$ reflection maps $\mathcal{J}$ and $\mathcal{I}$ map modes of 
a given boundary condition again to such a mode. Precisely, the formulas needed in this paper are: 
\begin{subequations}
\label{IJtrans}
\begin{align}
&\mathcal{I} ({}_{-2} \Upsilon_{\omega\ell m}^{\rm in, up})^* = {}_{-2} \Upsilon_{(- \omega^*)\ell(-m)}^{\rm in, up},\\
&\Psi_2^{-\frac{4}{3}} \mathcal{J} {}_{+2} \Upsilon_{\omega\ell m}^{\rm in, up} = {}_{+2} C{}_{-2} \Upsilon_{(-\omega)\ell (-m)}^{\rm out, down},
\end{align}
\end{subequations}
which may be demonstrated noting that both sides of each equation are annihilated by $\O^\dagger$ and satisfy the prerequisite boundary conditions \eqref{eq:R bcs}.
The second relation may be viewed as the definition of the out and down modes; in particular, it fixes the normalizations.

In more conventional terms, the relations \eqref{IJtrans} correspond to well-known symmetries of the radial and angular mode solutions which may easily be recovered from these relations by substituting the definitions of $\mathcal{J}$ and $\mathcal{I}$ in the Kinnersley 
frame. 

Eqs. \eqref{eq:R bcs} imply the
absence of incoming radiation from the past horizon and past null
infinity, respectively. Eqs. \eqref{eq:R bcs} are somewhat informal because 
they do not actually select uniquely a solution in the case ${\rm Im} \omega<0$: 
we may clearly add a multiple of the subdominant solution as $r_* \to \pm \infty$ and this will not change the asymptotic behavior. 
In the standard approach, mode solutions are typically obtained via series expansions \cite{leaver1986solutions}. These involving three-term recurrence relations for the series coefficients. 
If one selects a ``minimal solution'' in the sense of \cite{gautschi1967computational}, then the series representation converges at the horizon (in) or infinity (up), 
which is the actual technical definition of these modes.
Imposing both the in and up conditions simultaneously
 %
%
%
 determines the discrete set of QNM frequencies~\cite{leaver1986solutions} $\omega_{N\ell m} \in \mathbb{C}$.\footnote{To actually find the QNMs in practice, one may make the ansatz \cite{leaver1986solutions}
\begin{equation}
    R(r) = e^{i\omega r}(r-r_-)^{-1-s+i\omega+i\sigma_+}(r-r_+)^{-s-i\sigma_+} f(r) ,
\end{equation}
where $\sigma_+=(\omega r_+-am)/(r_+-r_-)$ and $f(r)=\sum_{n=0}^{\infty}d_n\left(\frac{r-r_+}{r-r_-}\right)^n$. In this ansatz, the coefficients $d_n$ are 
determined to be a ``minimal solution'' \cite{gautschi1967computational} to a three-term recursion relation~\cite{leaver1986solutions}. Then the series is uniformly absolutely convergent as $r\rightarrow\infty, r_+$, and thus characterizes the QNMs.}. We follow the labelling of \cite{leaver1986solutions}, where $N=0,1,2,\ldots$ are the so-called ``overtone'' numbers. The corresponding QNM mode functions will also be denoted by ${}_s \Upsilon_{N\ell m} \equiv {}_s \Upsilon_{\omega_N \ell m}$.  It is known that ${\rm Im}\, \omega_{N\ell m} < 0$ for QNM frequencies in Kerr \cite{Whiting1989}.
 
 As shown in \cite{Green2022a}, QNMs are orthogonal with respect to the scalar product defined in Sec. \ref{sec:bilinear}, i.e., there exists a constant ${}_s A_{N\ell m}$ such that
 \begin{equation}
 \label{norm}
 \llangle {}_s \Upsilon_{1}, {}_s \Upsilon_{2} \rrangle_t = {}_s A_{N_1\ell_1 m_1} \, \delta_{m_1m_2}\delta_{\ell_1\ell_2}
 \delta_{N_1N_2}, 
 \end{equation}
 where ${}_s \Upsilon_1 \equiv {}_s \Upsilon_{N_1 \ell_1 m_1}$ etc., and
where it is understood that an appropriate regulator must be chosen in order to define the integral over the constant BL time $t$ slice \cite{Green2022a}. 
This regulator can be implemented in different ways. In \cite{Green2022a}, the integration over the constant $t$ slice $\mathscr C$ entering the definition 
of the scalar product via Eq. \eqref{Pidef} was replaced by a complex contour, moving the radial BL coordinate $r$ into the complex plane (note that this procedure is 
unaffected by local changes of the complex contour since $\pi^a$ in Eq. \eqref{Pidef} is a conserved current). In this paper, we shall employ a ``minimal subtraction (MS)''
scheme \cite{cannizzaro2024relativistic}, which is formally equivalent. 

In the MS scheme, the integration over the constant $t$ slice in Eq. \eqref{Pidef} is first restricted to 
$r_+ + \delta < r < \delta^{-1}$, wherein $\delta>0$ is a regulator that we would like to take to zero. Eq. \eqref{Pidef}
is then replaced by
\begin{equation}
\label{Pidef1}
\Pi_{t}[ {}_s \psi, {}_{-s} \phi ] = {\rm F.P.}_{\delta \to 0} \int\limits_{\mathscr{C}(t,\delta)} \pi^a [{}_s \psi, {}_{-s} \phi] \, \dd S_a \ , 
\end{equation}
where the cutoff integration domain $\mathscr{C}(t,\delta) := \{x^0=t,r_+ + \delta < x^1 < \delta^{-1}\}$ 
is referring to BL coordinates $x^\mu = (t,r,\theta,\phi)$, and where $\mathrm{F.P.}$ means the finite part in a Laurent expansion. 
When ${}_s \psi, {}_{-s} \phi$ are solutions to the Teukolsky equations with appropriate analytic continuations in $r$ as 
described in \cite{Green2022a}, then taking the finite part is the same as a contour integral over an appropriate complex-$r$ contour. 
In the following, the scalar product \eqref{sprod} is understood with the MS regulated definition of \eqref{Pidef1}.

Eq. \eqref{p4T} implies a relationship 
between the normalization constants ${}_s A_{N\ell m}$ for QNMs
for opposite spins $s=\pm 2$. To see this, 
we use Eq. \eqref{TS} in the following sequence of equalities, for QNM labels $(\omega_1, \ell_1, m_1)$ and $(\omega_2, \ell_2, m_2)$:
\begin{equation}
    \begin{split}
         &16\llangle {}_{+2} \Upsilon_{\omega_1 \ell_1 m_1}, {}_{+2} \Upsilon_{\omega_2 \ell_2 m_2} \rrangle\\ 
         &= \llangle \thorn^4 {}_{-2} \Upsilon_{(-\omega^*_1) \ell_1 (-m_1)}^*, \thorn^4 {}_{-2} \Upsilon_{(-\omega^*_2) \ell_2 (-m_2)}^* \rrangle\\
         &= \llangle \zeta^4 \thorn'{}^{ 4} \zeta^{*4} \thorn^4 {}_{-2} \Upsilon_{(-\omega^*_1) \ell_1 (-m_1)},  {}_{-2} \Upsilon_{(-\omega^*_2) \ell_2 (-m_2)} \rrangle^*.
    \end{split}
\end{equation}

 It is well-known (see e.g., \cite[Ch.~5]{Price07}) that 
$\zeta^4 \thorn^{\prime 4} \zeta^{*4} \thorn^4$ is related to a Teukolsky-Starobinski (TS)  identity \cite{starobinskii1973amplification,teukolsky1974perturbations}, stating that, for a suitable TS constant ${}_2 D_{\omega \ell m}^2$, 
\begin{equation}
\label{DTS}
\begin{split}
  &\zeta^4 \thorn^{\prime 4} \zeta^{*4} \thorn^4 {}_{-2} \Upsilon_{\omega \ell m} \\
  =& \left(
  \zeta^4 \edth^{\prime 4} \zeta^{*4} \edth^4-
  9 M^{-\frac{2}{3}} \mathcal{L}_T^2 \right) {}_{-2} \Upsilon_{\omega \ell m}  
  \\
  =& \,
  ({}_2 D_{\omega \ell m})^2 {}_{-2} \Upsilon_{\omega \ell m}.
\end{split}
\end{equation}
In the second line, $\mathcal{L}_T$ is the GHP covariant Lie derivative \cite{edgar2000integration} with respect to $T^a = (\partial_t)^a$. The value of $({}_2 D_{\omega \ell m})^2$ may be found e.g., by making use of the representation in the second line and using the angular TS identities \cite{starobinskii1973amplification,teukolsky1974perturbations}. However, below, we will only need $({}_2 D_{\omega \ell m})^2$ for the nNHEK geometry, where a more direct route based on ladder operator properties of $\thorn,\thorn'$ in nNHEK, see App. \ref{radwave}, may be used. Either way, having determined $({}_2 D_{\omega \ell m})^2$, 
we find ${}_{+2} A_{N\ell m}$ via
\begin{equation}
\label{Apm}
    {}_{+2} A_{N\ell m} = \frac{1}{16}( {}_2 D_{(-\omega_N^*) \ell (-m)}^*){}^2 {}_{-2} A_{N\ell (-m)}^*.
\end{equation}

\subsection{Retarded Green's function and bilinear form}
\label{Greensf}

Below, we require the retarded Green's function $G^{\rm ret}$ for the spin $s=+2$ Teukolsky operator $\O \equiv {}_{+2} \O$.  A well-known  
mode expression for $G^{\rm ret}$ in BL coordinates $x^\mu= (t,r,\theta,\phi)$ is \cite{leaver1986spectral,casals2016high}
\begin{widetext}
\begin{equation}
\label{eq:Gretadv}
G^{\rm ret}(x,x') = \sum_{\ell,m}\; \int\limits_{-\infty + i0}^{\infty + i0} \dd \omega \, e^{-i\omega(t-t')} e^{im(\phi-\phi')}
g_{\omega \ell m}(r,r')
{}_{+2}S_{\omega \ell m}(\theta)   {}_{+2}S_{\omega \ell m}(\theta').
\end{equation}
\end{widetext}
Here and below, the sum stands for $\sum_{\ell = 2}^\infty \sum_{m=-\ell}^\ell$, and the response kernel is defined as
\begin{widetext}
\begin{equation}
\label{eq:g+}
    g_{\omega \ell m}(r,r') =  \, \frac{
    \Delta^2(r')}{
    W_{\omega \ell m}
    }\left[
\Theta(r-r')
{}_{+2}R_{\omega \ell m}^\textrm{in}(r')  {}_{+2}R_{\omega \ell m}^\textrm{up}(r)
    +
\Theta(r'-r)
{}_{+2}R_{\omega \ell m}^\textrm{in}(r)  {}_{+2}R_{\omega \ell m}^\textrm{up}(r')
    \right]
    .  
\end{equation}
\end{widetext}
We defined the step function as $\Theta(r)=1$ if $r\ge 0$, $\Theta(r)=0$ if $r<0$, and
the $\Delta$-scaled Wronskian $W \equiv W_{\omega \ell m}$ is defined by 
\begin{equation}
    W = \Delta^3 \left( 
    {}_{+2} R_{\rm up} \frac{\dd}{\dd r} {}_{+2} R_{\rm in} -
    {}_{+2} R_{\rm in} \frac{\dd}{\dd r} {}_{+2} R_{\rm up}
    \right)  .
\end{equation}
Properties of the radial and angular Teukolsky equations, notably the absence of QNMs in the upper complex frequency plane \cite{Whiting1989}, imply that the response kernel is analytic for $\Im \omega > 0$, which is reflected in the above choice of 
integration contour for $\omega$. The response kernel and spheroidal harmonics have poles and branch cuts in the lower complex $\omega$-plane. For $t>t_{\rm p}$ we may deform the integration contour for $\omega$ to a large semi-circle in the lower complex 
$\omega$-plane and a contour along the branch cut merging with $\omega=0$, at the expense of residue where the Wronskian 
$W \equiv W_{\omega \ell m}$ happens to vanish. These points in the complex $\omega$-plane correspond precisely to the QNM frequencies 
$\omega_{N\ell m}$. It is generally accepted that this procedure yields a decomposition \cite{leaver1986spectral}
\begin{equation}
G^{\rm ret} = G^{\rm qnm} + G^{\rm cut} + G^{\rm arc}.
\end{equation}
The pieces $G^{\rm arc}$ and $G^{\rm cut}$ are generally associated with the direct absorption by the black hole 
of gravitational waves emitted by a compact source respectively with the  ``Price tail'' \cite{price1972nonspherical}, respectively. In this work we assume to be in a dynamical 
regime where the latter two components are negligible, i.e., that one is in a dynamical era long before the Price tail but after the absorption. 
Thus, we shall generally approximate $G^{\rm ret}$ by the QNM piece $G^{\rm qnm}$, given by
\begin{widetext}
\begin{equation}
\label{eq:GretadvQNM}
G^{\rm qnm}(x,x') = 2\pi \sum_{\ell,m,N}\;  \frac{
    \Delta^2(r')}{
    \dd W_{\omega \ell m}/\dd \omega|_{\omega = \omega_{N\ell m}}
    } \, e^{-i\omega_{N\ell m}(t-t')} e^{im(\phi-\phi')}
{}_{+2}R_{N \ell m}(r)   {}_{+2}R_{N \ell m}(r')
{}_{+2}S_{N \ell m}(\theta)   {}_{+2}S_{N \ell m}(\theta').
\end{equation}
\end{widetext}
Now we consider the definition of the $t-\phi$ reflection $\mathcal J$ \eqref{Jdef}, \eqref{eq:boostParams}, the value 
$\Psi_2 = -M/(r-ia \cos \theta)^3$ in the Kinnersley frame, 
the QNM boundary conditions for $s=\pm 2$ \eqref{eq:R bcs}, and the relation between 
$ \dd W_{\omega \ell m}/\dd \omega|_{\omega = \omega_{N\ell m}} $ and the scalar product $_{+2} A_{N\ell m}$ \eqref{norm}
between the QNMs, see \cite[Lem. 5]{Green2022a}. Combining these, it follows that we can---usefully for later---represent $G^{\rm qnm}$ as
\begin{equation}
\label{Gqnm}
G^{\rm qnm}(x,x') = \sum_q \frac{1}{{}_{+2} A_q} \, {}_{+2} \Upsilon_q(x) \left( \zeta^4 \mathcal{J} \, {}_{+2} \Upsilon_q \right)(x'), 
\end{equation}
where from now on, we subsume all QNM indices into a multi-index
\begin{equation}
q = (N, \ell, m).
\end{equation}

\subsection{Dynamical system}

Consider a 1-parameter family of metrics depending on some parameter $\alpha$ of the form 
\begin{equation}
\label{gal}
g_{ab}(\alpha) = g_{ab} + \alpha h_{ab}(\alpha) ,
\end{equation}
where $g_{ab}$ is the metric of Kerr, and where $h_{ab}(\alpha)$ is a non-linear perturbation. 
We may think of $h_{ab}(\alpha)$ in terms of a formal perturbation series, 
\begin{equation}
h_{ab}(\alpha) \sim h^{(1)}_{ab} + \alpha h^{(2)}_{ab} + \dots
\end{equation}
though we will never consider the equations for the perturbation orders $h^{(n)}_{ab}$ individually 
in a naive perturbation theory, but instead work fundamentally with $h_{ab}(\alpha)$ up to a certain order.
Informally, we think of the $h^{(n)}_{ab}$ as being of order $O(1)$, and we think of $\alpha$ as small, but 
sufficiently large so as to induce weak non-linear effects when we impose the Einstein equation (EE)
$G_{ab}[g(\alpha)]=0$. 

In fact, we shall restrict ourselves to the leading effects of the non-linearity 
in the EE, which appear at $O(\alpha^2)$, and are described by the equation
\begin{equation}
\label{EE2}
\mathcal{E}_{ab}[h] = 8\pi \, \alpha \, \T_{ab}[h,h]. 
\end{equation}
Here, $\mathcal{E}_{ab}$ is the linear operator appearing in the linearized EE, 
\begin{align}\label{eq:linearE}
  \mathcal{E}_{ab}[h] \equiv \frac{1}{2}\Big[ &-\nabla^c\nabla_c h_{ab} - \nabla_a\nabla_bh + 2 \nabla^c\nabla_{(a} h_{b)c} \nonumber\\
 & + g_{ab}(\nabla^c \nabla_c h - \nabla^c\nabla^d h_{cd}) \Big],
\end{align}
whereas $8\pi \T_{ab}$ is minus the second order Einstein tensor, i.e.,
\begin{widetext}
\begin{equation}
\label{G2E}
\begin{split}
-8\pi \T_{cd}[h,h] = &-\frac{1}{2} (\nabla_b h^{ab}-\frac{1}{2} g^{ab}\nabla_b h^{})(2\nabla_{(d} h_{c)a)}^{}-\nabla_a h_{ cd}^{}) + 
\frac{1}{4} \nabla_c h^{ab} \nabla_d h^{}_{ab} \\
&+ \frac{1}{2} \nabla^b h^{a}{}_{c}(\nabla_b h^{}_{ad}-\nabla_a h^{}_{bd}) + \frac{1}{2} h^{ab}(\nabla_c \nabla_d h^{}_{ab}+\nabla_a \nabla_b h^{}_{cd}-2\nabla_{(d} \nabla_{|b|} h^{}_{c)a}).
\end{split}
\end{equation}
\end{widetext}
Eq. \eqref{EE2} is by itself of a comparable complexity as the full non-linear EE. We now describe how to turn it into a dynamical system for QNM amplitudes. 

First, in order to connect Eq. \eqref{EE2} to the Teukolsky formalism, we may employ the corrector tensor method (GHZ-approach) \cite{green2020teukolsky}. In the GHZ approach, 
one shows that -- in practice order-by-order in $\alpha$ --- that 
\begin{equation}
\label{GHZ}
g_{ab}(\alpha) = g_{ab} + \alpha  h_{ab}^{\rm IRG}(\alpha) + \alpha  x_{ab}(\alpha),
\end{equation}
modulo certain gauge pieces (which we ignore), and 
modulo certain algebraically special pieces (which we likewise ignore). Here, $h_{ab}^{\rm IRG}$ is the so-called reconstructed part in the ingoing radiation gauge (IRG), 
\begin{equation}
\label{hIRG}
   h_{ab}^{\rm IRG} = \Re \, \S^\dagger_{ab} \phi,
\end{equation}
where $\S^\dagger$ is given by Eq. \eqref{eq:Sdag}.
We should think of the Hertz potential $\phi(\alpha) \circeq \GHPw{-4}{0}$
and the corrector $x_{ab}(\alpha)$, as having formal expansions 
\begin{subequations}
\begin{align}
\phi(\alpha) =& \ \phi^{(1)} + \alpha \, \phi^{(2)} + O(\alpha^2)\\
x_{ab}(\alpha) =& \ \alpha \, x^{(2)}_{ab} 
+ O(\alpha^2),
\end{align}
\end{subequations}
noting that the corrector is zero at order $O(\alpha^0)$ \cite{green2020teukolsky,hollands2024metric}. Therefore, when the GHZ decomposition 
$h_{ab}(\alpha) = \Re [\S^\dagger \phi(\alpha)]_{ab} + x_{ab}(\alpha)$ is substituted into Eq. \eqref{EE2}, we may 
omit the corrector on the right hand side when working consistently to accuracy $O(\alpha^2)$ for the deviations off of Kerr. 
On the other hand, the corrector cannot be neglected on the left side of Eq. \eqref{EE2}, nor on the right side if we were to increase our 
accuracy to $O(\alpha^3)$. 

We now apply Teukolsky's source operator $\S$ [see Eq. \eqref{eq:S}] to Eq. \eqref{EE2}. As shown by \cite{green2020teukolsky,hollands2024metric}, 
$x_{ab}$ then drops out from the left side (to all orders in $\alpha$), and we obtain\footnote{Similar equations have appeared e.g., in \cite{ma2024excitation,Campanelli1999}. The main difference to our argumentation is that we maintain, in principle, control of the metric itself in the GHZ scheme \cite{green2020teukolsky,hollands2024metric}.}
\begin{equation}
\label{EE3}
\S \left\{\mathcal{E}\left[\Re(\S^\dagger \phi) \right] \right\} = 8\pi \, \alpha \, \S \left\{ \T\left[\Re(\S^\dagger \phi),\Re(\S^\dagger \phi)\right] \right\}. 
\end{equation}
The expression on the right side may further be simplified by using intertwining relations between the operators appearing in Teukolsky's equation and the TS identities; see App. \ref{discreteisometries}. One obtains 
\begin{equation}
\label{EE4}
\O( \th^4  \phi^*) = -16\pi \, \alpha \, \S \left\{ \T\left[\Re(\S^\dagger \phi),\Re(\S^\dagger \phi)\right] \right\}, 
\end{equation}
where $\O \equiv {}_{+2} \O$ is the $s=+2$ Teukolsky operator. For an essentially equivalent equation, see, e.g., \cite[eq. 48]{ma2024excitation}.

Eq.~\eqref{EE4} is a non-linear partial differential equation of order six for $\phi$. It does not appear to be of canonical type, so it is unclear whether it is amenable to a rigorous mathematical analysis or whether it is, in this form, practically useful. For us Eq. \eqref{EE4} will merely serve as the starting point for deriving our dynamical system for the QNM amplitudes. That dynamical system is not fully equivalent to Eq. \eqref{EE4} because we will consider only the QNM part (defined below) of $\phi$ as effectively contributing to the non-linear behavior. Our view is that even though Eq. \eqref{EE4} might not be amenable to a mathematical analysis, the truncation to the QNM part will capture relevant features of the weakly non-linear dynamics of the EE in our regime.

To this end, we first recall the retarded Green's function for the spin $s=+2$ Teukolsky operator $\O$ described in Sec. \ref{Greensf}. 
Defining $\psi := \frac{1}{4}\th^4 \phi^* \circeq \GHPw{4}{0}$ and using the Green's function property of $G^{\rm ret}$, we have (here $T^a = (\partial_t)^a$ is the asymptotically 
timelike normalized Killing field of Kerr)
\begin{widetext}
\begin{equation}
\label{G2}
\begin{split}
\psi(x) =& \int_{\sM} G^{\rm ret}(x,x') \O\psi(x') \dd V' \\
=& \int\limits_{-\infty}^t \dd t' \int\limits_{ \{ x^{\prime 0} = t'\} } G^{\rm ret}(x,x') \O\psi(x') T^{a \prime}\dd S_a'\\
\sim& \int\limits_{-\infty}^t \dd t' \int\limits_{ \{ x^{\prime 0} = t'\} } G^{\rm qnm}(x,x') \O\psi(x') T^{a \prime}\dd S_a'\\
\end{split}
\end{equation}
\end{widetext}
In the last step we have approximated the retarded Green's function by its QNM part, see Sec. \ref{Greensf}. This should be understood
as a condition on $\psi$, and therefore indirectly on the solution to EE and the regime that we consider. Effectively, we 
are assuming to be in an era where the solution can be described by the non-linear dynamics of QNMs, i.e. after 
the direct emission \textcolor{red}{of} gravitational waves exciting the spacetime, but long before the late-time tail behavior kicks in. 

At this stage, we substitute our previous expression \eqref{Gqnm} for $G^{\rm qnm}$. 
Setting 
\begin{equation}
\label{cdef}
c_q(t) := \llangle {}_{+2} \Upsilon_q, \psi \rrangle_t, 
\end{equation}
we thereby learn from relation \eqref{G2} that 
 \begin{equation}
 \psi(x) \sim \sum_q \frac{1}{{}_{+2} A_q} \ {}_{+2} \Upsilon_q(x) \int\limits_{-\infty}^t \dd t' \, \frac{\dd}{\dd t'} c_q(t'),
 \end{equation}
 where ${}_{+2} \Upsilon_q$ are the spin $s=+2$ QNMs, and where ${}_{+2} A_q$ are the 
 norms of their scalar product, see Eq. \eqref{norm}, and $\S^\dagger_{ab}$ is given in Eq. \eqref{eq:Sdag}. If we assume that, initially, the overlap \eqref{cdef}
 between $\psi$ and a QNM is small, we can neglect the lower boundary of the $t'$-integration and write
  \begin{equation}
  \label{Psisum}
    \psi \sim \sum_q \frac{1}{{}_{+2} A_q} c_q(t) \ {}_{+2} \Upsilon_q ,
 \end{equation}
 which expresses $\psi(x)$ as a sum of QNM of the homogeneous Teukolsky equation with time-dependent amplitudes, $c_q(t)$. 
 
 Our aim is to derive a dynamical system for these amplitudes. We will obtain this system from Eq. \eqref{EE4}. We first 
 need to obtain a relation similar to \eqref{Psisum} for $\phi$, where $\tfrac{1}{4}\th^4 \phi^* = \psi$.  Eq. \eqref{TS} implies that 
 \begin{equation}
  \label{Phisum}
  \begin{split}
\phi 
&= \sum_q \frac{1}{{}_{+2}  A_q^*} c_q^*(t) {}_{-2} \Upsilon_{-q^*} + O(\alpha),
\end{split}
 \end{equation}
 because if we apply $\th^4$ to this expression and use Eq. \eqref{TS}, then we find $\frac{1}{4}\th^4 \phi^* = \psi+ $(terms containing at least one time derivative of $c_q(t)$). 
 Such terms are of order $O(\alpha)$ by the dynamical equation and so may be neglected self-consistently at our approximation level. We have used a transformation formula \eqref{IJtrans}, using the notation
 
 \begin{equation}
 -q^* = (-\omega_{N\ell (-m)}^*, \ell, -m).
 \end{equation}
 We are now ready to derive this system: We start by taking the $t$-derivative of $c_q(t)$, see Eq. \eqref{cdef}, into which we substitute Eq. \eqref{dtid}. We obtain 
\begin{equation}
\frac{\dd}{\dd t} c_q(t) \sim \int\limits_{ \{ x^{0} = t\} } {}_{-2}\Upsilon_{-q} ( \O \psi) \, T^a \dd S_a, 
\end{equation}
using again a transformation formula \eqref{IJtrans}, and the notations ${}_{-2} \Upsilon_{-q} := \zeta^4 \mathcal{J} _{+2}\Upsilon_q$ (satisfying 
anti-QNM boundary conditions), where 
 \begin{equation}
 -q = (-\omega_{N\ell (-m)}, \ell, -m).
 \end{equation}
For $\O \psi$, we next substitute Eq. \eqref{EE4}, and then we substitute \eqref{Phisum}. We thereby get
\begin{widetext}
\begin{equation}
\label{G3}
\begin{split}
\frac{\dd}{\dd t} c_{q_1}(t) \sim& -4\pi \alpha \sum_{q_1,q_2} \int\limits_{ \{ x^{0} = t\} } T^a \dd S_a \times \\
&{}_{-2} \Upsilon_{-q_1}  \ \S \left\{ \T\left[\Re\left(\frac{1}{{}_{+2} A_{q_2}^* }\S^\dagger  \{c^*_{q_2}(t) _{-2} \Upsilon_{- q_2^*}\}  \right),\Re \left(
\frac{1}{{}_{+2} A_{q_3}^*}\S^\dagger\{ c_{q_3}^*(t) _{-2} \Upsilon_{- q_3^*}\} \right) \right] \right\}.
\end{split}
\end{equation}
\end{widetext}
The operators $\S^\dagger, \S, \T$ contain $t$ derivatives, but when these hit a coefficient $c_{q_2}(t)$ or $c_{q_3}(t)$, we may substitute the expression for this derivative
and get a term on the right side that is at least of order $O(\alpha^2)$. Such a term may be neglected self-consistently at our level of approximation. We may therefore pull out $c_{q_2}(t)$ or $c_{q_3}(t)$ and write the above equation in a neater form. For this, set 
\begin{widetext}
\begin{equation}
\label{G4}
\begin{split}
U_{123}(t) =& -   \frac{\pi \delta_{m_1,m_2+m_3}}{A_{2}  A_{3}} \int\limits_{ \{ x^{0} = t\} }  
 {}_{-2} \Upsilon_{-q_1} \ 
 \S \T\bigg[(\S^\dagger \bar {}_{-2} \Upsilon_{-q_2^*})^* , (\S^\dagger \bar {}_{-2} \Upsilon_{-q_3^*})^* \bigg]  T^a \dd S_a , \\
X_{123}(t) =& -   \frac{\pi \delta_{m_1,-m_2-m_3}}{ A_{2}^* A_{3}^*} \int\limits_{ \{ x^{0} = t\} }  {}_{-2} \Upsilon_{-q_1} \ 
\S  \T\bigg[\S^\dagger {}_{-2} \Upsilon_{-q_2^*} , \S^\dagger  {}_{-2} \Upsilon_{-q_3^*} \bigg]  T^a \dd S_a , \\
V_{123}(t) =& -   \frac{\pi \delta_{m_1,-m_2+m_3}}{A_{2}^*  A_{3} } \int\limits_{ \{ x^{0} = t\} }  
 {}_{-2} \Upsilon_{-q_1} \ 
 \S  \T\bigg[\S^\dagger  {}_{-2} \Upsilon_{-q_2^*} , ( \S^\dagger {}_{-2} \Upsilon_{-q_3^*})^* \bigg]  T^a \dd S_a  \\
&+ (q_2 \leftrightarrow q_3)
\end{split}
\end{equation}
\end{widetext}
using condensed notations such as $A_{1} \equiv {}_{+2} A_{q_1} \equiv {}_{+2} A_{N_1\ell_1 m_1}$.
These coefficients are defined in a GHP invariant way and can be computed in any frame, e.g. the Kinnersley frame. 
The selection rules implied by the Kronecker $\delta$'s in the magnetic quantum numbers $m_i$ are a consequence of the axisymmetry of the spacetime and the harmonic dependence $e^{\pm im\phi}$ etc. of the modes involved in the expressions.

Consistently neglecting contributions of $O(\alpha^2)$ to the 
excitation coefficients $c_q$, the dynamical system is 
\begin{equation}
\label{dyneq}
\frac{\dd}{\dd t} c_1 = \alpha \sum_{2,3} \left(U_{123} c_{2} c_{3} + V_{123}c_{2} c^*_{3}\right),  
\end{equation}
using condensed notations such as  $c_{1} \equiv c_{q_1} \equiv c_{N_1\ell_1 m_1}$. Based on Eq. \eqref{G4}, one might have expected the appearance of a term of the form $X_{123}c^*_2 c^*_{3}$. But the structures of the operators $\S^\dagger, \S, \T$ imply that $X_{123}=0$.\footnote{This may also be seen e.g., from \cite{ma2024excitation}, in the formula after Eq. (50), when using Eqs. (39), (43) and (46). The point is that here the Hertz potential appears with a complex conjugation and this leads to the absence of some terms.} 

Eq. \eqref{dyneq} is the main result of the section.
We expect that, in the dynamical range considered, a solution to this system will be accurate up to and including $O(\alpha)$. 
Then, substituting the values of the amplitudes $c_q$ as functions of $t$ into Eq. \eqref{Psisum}, we obtain $\phi(x)$, which we thereby expect to 
give an approximation up to and including $O(\alpha^2)$
of the metric through Eq. \eqref{GHZ}. More explicitly, 
let us define the reconstructed part of the metric perturbation
in Eq. \eqref{GHZ} as 
 \begin{equation}
  \label{Phisum2}
h_{ab}^{\rm IRG} 
= \sum_q \Re \S^\dagger_{ab} \left[  \frac{1}{{}_{2}  A_q^*} c_q^*(t) {}_{-2} \Upsilon_{-q^*} \right]. 
\end{equation}
After that, we define the GHZ corrector $x_{ab}$ as 
\begin{equation}
\label{xrecon}
    x_{ab} = \int\limits H_{ab}{}^{a'b'}
    \T_{a'b'}[h^{\rm IRG}, h^{\rm IRG}] \ \dd V'
\end{equation}
where $H_{ab}{}^{a'b'}(x,x')$ is the Green's function of the GHZ transport equations \cite{casals2024spin}. 
With this, the metric up to and including order $O(\alpha^2)$
is given by Eq. \eqref{GHZ}. We expect Eq. \eqref{GHZ} to give a good approximation of the metric in a spacetime region where the non-linear perturbation of Kerr can be considered as dominated by QNMs. 

Concretely, the structure of the ``overlap coefficients'' in Eq. \eqref{G4} is excessively complicated, since each operator $\S, \S^\dagger$ has a large number of terms when written out 
completely and explicitly going back to the definitions of the GHP operators in \eqref{eq:S}, \eqref{eq:Sdag}, 
as has the quadratic Einstein tensor $\T_{ab}$, see Eq. \eqref{G2E}. Nevertheless, they can, in principle, be computed numerically given a reasonable approximation for the QNMs. 

We will not attempt doing this here, but will instead consider in the next sections 
the near extremal regime $\varepsilon \ll 1$, in which these expressions, while still involved, 
simplify considerably. The simplifications include 
(a) the overlap coefficients become time independent, 
(b) the number of terms is reduced significantly, and 
(c) there appear selection rules in the sum over the QNM frequency labels $q_1, q_2, q_3$ which are not apparent.  

\section{QNMs and bilinear form in nNHEK}
\label{nNHEK_QNM}

\subsection{Matched asymptotic expansion}
It has been observed that, when the Kerr black hole is nearly extremal, there appears a sequence of long-lived QNMs \cite{DetweilerModes,andersson2000superradiance}. Their frequencies and the corresponding mode functions 
can be analyzed by a matched asymptotic expansion approach initiated by \cite{teukolsky1974perturbations}, and further developed and applied in this context by \cite{DetweilerModes, sasaki1990gravitational, Hod:2012bw, yang2013branching, Yang2013,sasaki1990gravitational,casals2016horizon,Gralla:2018xzo} and others. Some of the results of this type of analysis have been corroborated by rigorous mathematical investigations \cite{joykutty2022existence, hintz2021quasinormal, hintz2022quasinormal, gajic2021quasinormal}.

In the following, we will make the assumption that these long-lived modes give the dominant contribution to the dynamical evolution of the metric for a parametrically large time in the weakly non-linear regime. The matched asymptotic expansion analysis in this section will therefore be used below to simplify our dynamical equation \eqref{dyneq} for the excitation amplitudes of the QNMs.



In the matched asymptotic expansion approach which we now recall, one solves the radial equation in two overlapping asymptotic regions and matches the solutions in the region of overlap such that the boundary conditions are satisfied. For simplicity, we restrict to QNMs which are not axisymmetric $(m \neq 0)$, though a variant of the analysis applies also to the case $m=0$, see App. \ref{axisym} for further discussion. One introduces a dimensionless frequency parameter
\begin{align}
\label{bomom}
\bar\omega=\frac{ 2M}{\varepsilon}\left(\omega-\frac{m}{2M}\right),
\end{align}
such that 
\begin{equation}
e^{-i\omega t+im\phi}=e^{-i\bar\omega \bt+ im\bar\phi}.
\end{equation}
We make the approximation
\begin{align}\label{eq:approxes}
\varepsilon \ll 1 
\end{align}
but place no restriction on the size of $\bar\omega$. Given that $\Omega_H = 1/(2M)$ for an extremal Kerr black hole, this scaling basically amounts to the statement that $|\omega - m\Omega_H| = O(\varepsilon)$, i.e., one is considering frequencies near the superrradiant 
threshold. With these quantities, we define the following asymptotic regimes [recall the definition of $x = (r-r_+)/r_+$]
\begin{align}\label{eq:MAE}
    \textrm{near-zone: } & x \ll 1 \\
    \textrm{far-zone: } & x \gg  \varepsilon \bar \omega \\
    \textrm{overlap region: } &  \varepsilon \bar \omega  \ll x \ll 1.
\end{align}  
The overlap region corresponds to the intermediate scaling $x \sim \varepsilon^p$ for some chosen $0<p<1$ e.g., $p=1/2$. 

At the leading order in this approximation, the spin-weighted spheroidal functions are now evaluated $\omega=m\Omega_H=m/(2M)$,  greatly simplifying the analysis. We denote this leading-order contribution to the angular eigenfunctions by 
\begin{equation}
	{}_s S_{\ell m} := {}_s S_{\omega \ell m}\vert_{\omega=\frac{m}{2M}} \label{maxspheroidal}
\end{equation}
where ${}_s S_{\omega \ell m}$ are the standard spin-weighted spheroidal harmonics
\cite{Breuer1977, chandrasekhar1998mathematical,flammer2014spheroidal}. 
The angular equation becomes
\begin{widetext}
\begin{align}\label{eq:SphNHEK}
  \left[\frac{1}{\sin \theta} \frac{ \dd}{\dd \theta}\left(\sin \theta \frac{\dd \,}{\dd \theta} \right) \right. \left. + \left( {}_s E_{\ell m}  -  \frac{m^2+s^2+2 m s \cos \theta}{\sin^2 \theta}  + \frac{m^2}{4} \cos^2 \theta 
  -m s \cos \theta \right) \right]{}_s S_{\ell m} = 0,
\end{align}
\end{widetext}
where ${}_s E_{\ell m} := {}_s E_{\ell m}(a\omega = m/2M)$ is a separation constant, determined again by demanding regularity of the solutions at $\theta = 0,\pi$.
The corresponding radial functions in the far and near asymptotic regions will be denoted by $R^{\rm far}$ and $R^{\rm near}$, respectively. The full mode solutions are correspondingly denoted by 
\begin{subequations}
    \begin{align}
    \label{nearQNM}
        {}_s \Upsilon_{\omega \ell m}^{\rm near}(\bx^\mu)
        =&\  {}_s 
        R^{\rm near}_{\omega \ell m}(\bx)
        {}_s S_{\ell m}(\bar \theta) e^{-i\bar \omega \bt + im\bphi}\\
        {}_s \Upsilon_{\omega \ell m}^{\rm far}(x^\mu)
        =&\  {}_s 
        R^{\rm far}_{\omega \ell m}(x)
        {}_s S_{\ell m}(\theta) e^{-i\omega t + im\phi},
    \end{align}
\end{subequations}
where we note that the two solutions refer to different coordinates, see Eqs. \eqref{scaledcts}.

\subsection{Near-zone solution}\label{sec:near sol}
In the near horizon limit, we first perform a change of variables in the Teukolsky master equation to the coordinates $\bar x^\mu = (\bt, \bx, \bar \theta, \bphi)$ of Eq. \eqref{scaledcts}. Then we transform the master field 
as a GHP scalar of weights $\circeq \GHPw{2s}{0}$, as in Eq. \eqref{GHPtrafo}, 
when we apply the $\varepsilon$-dependent boost to the Kinnersley frame resulting in the frame \eqref{nNHEK_NP} in the limit as $\varepsilon \to 0$. In this frame, coordinates, and limit keeping $\bar x^\mu$ fixed, we drop the subleading terms of order $O(\varepsilon)$ in the potential in the radial Teukolsky 
equation \ref{eq:radial}, resulting in 
\begin{equation}
\left[ f^{-s} \frac{\mathrm{d}}{\mathrm{d} \bx}\left(f^{s+1} \frac{\dd}{\dd \bx} \right)-{}_s V_{k\ell m}^{\rm near}(\bx) \right] {}_s R^{\rm near}_{k\ell m}=0, 
\label{radialeqnnear}
\end{equation}
where $f=\bx(\bx+2)$,
\begin{equation}
k := \bar\omega+m.
\end{equation}
and the potential is given by
\begin{widetext}
	\begin{equation} 
\label{Vklm}
	{}_s V_{k \ell m}^{\rm near}(\bx)=-\frac{3}{4}m^2-s(s+1)+{}_s E_{\ell m}-2ism+\frac{(m\bx +k)(2is-k+2is\bx -m\bx)}{f}.
	\end{equation}
\end{widetext}

The near-zone solution which is ingoing at the horizon is given by the hypergeometric function \cite{Compere2018}  
\begin{widetext}
\begin{equation}
\label{nearradial}
{}_{s} R^{\rm near}_{\rm in} 
	= {}_{s} C \ {{\bx}}^{-s-\frac{i k}{2}} \left(\frac{{\bx}}{2}+1\right)^{-s+i \left(\frac{k}{2}-m\right)} {}_{2}F_1\left({}_{s} h_{+}-i m-s,{}_{s}h_--i m-s;1-i k-s;-\frac{{\bx}}{2}\right).
	\end{equation}
\end{widetext}
where ${}_s C_{\omega \ell m}$ is a prefactor 
that is dictated by our normalization conventions for the modes. 
It is ${}_{-2}C_{\omega\ell m}=1$, 
which using the ladder operator method for $\thorn^4$ described in App.\ref{radwave} implies that 
\begin{equation}
\label{Ncoeff}
    {}_{+2}C_{\omega\ell m}= \frac{(-1)^m}{4}\prod_{j=0}^3\left[-i k+(2-j)\right].
\end{equation} 
${}_s h_\pm$ is given by 
\begin{equation}\label{eq:hpm}
	{}_s h_\pm =\frac{1}{2} \pm \frac{1}{2} {}_s \eta, \quad {}_s \eta_{\ell m} \equiv \sqrt{1-7 m^2+4 {}_s E_{\ell m}},  
\end{equation}
noting that ${}_s h_\pm = {}_{-s}h_\pm$.\footnote{We will often omit the indices $(s,\ell,m)$ from ${}_s h_{\pm \ell m}$ to lighten the notation in the following.}
The asymptotic behaviors are 
\begin{align}
R_{\rm in}^{\rm near} \sim 
{}_{s}C_{\omega\ell m} \bar x^{-\frac{i k}{2}-s}, \qquad ~ \bx\to 0\,,\label{Rnearin0} 
\end{align}
at the horizon, and 
\begin{equation}
\label{eq:near in buffer}
{}_s R_{\rm in}^{\rm near} \sim {}_{s}C\left({}_s a_-\, \bar x^{-h_--s}+ {}_s a_+\, \bar x^{-h_+-s}\right), \qquad~ 
\bar x \to \infty\ , 
\end{equation}
at infinity (the buffer region). Here, the asymptotic coefficients are given by

\begin{equation}
\begin{split}\label{eq:Apm}
{}_sa_+ & =\frac{2^{h_+-\frac{ik}{2}}\Gamma(1-2h_+)\Gamma(1-ik-s)}{\Gamma(1-h_+-im-s)\Gamma[1-h_+-i(k-m)]}, \\
{}_sa_- &= {}_{s}a_+\vert_{h_+\to h_-}.
\end{split}
\end{equation}

\subsection{Far solution}\label{sec:far sol}
The far-zone radial equation is given by the separated 
Teukolsky equation \eqref{eq:radial} in extremal Kerr 
at $\omega=m\Omega_H = m/(2M)$, keeping only 
the dominant terms in the regime $x \gg \varepsilon \bar \omega$, see e.g., \cite[A.4]{casals2016horizon} for $s=0$. 
For general $s$, the equation is 

\begin{equation}
\left[ (x^2)^{-s} \frac{\mathrm{d}}{\mathrm{d} x}\left( (x^2)^{s+1} \frac{\dd}{\dd x} \right)-{}_s V_{\ell m}^{\rm far}(x) \right] {}_s R^{\rm far}_{\ell m}=0, 
\label{radialeqn}
\end{equation}
(keeping  in mind the definition \eqref{xdef} of $x$),
where the potential is given by
\begin{widetext}
	\begin{equation}
\label{Vklmfar}
	{}_s V_{\ell m}^{\rm far}(x)=
    -m\left[\frac{1}{4} m(x+2)^2 +is(x+2) + \frac{3m}{4}
    -2is \right] + {}_s E_{\ell m}-s(s+1).
	\end{equation}
\end{widetext}
The solution which is outgoing at infinity is given by a combination of confluent hypergeometric functions, $M(a,b;x)$ \cite[Sec. 13.14]{nist}. It is
\begin{equation}
\label{Rfar}
\begin{split}
{}_s R^{\rm far}_{\rm up} =&  e^{- \frac{imx}{2}} \left[ x^{-h_--s} M(1-h_-+i m - s,2h_+;imx)\right. \\ 
&+\left. {}_s Q x^{-h_+-s}  M(1-h_++i m - s,2h_-;imx)\right],
\end{split}
\end{equation} 
with
\begin{equation}\label{eq:script }
{}_s Q \equiv (-im)^{h_+-h_-} \frac{ \Gamma(2h_--1)\Gamma(h_+-im+s)}{\Gamma(2h_+-1) \Gamma(h_--im+s)}.
\end{equation}
This ratio defines the far ``up'' solution up to an overall normalization.

\subsection{Near horizon QNMs}\label{sec:NHMs}

A QNM is by definition a frequency $\omega$ at which the solutions are purely outgoing at infinity and ingoing at the horizon, see e.g., \cite{Berti:2009kk,Kokkotas:1999bd} for reviews on QNMs. In the matched asymptotic expansion, this occurs at the frequencies where $R_{\rm in}^{\rm near}$ matches onto the far-zone outgoing solution, or $a_-/a_+=Q$.  Explicitly, the QNM condition is found \cite{DetweilerModes} to be
\begin{widetext}
\begin{align}
\label{eq:QNM condition}
\frac{\Gamma^2(h_+-h_-) \Gamma(h_--im-s)\Gamma(h_- -im+s)\Gamma\left(h_- -i(k-m)\right)}{\Gamma^2(h_--h_+) \Gamma(h_+-im-s)\Gamma(h_+-im+s)\Gamma\left(h_+-i(k-m)\right)}(- i m \varepsilon)^{h_--h_+}=1,
\end{align}
\end{widetext}
where we recall that $m\neq 0$ (the axisymmetric modes are treated separately in App. \ref{axisym}).

Condition \eqref{eq:QNM condition} was analyzed by 
\cite{sasaki1990gravitational} for $\ell = m$,
by \cite{Hod:2012bw} for ${}_s \eta_{\ell m} \in i\mathbb{R}$ [see Eq. \eqref{eq:hpm}] and later by \cite{yang2013branching, Yang2013} for general ${}_s \eta_{\ell m}$, thereby clarifying some aspects of the analysis by \cite{Hod:2012bw} pertaining to the distinction between ``damped QNMs'' and ``zero-damped QNMs'', see Sec. \ref{largeell}.  

Consider first a real $h_+ > 1/2$ [see Eq. \eqref{eq:hpm}], and $m>0$ -- the case $m<0$ 
can be obtained from the symmetry $\omega_{N\ell m} = -\omega_{N(-m)\ell}^*$. If we assume e.g., that $m\varepsilon<1$,
as will certainly be the case e.g., if $\ell \varepsilon \ll 1$, and as we will be assuming in the following, the quantity $(-im\varepsilon)^{h_--h_+}$ is growing as $(m\varepsilon)^{1-2h_+}$. To compensate this in Eq. \eqref{eq:QNM condition}, either the argument of the gamma function $\Gamma(h_+-i(k-m))$, or of $\Gamma(h_--h_+)$ must land parametrically close, in $m \varepsilon \ll 1$, to a pole i.e., a non-positive integer $-N$. 

However, it turns out that, if $h_--h_+ \approx -N$ is parametrically close, in $m \varepsilon  \ll 1$, to a non-positive integer $-N$, then the two linearly independent near zone solutions degenerate, which invalidates the derivation of Eq. \eqref{eq:QNM condition} \cite{yang2013branching}. Indeed, \cite{yang2013branching} found no long lived QNMs in 
this regime. Below (see Sec. \ref{largeell}), we will consider a scaling regime 
in which $\ell \gg 1$, but still $\ell \varepsilon \ll 1$. 
In such a regime, we have Eq. \eqref{Eexpand}, implying 
by Eq. \eqref{eq:hpm} that $h_+ = \ell+1-15m^2/(16 \ell) + O[(m^2/\ell)^2]$. Thus, in that scaling regime, it is 
impossible for $h_--h_+=1-2h_+$ to be parametrically close, in $m \varepsilon  \ll 1$, to a non-positive integer. 

On the other hand, if $h_+-i(k-m)$, but not $h_--h_+$, is parametrically close to a non-positive integer $-N$, then the analysis remains valid, suggesting that we obtain $k$, and thereby the scaled QNM frequency \eqref{bomom} $\bar \omega = k-m$, up to an error of order $O(\varepsilon^{2h_+-1})$. 

Actually, for a more honest estimate of the error, we should recall that we made a simplification \eqref{maxspheroidal} setting $a\omega := m/(2M)$ in the angular equation \eqref{eq:Sph eq}, anticipating that $\omega = m\Omega_H + O(\varepsilon)$, and thereby effectively ignoring order $O(\varepsilon)$-corrections in ${}_s E_{\ell m}(a\omega) = {}_s E_{\ell m}|_{a\omega = m/2M} + O(\varepsilon)$. Including these as perturbative corrections to the angular equation \eqref{eq:Sph eq} using standard results in perturbation theory of self-adjoint operators \cite{kato2013perturbation}, 
we can, in fact, only claim that we determined $h_{\pm}$, hence the scaled QNM frequency \eqref{bomom} $\bar \omega = k-m$, up to an error of order $O(\varepsilon)$.

For complex $h_\pm$, a similar argument can be made, provided that $(-i)^{h_--h_+} = e^{\pi|h_+-h_-|}$
becomes large, as will be the case e.g., in a large $\ell$ limit and for $m/(\ell+1/2) > 0.74...$ \cite{Hod:2012bw, yang2013branching}. For complex 
$h_\pm$ but $|h_+-h_-|$ not necessarily large, 
the argument appears to be more subtle \cite{yang2013branching}, but mathematical analysis \cite{joykutty2022existence} suggest that the second case of 
\begin{equation}
\label{Hod}
\bar\omega_{N\ell m} =  
\begin{cases}
-i(h_++N) + O(\varepsilon),  & \text{if $h_+ \in \mathbb R_{+}$}\\
-i (h_++N) + o(1), & \text{if $h_+ \in 1/2+ i \mathbb{R}$}
\end{cases}
\end{equation}
where $N$ is a non-negative integer, is still correct. 

$N$ is called the overtone number. Accordingly, we label the QNM solutions of the radial equation \eqref{radialeqnnear} that we consider in this article as ${}_{s}R^{\rm near}_{\omega_N \ell m}\equiv {}_{s}R^{\rm near}_{N \ell m}$. Actually, in our applications, we will consider below the regime $\ell \gg m^2$, in which case  $h_\pm$ is seen to be real and parametrically large, $h_\pm \sim \ell$, i.e. we will be in the first case; see Sec.~\ref{largeell}.  

\subsection{Bilinear form for near-horizon modes}
\label{sec:BilinearNnHEK}
Here we compute the scalar product of two spin $s=-2$ QNMs [see Eq.~\eqref{nearQNM}] $\Upsilon_1 = {}_{-2} \Upsilon_{N_1\ell_1 m_1}$, $\Upsilon_2 = {}_{-2} \Upsilon_{N_2\ell_2 m_2}$, assuming that $m_1,m_2$ both are nonzero, see App.~\ref{axisym} for the treatment of the remaining case of axisymmetric modes. 
In the matched asymptotic expansion, where the radial and angular integrations decouple (see App.\ref{normalization}), it is natural to split the bilinear form into near and far zone contributions
\begin{equation}
\label{eq:near-far split}
\llangle \Upsilon_1, \Upsilon_2 \rrangle = \llangle \Upsilon_1, \Upsilon_2 \rrangle_{\rm near} + \llangle \Upsilon_1, \Upsilon_2 \rrangle_{\rm far}. 
\end{equation}
We may choose to split the integral in the scalar product e.g., with the intermediate scaling $x=c\sqrt{4\varepsilon}$, $c>0$ (or equivalently, $\bx = c/\sqrt{4\varepsilon}$) where the two solutions match. 
We may use the orthogonality of the spheroidal functions ${}_{-2} S_{\ell m}$ to put
 the QNM frequencies $\bar \omega_{N_1\ell_1 m_1}, \bar \omega_{N_2\ell_2m_2}$ [see Eq. \eqref{Hod}] at the same $\ell$ and $m$. 
 We next use relations (15.8.1) and (15.2.4) of \cite{nist} to express the near solution in terms of a finite polynomial
\begin{align}
{}_{-2}R_{N\ell m}^{\rm near} = \ &\bx^{2-\frac{i k_N}{2}} \left(1+\frac{\bx}{2}\right)^{-i \left(\frac{k_N}{2}-m\right)} \times  \no 
&\sum_{j=0}^N {}_{-2} P^{(N)}_j \left(-\frac{\bx}{2}\right)^j \label{radialwaves}
\end{align} 
where $k_N=\bar \omega_{N\ell m}+m$ [see Eq. \eqref{Hod}], 
\begin{equation}
 {}_{-2} P_j^{(N)}= \frac{ (-N)_j(1-2h_+-N)_j}{j! (1-h_+-N-im+2)_j},
\end{equation}
and where $(x)_n = \Gamma(x+n)/\Gamma(x)$ is the Pochhammer symbol.
Products of near-zone modes then reduce to the sum
\begin{widetext}
\begin{equation}
\label{product}
R_1 R_2
= \bx^{-i\frac{k_1+k_{2}}{2}+4} \left(1+\frac{\bx}{2}\right)^{-i \left[\frac{k_1+k_2}{2}-m_1-m_2\right]}\sum_{j=0}^{N_1+N_2} {}_{-2} P_j^{(N_1,N_2)}\left(-\frac{\bx}{2}\right)^j,
\end{equation}
\end{widetext}
using the shorthands $R_i = {}_{-2}R_{N_i\ell_i m_i}^{\rm near}$.
This formula defines ${}_{-2}P_i^{(N_1,N_2)}$. By a calculation outlined in App. \ref{normalization}, we find
\begin{widetext}
\begin{align}
	\langle\langle\Upsilon_1,\Upsilon_2 \rangle \rangle_{\rm near} &=- \delta_{\ell_1 \ell_2}\delta_{m_1m_2} \, 2\sqrt{2}M^\frac{10}{3} 2^{-\frac{i k_1}{2}-\frac{i k_2}{2}} \times	
	\\
	&\sum_{j=0}^{N_1+N_2} {}_{-2} P^{(N_1,N_2)}_j\frac{(-1)^j \Gamma (j+\hat \alpha -2) \Gamma (-j-\hat \alpha -\hat \beta +1) [\hat \gamma  (\hat \alpha +\hat \beta +j-1)+\hat e  (\hat \alpha +j-2)]}{\Gamma (-\hat \beta )}+ 
	O(\varepsilon^{p}),\nonumber
\end{align}\label{normalizationUpsilon}
\end{widetext}
where $p=\Re(h_{1+})/2+\Re(h_{2+})/2>0$, which is either $=1$ or $1+\eta_1/2+\eta_2/2$, depending on whether $\eta$ is imaginary or real [see Eqs. \eqref{Hod}, \eqref{eq:hpm}],
where
	\begin{equation}
	\begin{split}
		&\hat \alpha= 4-\frac{i}{2}(k_1+k_2), \ \
        \hat \beta= -3-\frac{i}{2} \left(k_1+k_2-4m\right),\\
		& \hat e= -4 (2-i m), \ \ \hat \gamma=4-i k_1-ik_2,
		\end{split}\label{parameters}
	\end{equation}
 and where we use the shorthands $k_1 = k_{N_1 \ell m},$ etc.
The term on the right side without the $O(\varepsilon^{p})$-contribution is already vanishing for $N_1 \neq N_2$. We have checked this for a range of $N_1,N_2$ values numerically, since it does not appear to follow straightforwardly from the complicated expression for the sum in 
Eq. \eqref{normalizationUpsilon}. It may be demonstrated by noting that $\langle\langle\Upsilon_1,\Upsilon_2 \rangle \rangle_{\rm near}$ is 
to leading order in $\varepsilon$ equal to the bilinear form of the near solution in the nNHEK geometry, where the orthogonality of the modes can be shown by the same arguments as given in \cite{Green2022a} in the case of Kerr. 

At any rate, this suggests that $O(\varepsilon^{p})$ is 
actually the value of $\langle\langle\Upsilon_1,\Upsilon_2 \rangle \rangle_{\rm far}$, which hence appears to be negligibly small for $\varepsilon \ll 1$, which we have 
also tested numerically in several cases. Assuming that this is the case generally, the 
normalization factor for the QNM modes is
found to be
\begin{equation*}
{}_{-2} A_{N\ell m} := \llangle {}_{-2}\Upsilon_{N\ell m}, {}_{-2}\Upsilon_{N\ell m} \rrangle = {}_{-2}  A_{N\ell m}^{\rm near} + O(\varepsilon^{p}), 
\end{equation*}
where
\begin{equation}\begin{split}
    {}_{-2} A_{N\ell m}^{\rm near} = & \ (-1)^{N} M^\frac{10}{3} 2^{-ik_N+\frac{7}{2}}N! \times\\& \frac{ \Gamma (2 h_++N) \Gamma (-h_+-i m-N+3)}{\Gamma (h_+-i m+2) (h_++i m-2)_N}.
    \label{PZsimply}
    \end{split}
\end{equation}
Some details of this computation may be found in App. \ref{normalization}, and in App. \ref{SL2R} from the viewpoint of $\SLg$ representations.

We may similarly compute the scalar product
of two spin $s=+2$ QNMs $\Upsilon_1 = {}_{+2} \Upsilon_{N_1\ell_1 m_1}$, $\Upsilon_2 = {}_{+2} \Upsilon_{N_2\ell_2 m_2}$, again assuming that $m_1,m_2$ are both non-zero. ${}_{+2}R_{N \ell m}^{\rm near}$ is now given by
\begin{equation}
\begin{split}
    {}_{+2}R_{Nlm}^{\rm near} =\  & {}_{+2}C_{N\ell m}{\bx}^{-2-\frac{i k}{2}} \left(1+\frac{\bx}{2}\right)^{-i \left(\frac{k}{2}-m\right)}  \times \\
    &\sum_{j=0}^N {}_{+2} P^{(N)}_j \left(-\frac{\bx}{2}\right)^j,
    \end{split}
\end{equation}
where now
\begin{equation}
    {}_{+2} P^{(N)}_{j}=\frac{(-N)_j (1-2h_+-N)_j}{(1- h_+-N-im-2)_j j!},
\end{equation}
and where ${}_{+2}C_{N\ell m}$ has been defined in Eq. \eqref{Ncoeff}.
From this, the computation of ${}_{+2} A_{N\ell m}^{\rm near}$ proceeds practically along the same lines as above. Alternatively, we may use \eqref{Apm}. In the near zone, the computation of the prerequisite TS constant \eqref{DTS} is most straightforward using the ladder operator formalism for $\thorn,\thorn'$ in nNHEK 
described in App. \ref{radwave}. One finds, in nNHEK
\begin{equation}
\label{Dnear}
    ({}_2 D^{\rm near}_{N\ell m})^2 = (2M^{\frac{2}{3}})^{-4}
    \prod_{j=-1}^2
    (h_+-i m-j) (h_++i m+j-1)
\end{equation}
and thereby, using \eqref{Apm}
\begin{equation}
\begin{split}
    &{}_{+2} A_{N\ell m}^{\rm near} = \frac{1}{16}( {}_2 D_{-\omega_N^* \ell (-m)})^*{}^2 ({}_{-2} A_{N\ell (-m)}^{\rm near})^*\\
    &=(-1)^N M^{\frac{2}{3}} 2^{-i k_N}  N! \times \\&\frac{\Gamma (h_++i m+2) \Gamma (2h_++N) \Gamma (-h_+-i m-N+3)}{\sqrt{2}\Gamma (h_+-i m-2) \Gamma (h_++i m+N-2)}.
    \end{split} \label{PZsimply2}
\end{equation}

\section{Dynamical system in near extremal scaling regime}
\label{sec:dynsysnNHEK}

\subsection{Small extremality parameter \texorpdfstring{$\varepsilon \ll 1$}{} }

In this section, we exploit the simplifications for the overlap coefficients \eqref{G4} $U_{123}, V_{123}$ in our dynamical system \eqref{dyneq} arising in the near extremal regime $\varepsilon \ll 1$.

We split the integrals in Eqs. \eqref{G4} into the near and far zone as delimited by the value $\bar x = c/\sqrt{4 \varepsilon}$ of the scaled coordinate $\bar x$, see Eqs. \eqref{scaledcts}. In the near zone, 
we substitute the near zone QNM $\Upsilon^{\rm near}_q$, whereas in the 
far zone, we substitute the far zone QNM $\Upsilon^{\rm far}_q$, see the discussion in Sec. \ref{sec:NHMs}. Correspondingly, 
we split
\begin{subequations}
    \begin{align}
        U_{123} &= U^{\rm near}_{123} +
        U^{\rm far}_{123},\\
        V_{123} &= V^{\rm near}_{123} +
        V^{\rm far}_{123},
    \end{align}
\end{subequations}
where ``near'' and ``far''
mean the part of the radial integrals implicit in $U_{123}, V_{123}$ in the near- and far zone. 

Our central hypothesis from now is that the far zone contributions are parametrically small in $\varepsilon$. To support this claim, we could argue as in the similar case of the scalar products between the QNMs, see App. \ref{normalization}, where we give evidence that the portion coming from the near zone dominates. In fact, 
the analysis of the near zone 
radial integrals in App. \ref{radoverlap} shows that their integrand decays as $\bar x \to \infty$, indicating that the far zone contribution is indeed negligible.

According to our hypothesis, we therefore rewrite our dynamical system \eqref{dyneq} as
\begin{equation}
\label{dyneq1}
\frac{\dd}{\dd \bt} c_1 = \alpha \sum_{2,3} \left( U_{123}^{\rm near} c_{2} c_{3} 
+ V_{123}^{\rm near} c_{2} c^*_{3} \right).  
\end{equation}
Note that, from now, we write our dynamical system in terms of the slow time $\bar t$ of Eq. \eqref{scaledcts}, and correspondingly, we have to use the 
integration element $\bar T^a \dd \bar S_a$
in the overlap coefficients \eqref{G4}, 
where $\bar T^a = (\partial_{\bt} )^a$ is the 
timelike Killing field corresponding to the slow time $\bar t$, 
and $\dd \bar S_a$ is the induced volume element on a constant $\bar t$ surface in nNHEK.\footnote{
In the scaled coordinates $\bar x^\mu = (\bt, \bx, \bphi, \bar \theta)$ [see Eqs. \eqref{scaledcts}] covering nNHEK, this volume element is given by Eq. \eqref{TdS}.}


We next seek ways to simplify the complicated expressions for the overlap coefficients, Eq. \eqref{G4}, exploiting that $\varepsilon \ll 1$ and the fact that we are working from now on only in the near zone i.e., the nNHEK geometry, see Sec. \ref{nhekgeo}. An object that appears in both overlap coefficients \eqref{G4} is the bilinear expression $\S\T[\hat h_1, \hat h_2]$, which is being applied to various complex  symmetric tensors, $\hat h_{1 \, ab}, \hat h_{2 \, ab}$. By construction, all of these contain the reconstruction operator $\S^\dagger$ [see Eq. \eqref{eq:Sdag}], and hence are automatically in IRG, meaning 
$l^a \hat h_{i \, ab} = 0 = g^{ab} \hat h_{i\, ab}, i=1,2$. Using this simplification, 
the fact that, in nNHEK, the NP coefficients $\rho, \rho',\kappa, \kappa', \sigma, \sigma'$ all vanish, and a Mathematica notebook for $\mathcal{T}_{ab}$ due to \cite{npnotebookspiers,npnotebook} automating GHP calculus and the form of $\T_{ab}$, we find the, still complicated, yet considerably simplified expression compared to the full expression in Kerr. Since it is lengthy we have moved it into App. \ref{app:ST}.

To compute, using the formulas in App. \ref{app:ST}, the overlap coefficients \eqref{G4} in the near zone we require the cases $\hat h_{i \, ab} = \S^\dagger_{ab} \phi_i$
or $\hat h_{i \, ab}^* = (\S^\dagger_{ab} \phi_i)^*$, where $i=1,2$, $\phi_i \circeq \GHPw{-4}{0}$ solves $\O^\dagger \phi_i = 0$. The $\phi_i$'s will be taken to be equal to a suitable near zone QNM $\phi = {}_{-2}\Upsilon^{\rm near}_q$ momentarily. This brings about a number of further simplifications, resulting in 
\begin{widetext}
    \begin{equation}
\begin{split}
\hspace{-1.5cm}
8\pi \S\T[(\S^\dagger \phi_{1})^*,\S^\dagger \phi_2] =	&-\thorn^2 \psi_1 (2 \tau{}+\edth) \edth\phi_2-6\tau{} \thorn\psi_1  \edth \thorn \phi_2\\&+2 \edth \thorn\psi_1( \tau{}+\edth) \thorn \phi_2  +(\thorn^2 \phi_2)  (6 \tau{} -\edth)\edth \psi_1+(1\leftrightarrow2) \label{secondeinstein}
\end{split}
\end{equation}
\end{widetext}
as well as 
\begin{widetext}
    \begin{equation}
\begin{split}
\hspace{-1.5cm}	
8\pi \S\T[(\S^\dagger \phi_1)^*,(\S^\dagger \phi_2)^*] =	&-\thorn^2 \psi_1 (2 \bar\tau{}+\edth') \edth' \phi_2^*  - \thorn\psi_1 (6 \bar\tau{}+4\edth') \edth'\thorn\phi_2^* \\&+4\edth'\thorn^2\phi_2^*  (2 \edth'-\bar\tau{})\psi_1-6 \psi_1 \edth'{}^2\thorn ^2\phi_2^* +\thorn^2\phi^*_2 (6 \bar\tau{}-\edth')   \edth'\psi_{1} \\& +2\edth'\thorn\psi_1( \bar\tau{}+\edth') \thorn\phi_2^*  +\frac{3}{2} \edth'\thorn^3\phi_1^* \edth'\thorn^3\phi_2^*  +\thorn^3\phi_1^*  (\bar\tau{}-\edth') \edth'\thorn^3\phi_2^*+(1\leftrightarrow2). \label{firsteinstein}
\end{split}
\end{equation}
\end{widetext}
In both expressions, we used the shorthand 
$\psi := \frac{1}{4} \thorn^4 \phi^*$. This arrangement of terms is convenient not only because  it reduces the total number of derivatives appearing explicitly to at most four, but also allows us to use $\psi = {}_{+2} \Upsilon^{\rm near}_{q}$ in case that $\phi = {}_{-2}\Upsilon^{\rm near}_{-q^*}$ in view of the TS identity, Eq. \eqref{TS}.

Using these simplifications, the near zone approximations of the overlap coefficients \eqref{G4} become 
\begin{widetext}
\begin{equation}
\label{G41}
\begin{split}
U_{123}^{\rm near}(\bar t) =& -   \frac{\delta_{m_1,m_2+m_3}}{A_{2}  A_{3}} \int\limits_{ \mathscr{C}(\bt,\varepsilon) }  
 \bar T^a \dd \bar S_a
  \times \\ 
({}_{-2}^{} \Upsilon_{-q_1}) \Big[&
-\thorn^2 ({}_{+2}^{} \Upsilon_{q_2}) (2 \bar\tau{}+\edth') \edth' ({}_{-2}^{} \Upsilon_{-q_3^*}^*) - \thorn ({}_{+2}\Upsilon_{q_2}) (6 \bar\tau{}+4\edth') \edth'\thorn ({}_{-2}^{}\Upsilon_{-q_3^*}^*) \\&+4 \edth'\thorn^2( {}_{-2}^{} \Upsilon_{-q_3^*}^*)  (2 \edth'-\bar\tau{})({}_{+2}^{} \Upsilon_{q_2}) -6 \, {}_{+2}\Upsilon_{q_2} \edth'{}^2\thorn^2 ({}_{-2}^{} \Upsilon_{-q_3^*}^*)\\& +\thorn^2({}_{-2}\Upsilon^*_{-q_3^*}) (6 \bar\tau{}-\edth')   \edth'({}_{+2}\Upsilon_{q_2})+2\edth'\thorn ({}_{+2}^{}\Upsilon_{q_2})( \bar\tau{}+\edth') \thorn ({}_{-2}^{} \Upsilon_{-q_3^*}^*) \\& +\frac{3}{2} \edth'\thorn^3 ({}_{-2}^{} \Upsilon_{-q_3^*}^* )\edth'\thorn^3( {}_{-2}^{} \Upsilon_{-q_2^*}^* ) +\thorn^3({}_{-2} \Upsilon_{-q_2^*}^*)  (\bar\tau{}-\edth') \edth'\thorn^3({}_{-2}^{} \Upsilon_{-q_3^*}^*) + (2 \leftrightarrow 3)
\Big],  \\
V_{123}^{\rm near}(\bar t) =&-\frac{\delta_{m_1,m_2-m_3}}{A_2^* A_3}\int\limits_{ \mathscr{C}(\bt,\varepsilon) }  
 \bar T^a \dd \bar S_a
  \times \\ 
({}_{-2}^{} \Upsilon_{-q_1}) \Big[&
-\thorn^2 ({}_{+2}\Upsilon_{q_2}) (2 \tau{}+\edth) \edth({}_{-2}\Upsilon_{-q_3^*})-6\tau{} \thorn({}_{+2}\Upsilon_{q_2})  \edth \thorn ({}_{-2}\Upsilon_{-q_3^*})\\&+2 \edth \thorn({}_{+2}\Upsilon_{q_2})( \tau{}+\edth) \thorn ({}_{-2}\Upsilon_{-q_3^*})  +\thorn^2 ({}_{-2}\Upsilon_{-q_3^*})  (6 \tau{} -\edth)\edth ({}_{+2}\Upsilon_{q_2})
\Big],
\end{split}
\end{equation}
\end{widetext}
where the near zone constant $\bt$ slice is 
\begin{equation}
\mathscr{C}(\bt,\varepsilon)=
\{ \bar x^\mu: \bar x^{0} = \bt, 0 \le \bar x^1 \le c/\sqrt{4\varepsilon} \}. 
\end{equation}
To lighten the notation, we have suppressed the superscript `near' as in ${}_2 A_q \equiv {}_2 A_q^{\rm near}$ for the scalar products,
and we also use the previously introduced notation, meaning e.g., that ${}_{-2} \Upsilon_{-q_2^*} \equiv {}_{-2} \Upsilon_{N_2\ell_2-m_2}^{\rm near}$ and  ${}_{- 2} \Upsilon_{-q_1} \equiv \zeta^4 \mathcal J {}_{2} \Upsilon_{N_1 \ell_1 m_1}^{\rm near}$.

In order to compute the overlap coefficients we now need to substitute the values of $\tau, \tau'$ in nNHEK (see App. \ref{appA}), the QNMs ${}_{\pm 2} \Upsilon_q, q=(N,\ell,m)$ in their near zone approximations (see Sec. \ref{nNHEK_QNM}), and the action of the GHP operators $\thorn, \thorn',\edth,\edth'$ (see Apps. \ref{edthladder}, \ref{radwave}) on these quantities. 

The actions of $\thorn,\thorn'$ on the radial parts of the QNMs simplify in nNHEK compared to Kerr, since, as we will show in App.~\ref{radwave}, these operators act as ladder operators . Another non-trivial, and welcome, simplification is that in the above expressions, the $\bar x$ and the $\bar \theta$ integrals can be seen to factorize. Since the radial QNM functions defined in Sec. \ref{sec:NHMs} are polynomials in $\bar x$ times some complex power of $\bar x$ [see Eq.\eqref{radialwaves}], there is no difficulty, in principle, to carry out the $\bx$ integral, though we must of course consider its precise definition which involves a regulator. 

Unfortunately analogous simplifications do not occur for the actions of $\edth$ and $\edth'$ on the angular parts of the QNMs: these operators do not act as ladder operators in nNHEK nor in Kerr. A related difficulty is that the angular QNM functions are spin-weighted spheroidal harmonics whose eigenvalues are not known in closed form, even in the nNHEK limit. It is therefore not clear a priori how to explicitly evaluate the $\bar\theta$ integrals which involve triple products of spin-weighted spheroidal harmonics (and other trigonometric functions). 

One approach to this problem would be to use numerical methods to evaluate the angular integrals to the extent needed. Another approach, both in Kerr and in nNHEK, and valid in the regime $m^2 \lesssim \ell$, is to use the well-known fact that the spin-weighted spheroidal harmonics may be expanded in a rapidly converging series of spin-weighted spherical harmonics \cite{Kavanagh26}, see Eqs. \eqref{Eexpand}, \eqref{YexpandL} 
for the first non-trivial terms in such expansions. 

Furthermore, as we recall in App.~\ref{edthladder}, when acting on spin-weighted \emph{spherical} harmonics\footnote{This is also true to some extent also on their perturbations in $m^2/\ell$ \cite{Shah_2016}.}, the operators $\edth$ and $\edth'$ are related to ladder operators. Using these algebraic relations, the angular $\bar \theta$ integrals of triple products of spin-weighted spherical harmonics may be computed. 

In the next section, we will consider a limit when $m^2 \ll \ell$ for independent reasons. This will make the explicit computation 
of the overlap coefficients possible.

\subsection{Large angular momentum \texorpdfstring{ $\ell \gg 1$}{} }
\label{largeell}

Consider again the dynamical system \eqref{dyneq1} for the QNM mode amplitudes in the near extremal regime $\varepsilon \ll 1$.
It is clear that the overlap
coefficients $U^{\rm near}_{123}$ and $V^{\rm near}_{123}$ \eqref{G41} for three modes $q_1, q_2, q_3$ [recall the QNM labels $q=(\ell, m, N)$] in our dynamical system \eqref{dyneq1}
have the time dependence 
\begin{subequations}
\label{overlaptdepnear}
    \begin{align}
      U^{\rm near}_{123}(\bar t) =& e^{i(\bar \omega_1-\bar \omega_2 - \bar \omega_3)\bt}  (\dots),\\
      V^{\rm near}_{123}(\bar t) =& e^{i(\bar \omega_1-\bar \omega_2 - \bar \omega_3)\bt}  (\dots).
    \end{align}
\end{subequations}
We expect that, when  
\begin{equation}
\label{nres0}
    |\Re(\bar \omega_1-\bar \omega_2 - \bar \omega_3)| \gg 1,
\end{equation}
oscillations will tend to cancel over a (slow-) time scale $\bar t$ of order unity, or equivalently, over a BL time scale $t$ of order $M/\varepsilon$.
It is thus natural to look for QNM frequencies such that 
\begin{equation}
\label{nres}
    |\Re(\bar \omega_1-\bar \omega_2 - \bar \omega_3)| \lesssim 1.
\end{equation}
However, unless 
\begin{equation}
\label{nres1}
    \Im(\bar \omega_1-\bar \omega_2 - \bar \omega_3) \lesssim 1,
\end{equation}
the coupling coefficients \eqref{overlaptdepnear} in the dynamical 
system \eqref{dyneq1} will become large on a (slow-) times scale $\bar t$ of order unity. This would invalidate our small amplitude approximation underlying the derivation of \eqref{dyneq1}.

 It is not easy to identify a regime where both conditions 
\eqref{nres} and \eqref{nres1} are satisfied. This is because the [scaled, see Eq. \eqref{bomom}] QNM frequencies $\bar\omega_{N\ell m}$ are in general very complicated functions of $m,\ell$ even in nNHEK, given that $h_+$ [see Eq. \eqref{eq:hpm}] in $\bar \omega_{N\ell m}$ [see Eq. \eqref{Hod}] depends on the angular eigenvalue for the spheroidal differential equation, $E_{\ell m}$. However, matters simplify in a regime of large angular momenta, $\ell_i\gg 1$.


The regime where $\ell \gg 1$ supports the ``eikonal approximation'' for QNMs, even without a smallness assumption 
about the extremality parameter, $\varepsilon$. In fact, the eikonal approximation was first considered for Schwarzschild black holes \cite{ferrari1984new}, and later by \cite{Yang:2012he} in Kerr for arbitrary spin. 
In \cite{Yang:2012he}, the large $\ell$ asymptotics of the QNM frequencies was associated with trapped null geodesics, corresponding to unstable critical points of the effective potential in the radial Teukolsky equation. The observations by \cite{Yang:2012he} were analyzed by \cite{Hod:2012bw}
in the extremal limit. In that limit, one observes a qualitative difference depending on when $h_+$ \eqref{eq:hpm}
is real or imaginary i.e., whether one is in the first or second case of Eq. \eqref{Hod}. The real case occurs 
for $m/(\ell + 1/2) < 0.74...$, and the complex case for 
$m/(\ell + 1/2) > 0.74...$ \cite{Hod:2012bw}. 

As later clarified by \cite{yang2013branching,Yang2013}, in the case $m/(\ell + 1/2) > 0.74...$, all modes captured by the eikonal analysis are zero-damped modes, i.e. have a (scaled) QNM frequency [see Eq. \eqref{bomom}] given by the second case in Eq. \eqref{Hod}. 
On the other hand \cite{yang2013branching,Yang2013}, for $m/(\ell + 1/2) < 0.74...$, the eikonal analysis captures not
only the zero-damped QNMs given by the first case in Eq. \eqref{Hod}, but also certain ``damped'' QNMs, whose frequencies do not pile up at the superradiant bound, as assumed from the outset in our analysis through  \eqref{bomom}. These damped QNMs are not captured by our analysis.  They correspond to unstable critical points of the effective potential strictly outside the horizon even in the extremal limit, whereas the zero-damped QNMs correspond to critical points on the horizon \cite{yang2013branching,Yang2013}. 

Due to the complicated dependency of the real part of $\bar \omega_{N\ell m}$ on $\ell, m$ in the case $m/(\ell + 1/2) > 0.74...$ stemming from the presence of the angular eigenvalue ${}_s E_{\ell m}$, it seems hard to make general statements. Hence, from now on, we will consider the case $m/(\ell + 1/2) < 0.74...$, in which $\bar \omega_{N\ell m}$ is imaginary, i.e. \eqref{nres} holds for the zero-damped QNMs by the first case of Eq. \eqref{Hod}. As for the damped modes, it appears that there are only finitely many \cite{yang2013branching,Yang2013} and we exclude them from our analysis\footnote{Since the damped modes do not pile up at the superradiant bound $m\Omega_H$, one might expect that it would be hard for them to satisfy a resonance condition.}. 

It is possible to characterize the angular eigenvalues ${}_s E_{\ell m}$ associated with the angular equation \eqref{eq:SphNHEK}, in the scaling regime
$\ell \gg 1$ and, say, even $m^2/\ell \lesssim 1$, such 
$h_+$ is real and consequently $\bar \omega_{N\ell m}$ is imaginary. The point is that, in this regime, the angular equation is a small perturbation of the equation for spin-weighted spherical harmonics, 
${}_s Y_{\ell m}$, which is 
\begin{widetext}
\begin{align}\label{eq:Yph eq}
  \left[\frac{1}{\sin \bar \theta} \frac{ \dd}{\dd \bar \theta}\left(\sin \bar \theta \frac{\dd \,}{\dd \bar \theta} \right) \right. \left. + \left( {}_s \bar E_{\ell} -  \frac{m^2+s^2+2 m s \cos \bar \theta}{\sin^2 \bar \theta}  \right) \right] {}_s Y_{\ell m}(\bar \theta) = 0,
\end{align}
\end{widetext}
where ${}_s \bar E_{\ell} = \ell (\ell +1)$ in our conventions. 
Indeed, if we compare this operator with that \eqref{eq:SphNHEK}  for the spin-weighted spheroidal harmonics, we see that the additional terms in 
\eqref{eq:SphNHEK} are given by bounded functions of $\bar \theta$
times constants of order $\lesssim m^2$. 
The spacing between subsequent eigenvalues ${}_s \bar E_{\ell} = \ell (\ell +1)$ of the operator for spin-weighted spherical harmonics is $2\ell + 2=O(\ell)$. By standard methods for bounded perturbations of self-adjoint operators with a pure point spectrum (such as the resolvent method, see e.g., \cite{kato2013perturbation}), one can obtain a convergent perturbation series both of 
eigenvalues and eigenfunctions for a bounded perturbation, as long as the norm of the perturbing operator is less than the spacing between the eigenvalues. 

The first relevant terms in the expansion of the eigenvalue may e.g., be found setting $a\omega=am\Omega_H=m/2$ in the small $a\omega$-expansions \cite{Kavanagh26, TeukolskyII,Berti06} [see Eq. \eqref{maxspheroidal}],
\begin{equation}
\label{Eexpand}
\begin{split}
    {}_s E_{\ell m} =& {}_s \bar E_{\ell} 
  -\frac{m^2}{8} +\frac{4 m^4-24 m^2 s^2-m^2}{32 \ell^2}\\&+\frac{-4 m^4+24 m^2 s^2+m^2}{32 \ell^3}+\dots
  \end{split}
\end{equation}
where the dots are terms that we will be able to neglect in our scaling regime \eqref{doublescalingregim}.\footnote{The leading term is already a very good approximation even for $m^2 \lesssim \ell$, in practice, but in App. \ref{statphase}, we also require subleading terms.} This expression assumes $m\neq 0$; the axisymmetric case $m=0$ requires a special treatment which we give in App. \ref{axisym}.

Likewise, under the same assumption, the first relevant terms in the series of the angular eigenfunction are obtained by setting $a\omega=am\Omega_H=m/2$ in the small $a\omega$-expansion \cite{Kavanagh26}
\begin{equation}
\label{YexpandL}
\begin{split}
    &{}_s S_{\ell m} = {}_s Y_{\ell m} + \\
    & 
    \frac{ms}{2} \left[ 
    \frac{{}_s \alpha_{\ell m}}{\ell} {}_s Y_{(\ell-1) m}
    -
    \frac{{}_s \alpha_{(\ell+1)m}}{\ell+1} {}_s Y_{(\ell+1) m}
    \right] + 
    \dots,
    \end{split}
\end{equation}
where the dots are terms that we will be able to neglect in our scaling regime \eqref{doublescalingregim}, and where
\begin{equation}
\label{aldef}
{}_s \alpha_{\ell m}= \frac{1}{\ell} \sqrt{\frac{(\ell^2-m^2)(\ell^2-s^2)}{(2\ell-1)(2\ell+1)}}.
\end{equation}
Again, the axisymmetric case $m=0$ requires a special treatment, see App. \ref{axisym}.

As a consequence of Eqs. \eqref{Eexpand}, \eqref{eq:hpm}, \eqref{Hod}, the QNM frequencies in the near extremal scaling limit conjugate to the slow time $\bt$ [see Eq. \eqref{bomom}] have the large $\ell$ asymptotics
\begin{equation}
\label{omegadoublescal}
    \bar \omega_{N\ell m} = - i\left(N+\ell+1-\frac{15 m^2}{16 \ell}+\frac{15 m^2}{32 \ell^2}\right) + \dots . 
\end{equation}
Again, the axisymmetric case $m=0$ requires a special treatment, see App. \ref{axisym}.

In order to be self-consistent with the simplification which we made in \eqref{maxspheroidal} setting $\omega := m/(2M)$ in the angular equation  \eqref{eq:SphNHEK}, anticipating that $\omega = m\Omega_H + O(\varepsilon)$, we must demand in view of 
Eqs. \eqref{bomom} and \eqref{omegadoublescal} that $\varepsilon \ell \ll 1$ i.e., in total that
\begin{equation}
\label{doublescalingregim}
   m^2 \ll \ell \ll \frac{1}{\varepsilon}.
\end{equation}
In the following, we assume that these conditions hold.

It is clear from Eq. \eqref{omegadoublescal} that condition \eqref{nres}
is satisfied but that there is no reason why condition \eqref{nres1} should be satisfied. Barring any exact ``selection rules'' for the angular momenta $\ell_i$ and overtone numbers $N_i$ in the overlap coefficients, which as we shall show do not occur, we must therefore resort to a regime where the prefactor $e^{i(\bar \omega_1-\bar \omega_2 - \bar \omega_3)\bt}$
in Eq. \eqref{overlaptdepnear} is trivially $\approx 1$. 

Since we have $\bar \omega \sim -i(\ell+N)$
for large $\ell$, this means for low $N_i$
that we should only 
consider times such that $\bar t L \lesssim 1$, where $L$
is the order of the $\ell_i$ considered. In view of $t \sim M\bar t/\varepsilon$, 
this means that 
\begin{equation}
    tL\varepsilon/M \lesssim 1,
\end{equation}
so the BL time $t$ may still be very large\footnote{Below, we will consider the scaling $L \sim 1/\sqrt{\varepsilon}$. Then the above condition means that we consider 
intermediate times $\tilde t = \sqrt{\varepsilon}t/(2M) \lesssim 1$.} in units of $M$ in our regime where $m^2 \ll L\ll \varepsilon^{-1}$.

To summarize and simplify this discussion, 
in our scaling regime \eqref{doublescalingregim} and assuming 
$\bar t \ell \lesssim 1$ we can basically\footnote{The next-to-leading contributions to these relations are, in fact, also important for the calculations in App. \ref{radoverlap}.} take 
\begin{equation}
\label{largelsubst}
    {}_s S_{\ell m} \to {}_s Y_{\ell m}, \quad
    \bar \omega_{N\ell m} \to -i(N+\ell+1), \quad 
\end{equation}
and we take $e^{i(\bar \omega_1-\bar \omega_2 - \bar \omega_3)\bt} \approx 1$.
The simplifications for the overlap coefficients resulting from these substitutions will be described in detail in the next section. 

\subsection{The dynamical system in the regime \texorpdfstring{  $m^2 \ll \ell \ll 1/\varepsilon$}{}}
\label{highlowelloverlap}

Now we consider in more detail the simplifications in our dynamical system \eqref{dyneq1} arising from the
substitutions \eqref{largelsubst} in the eikonal regime where one or more $\ell_i \gg m^2$. These substitutions are made in the overlap coefficients \eqref{G41} in Eq. \eqref{dyneq1}, which in their turn, already incorporate the simplifications due to $\varepsilon \ll 1$. In the scaling regime
\eqref{doublescalingregim}, we write 
\begin{equation}
    \ell = L \bar \ell, 
\end{equation}
where it will be understood that $\varepsilon^{-1} \gg L \gg m^2$ and $\bar \ell \ge c > 0$. 

First of all, this means that the near zone QNMs get replaced by 
\begin{equation}
\label{Upsaxi1}
\begin{split}
    &{}_s \Upsilon^{\rm near}_{N\ell m}(\bar x^\mu)
    \to \\
    & \  \ e^{-(\ell+N+1) \bar t+im\bphi}
    {}_s Y_{\ell m}(\bar \theta) \  {}_{s,0} R_{N \ell m}(\bar x),
\end{split}
\end{equation}
where the functions ${}_{s,\nu} R_{N\ell m}$ are defined in Eq. \eqref{tildeR1}, 
wherein $k_N = m-i(\ell+N+1)$, $h_+=\ell+1$ in the eikonal regime.
For $m=0$ we should instead set, according to our discussion in App. \ref{axisym}, 
\begin{equation}
\label{Upsaxi2}
\begin{split}
    &{}_s \Upsilon^{\rm near}_{N\ell 0}(\bar x^\mu)
    \to \\
    & \ \ e^{-(\ell+N+1) \bar t}
    {}_s Y_{\ell 0}(\bar \theta) \ {}_{s,0} R_{N \ell (-i\varepsilon (\ell+N+1))}(\bar x).
\end{split}
\end{equation}
Since we now have spin-weighted spherical harmonics ${}_s Y_{\ell m}$ instead of spheroidal harmonics in the near zone QNMs, the GHP operators $\edth, \edth'$ may be expressed in terms of spin-raising and lowering ladder type operators and their action on ${}_s Y_{\ell m}$, see App. \ref{edthladder}. Also, we use that $\thorn,\thorn'$ act as 
ladder operators on the functions ${}_{s,\nu} R_{N\ell m}$, see App. \ref{radwave}. The formulas for the explicit GHP scalars $\tau, \tau'$ in nNHEK are imported from App. \ref{appA}. Furthermore, we use the near zone approximations \eqref{PZsimply}, \eqref{PZsimply2} ($m\neq 0$), \eqref{+2Aapp}, \eqref{-2Aapp} ($m=0$) for the normalization factors ${}_{\pm2}A_{N\ell m}$ and their limits 
\eqref{normdecay-}, \eqref{normdecay+} ($m \neq 0$), \eqref{normalization:axisymmetric} ($m=0$) for $\ell \gg 1$.
Finally, the surface integration element in Eqs. \eqref{G41} is
\begin{equation}
\label{TdS}
    \bar T^a \dd \bar S_a = 2M^3 \sin \bar \theta (1+\cos^2 \bar \theta)
    \dd \bx \dd \bphi \dd \bar \theta
\end{equation}
in the nNHEK coordinates \eqref{scaledcts}.

It is a non-trivial fact that, after these substitutions---and for times such that $\bar t L \lesssim 1$---the overlap coefficients \eqref{G41} take a  term-by-term factorized form, of the following schematic form:
\begin{subequations}
\label{overlaptdep}
    \begin{align}
       U^{\rm near}_{123} =& \
       \delta_{m_1,m_2+m_3} 
       \sum_j u_{123}^{(j)} [123]^{(j)}_{} \{123\}^{(j)}_{} \\
    V^{\rm near}_{123} =& \ 
    \delta_{m_1,m_2-m_3}  
      \sum_j v_{123}^{(j)} [123]^{(j)}_{} \{123\}^{(j)}_{}
    \end{align}
\end{subequations}
When all $\ell_i$ are in the eikonal regime, 
each symbol $[123]^{(j)}$ stands for a particular member (i.e., particular values of $m_i, s_i, \ell_i$ for each $j$) of the class of angular integrals of the form
\begin{equation}
\label{a123low}
    [123] = \int\limits_0^\pi \dd \bar \theta \, \sin \bar \theta \, 
    \frac{1-i\cos \bar \theta}{1+i\cos \bar \theta}
    Y_1 Y_2 Y_3 
\end{equation}
where $Y_i = {}_{s_i} Y_{\ell_i m_i}$ are spin-weighted 
spherical harmonics with certain mode and spin indices. Similarly, 
when e.g., only $\ell_2, \ell_3$ are in the eikonal regime, the 
angular integrals to be considered are of the form
\begin{equation}
\label{a123}
    [123] = \int\limits_0^\pi \dd \bar \theta \, \sin \bar \theta \,    S_1 Y_2 Y_3,
\end{equation}
where $S_1$ is an expression involving a spin-weighted spheroidal harmonic acted upon by spin-raising and lowering operators and GHP scalars, etc.

The symbol $\{123\}^{(j)}$ stands for  (regularized) radial overlap integrals of the form 
\begin{equation}
\label{r123}
\{123\} = \int\limits_0^{c/\sqrt{4\varepsilon}} \dd \bar x \, f^2\,
R_1 R_2 R_3 
\end{equation}
involving a triple product of the special functions $R_i = {}_{s_i,\nu_i} R_{N_i\ell_i m_i}$ for certain mode indices and values of 
the parameter $\nu_i$ [see Eq. \eqref{tildeR} or \eqref{tildeR1}]. $u_{123}^{(j)}, v_{123}^{(j)}$ are numerical 
factors that are determined by writing out the overlap integrals \eqref{G41} in full detail. They depend on $s_i, \nu_i, m_i, \ell_i, N_i$, and their values are summarized in the tables in App. \ref{tables} in the case that all $\ell_i$ are large.

The leading behavior in the regime $\varepsilon^{-1} \gg L \gg m^2$ of the angular overlap integrals $[123]$ is found by 
combining an asymptotic formula for the spin-weighted spherical harmonics  with the method of stationary phase, as we show in App. \ref{statphase}. In fact, the final result for $[123]$, is found by combining Eqs. \eqref{a123}, \eqref{expansionYLM}, \eqref{brac0}, \eqref{brac2}. These formulas simplify using the selection rules on the $m_i$'s in 
Eq. \eqref{overlaptdep}. 

Next, the (regulated) radial overlap integrals  \footnote{They are related to Wigner $3j$-symbols of certain $\SLg$ representations \cite{Barut:1965, Stignerthesis}; see also App. \ref{SL2R}.} $\{123\}$ \eqref{r123} in Eq. \eqref{overlaptdep} are carried out by the same method, detailed in App. \ref{normalization}, that we used to find the scalar products of QNMs, using $\varepsilon^{-1} \gg L \gg m^2$ to further approximate the results. These computations are described in App. \ref{radoverlap}.

We now combine all these formulas, and renormalize the QNM amplitudes in order to shorten the expressions for $\ell \ge cL, c > 0$ as follows.
\begin{subequations}
\begin{align}
       c_{N\ell m}
       &
       \to  \left(
       \sqrt{N!}
       L^{4-\frac{N}{2}}
       \frac{
       \pi  i^{\ell+1}  2^{\frac{\ell}{2}-im} \bar \ell^{7-\frac{N}{2}}
       }{\sinh(\pi m)}
       \right)^{-1} 
       c_{N\bar\ell m}\\
       c_{N\ell 0}
 &\to - \left( \sqrt{N!} L^{3-\frac{N}{2}}\frac{i^{\ell}   2^{\frac{\ell}{2}} \bar\ell^{6-\frac{N}{2}}}{  \varepsilon }\right)^{-1}
 c_{N\bar\ell 0}
\end{align}
\label{renormc}
\end{subequations}
When $\ell \ll L$, we renormalize the amplitudes as follows:
\begin{subequations}
\begin{align}
   &c_{N\ell m}\to
   L 2^{\frac{ {}_2 h_{N\ell m+} }{2}}\left({}_{+2}A_{N \ell m}\right)^{-1} c_{N \ell m},\\
   & c_{N \ell 0}\to
   L (-1)^{\ell+1} 2^{\frac{\ell+1}{2}} \left({}_{+2}A_{N \ell0}\right)^{-1}  c_{N\ell 0},
   \end{align}
\end{subequations}
see Eqs. \eqref{PZsimply2}, \eqref{+2Aapp} for the definition of ${}_2 A_{N\ell m}, {}_2 A_{N\ell 0}$, respectively.
Finally, 
we renormalize the amplitude parameter $\alpha$ 
as 
\begin{equation}
\alpha \to  
2^{\frac{15}{2}}  M^{-\frac{2}{3}}  \alpha.
\end{equation}
By an abuse of notation, we denote the new coupling coefficients arising from $U_{123}^{\rm near}, V_{123}^{\rm near}$ after these renormalizations by the same symbols.
When {\it all} 
modes have a large $\ell$ e.g.,
\underline{$\bar \ell_i \ge c>0$ for all $i=1,2,3$}, and when all $N_i=0$, we find 
\begin{widetext}
\begin{equation}
\label{UVdefn}
    \begin{split}
        &  U^{\rm near}_{123}=\frac{\delta_{m_1,m_2+m_3}q_1}{4\bar\ell_S^{\frac{1}{2}} \bar\ell_1^{\frac{5}{2}}}\bigg[ \Theta(\bar \ell_1 - |\bar \ell_2-\bar \ell_3|) -
    \Theta(\bar \ell_1 - |\bar \ell_2+\bar \ell_3|)
    \bigg]\left[
        -6\bar\ell_2^2\bar\ell_3^2
        \csc^2\left(\frac{\chi_1}{2}\right)+(\bar\ell_2+\bar\ell_3)^4 
        \right]\tan \left(\frac{\chi_{1}}{2}\right)
        \\&\times\Big[\mathrm{odd}(\ell_S)
        \cos (\chi_{2}-\chi_{3}) \cos (m_3 \chi_{2}-m_2 \chi_{3})
        -\mathrm{ev}(\ell_S)\tanh(\pi m_1)
        \sin (\chi_{2}-\chi_{3}) \sin (m_3 \chi_{2}-m_2 \chi_{3})
        \Big],\\
        &V^{\rm near}_{123}= \frac{\delta_{m_1,m_2-m_3}q_1 (-1)^{\ell_3+m_3}\bar\ell_2^{4}}{\bar\ell_S^{\frac{1}{2}} \bar\ell_1^{\frac{5}{2}}} \bigg[ \Theta(\bar \ell_1 - |\bar \ell_2-\bar \ell_3|) -
    \Theta(\bar \ell_1 - |\bar \ell_2+\bar \ell_3|)
    \bigg] \cot \left(\frac{\chi_{1}}{2}\right)
    \\&\times\Big[\mathrm{odd}(\ell_S)
        \cos (3 \chi_3+\chi_{2}) \cos (m_3 \chi_{2}+m_2 \chi_3)
        -\mathrm{ev}(\ell_S)\tanh(\pi m_1)
        \sin (3 \chi_3+\chi_{2}) \sin (m_3 \chi_{2}+m_2 \chi_3)
        \Big].
    \end{split}
\end{equation}
\end{widetext}
Here, the step functions $\Theta$ impose the condition that the $\ell_i$ form the sides of a triangle, i.e., that
\begin{equation}
|\bar\ell_2 - \bar\ell_3| \le \bar\ell_1 \le \bar\ell_2 + \bar\ell_3,
\end{equation}
with angles $\chi_i$ opposite to the sides $\bar\ell_i$. To take into account a qualitative difference for axisymmetric modes, we defined
\begin{equation}
 q_j
 =\begin{cases}
      2\frac{\bar\ell_j}{\bar\ell_S} & \text{if $m_1=0$},\\
     1 & \text{if $m_1\neq 0$}.
 \end{cases}
\end{equation}
The formula for the angles $\chi_i$ is given by the cosine theorem e.g.,
\begin{equation}
\chi_1
    =\arccos\left[\frac{\bar \ell_2^2+\bar\ell_3^2-\bar\ell_1^2}{2\bar\ell_2\bar \ell_3}\right]
\end{equation}
and its cyclic permutations of (123), see Fig. \ref{fig:triangle}. We use the notation $\ell_S=\ell_1+\ell_2+\ell_3$ and $\ell_i/L=\bar \ell_i$.
\begin{figure}[h]
	\centering
	\begin{tikzpicture}[
	my angle/.style = {draw,
		angle radius=5mm, 
		angle eccentricity=1.8, 
		right, inner sep=1pt,
		font=\footnotesize} 
	]
	\draw   (0,0) coordinate (a) --
	(4,0) coordinate (c) --
	(3,2) coordinate (b) -- cycle;
	\pic[my angle, "$\chi_1$"] {angle = c--a--b};
	\pic[my angle, "$\chi_2$"] {angle = b--c--a};
	\pic[my angle, "$\chi_3$"] {angle = a--b--c};
	\coordinate [label=left:$\bar\ell_1$] (A) at (4.3,1);
	\coordinate [label=left:$\bar\ell_3$] (A) at (2.5,-.4);
	\coordinate [label=left:$\bar\ell_2$] (A) at (1.5,1.2);
	\end{tikzpicture}
	\caption{} \label{fig:triangle}
\end{figure}

The functions ev respectively odd in Eqs. \eqref{UVdefn} are one if and only if the argument is an even respectively odd number, and zero otherwise. 
In our expressions \eqref{UVdefn} for the overlap coefficients, and in the following, we are discarding consistently terms suppressed by higher inverse powers of $L$, in this case order $O(L^{-1/2})$.

When {\it only} \underline{$\ell_1, \ell_2$ are large e.g., $\bar \ell_i \ge c>0$ for $i=1,2$}, but $\ell_3$ is {\it not} large, we find the overlap coefficients to leading order in $L$ by again heavily using the results of Apps. \ref{radoverlap}, \ref{statphase}, \ref{tables}. The results are quite lengthy and presented in App. \ref{sec:exploverlap}.

When {\it only} \underline{$\ell_2, \ell_3$ are large e.g., $\bar \ell_i \ge c>0$ for all $i=2,3$}, but $\ell_1$ need not be large, a similar analysis can be made and 
the results are presented in App. \ref{sec:exploverlap}. 

A notable nontrivial feature of all overlap coefficients is that they do not depend explicitly on the large parameter $L$. The overlap coefficients for the higher overtone numbers ($N_S>0$) are discussed in App. \ref{app:higher N} in the case when all $\ell_i$'s are large. It is remarkable that we generally find them to be of the same order in $L$ and in fact equal to the $N_S=0$ overlap coefficients times certain $(N_i, \bar \ell_i, m_i, s_i)$-dependent dressing factors. We believe that this is also true in the other cases where some of the $\ell_i$'s are not large but we do not present these results here.

\section{Equilibrium distribution of interacting QNMs}
\label{sec:randomphase}

\subsection{Truncated dynamical system}

In this section, we carry our analysis of the dynamical system \eqref{dyneq1} further by
analyzing possible ``equilibrium'' solutions, by which we mean ones with 
\begin{equation}
    \frac{\dd}{\dd \bar t} c_q^{\rm eq}  = 0
\end{equation}
for all QNM labels $q=(N,\ell,m)$. By \eqref{dyneq1} this is equivalent to finding 
$c^{\rm eq}$ such that
\begin{equation}
\label{dyneq3}
 \sum_{2,3} \left( U_{123}^{\rm near} c_{2}^{\rm eq} c_{3}^{\rm eq} 
+ V_{123}^{\rm near} c^{{\rm eq}}_{2} c_{3}^{{\rm eq}*} \right) = 0 
\end{equation}
for all $q_1$, where we recall the condensed notation $c_1=c_{q_1}$ etc.\footnote{An equilibrium solution is similar to a QNM associated with the linearized EE in that its amplitude is constant in time, i.e. quadratically non-linear effects precisely cancel each other out.}

Since our dynamical system is just an approximation, in that, we restrict our attention to the near zone, to only QNM contributions to the field, and to at most quadratically non-linear effects, the equilibrium condition only means, in effect, that the $c_q$'s are time-independent to within these approximations. Even so, it is quite difficult to solve the system \eqref{dyneq3} directly. 
We therefore introduce further simplifications which allow us to find a solution.

To this end, the first simplification that we now introduce is to consider {\it infinitely many} QNMs having corresponding amplitudes $c_{N\ell m}^{\rm high}$ with a high $\ell \gtrsim L$ for some large $L$ (in practice e.g., $L=10^2-10^3$) and a {\it few} QNMs and their corresponding amplitudes $c_{N\ell m}^{\rm low}$. The truncation is that we simply neglect all midsize $\ell$'s in between. We think of the few low $\ell$ QNMs as being the driver of the high $\ell$ QNMs giving rise to the aforementioned pumping effect.  

Under this simplification, we may use the exact formulas for the scaling limits of the overlap coefficients discussed in Sec. \ref{highlowelloverlap} and App. \ref{sec:exploverlap}. Omitting at this stage any channel such as 
(high,low) $\to$ (low) and (low,low) $\to$ (low), which is suppressed by an inverse power of $L$ according to these formulas, the dynamical system \eqref{dyneq1} takes the following schematic form:
    \begin{subequations}
    \label{truncdyn}
\begin{align}
    \frac{\dd}{\dd \bar t} c^{\rm high}  =& \ (\dots)c^{\rm high} c^{\rm high} + (\dots)c^{\rm high} c^{\rm low}
    \\
       \frac{\dd}{\dd \bar t} c^{\rm low}  =& \ L^{\frac{1}{2}}[(\dots)c^{\rm high} c^{\rm high}
       + (\dots)c^{\rm high} c^{\rm low}].
\end{align}        
    \end{subequations}
In these equations, $(\dots)$ represents overlap coefficients, as presented in Sec.~\ref{highlowelloverlap} and App.~\ref{sec:exploverlap}, and a summation/integration over the QNM mode numbers understood. We shall give a more explicit version of this system in the next section.

\subsection{Equilibrium distributions}\label{sec:eqm dists}

Unfortunately, even the truncated system \eqref{truncdyn} is still quite intractable due to the complicated forms of the overlap coefficients as indicated by $\dots$ in Eq. \eqref{truncdyn}. With the goal of identifying an equilibrium distribution, we now make further simplifications
to Eqs.~\eqref{truncdyn}. We restrict to:
\begin{itemize}
    \item axisymmetric QNMs ($m=0$),
    \item zero overtone number $N=0$,
    \item only odd $\ell$'s.
\end{itemize}

The first and last simplifications are self-consistent: The first one because of the Kronecker delta functions of the $m$'s in the overlap coefficients as described in Sec.~\ref{highlowelloverlap}, App.~\ref{sec:exploverlap}, and the last one because, by inspection, the formulas in Sec.~\ref{highlowelloverlap}, App.~\ref{sec:exploverlap} imply that two odd $\ell$ QNMs may never excite an even $\ell$ QNM.  
Actually, the last simplification could be avoided at the expense of treating the even $\ell$ QNMs in the high $\ell$ sector independently, which we avoid for simplicity. The $N=0$ restriction on the overtone number is not self-consistent (a pair of $N=0$ QNMs can excite an $N>0$ QNM), but is made to arrive at a manageable truncated dynamical system. Nevertheless, the results that we present for the $N \ge 0$ modes in App. \ref{app:higher N} indicate that the general structure relevant for our subsequent arguments below is preserved also in this case, so the second restriction is likely not an essential one.

To write down the dynamical system under the above three simplifying assumptions, we  combine all the QNM amplitudes in the low $\ell$ sector, and we view the QNM amplitudes in the high $\ell$ sector as smooth functions in the rescaled angular momentum $\bar \ell = \ell/L$, which leads to the definitions
\begin{widetext}
\begin{subequations}
\begin{align}
    C^{\rm low} :=& \ \sqrt{\pi}\sum_{2\leq \ell\ll L}i^{\ell}   \sqrt{(2\ell+1)(\ell+2)(\ell+1)\ell(\ell-1)} c_{0\ell0}^{\rm low}\\
    c^{\rm high}_{\bar \ell} :=& \ c_{0\ell 0}^{\rm high}.
\end{align}
\end{subequations}
\end{widetext}
We now consider the scaling $L \sim 1/\varepsilon^{1/2}$, $c L \le \ell \le 1/\varepsilon^q$ 
for some $1/2 < q < 1$, so that 
$1 \ll L, \ell \varepsilon \ll 1$, as is required for our approximations in the high $\ell$
sector. Since our approximations require that $\ell \bar t \lesssim 1$, our dynamical system 
is valid for slow times or order\footnote{It would thus be sensible to write our dynamical system \eqref{DYNfin}
in terms of a medium time $\tilde t = \epsilon^{-q} \bar t$ of order $\tilde t \lesssim 1$, 
and correspondingly rescale the QNM amplitudes as 
$c_\ell \to \varepsilon^q c_\ell$ to maintain the form of  \eqref{DYNfin}.} 
$\bar t \lesssim \varepsilon^q$, or BL times $t \lesssim M\varepsilon^{q-1}$. 

We convert the summations over the high $\ell$'s to integrals  
with respect to the rescaled angular momentum $\bar \ell = \ell/L$
as in $\sum_\ell \to L \int_c^{\varepsilon^{1/2-q}} \dd \bar \ell$.
Furthermore, we use the substantial simplifications due to $m_i=0$ described in App. \ref{sec:exploverlap}.

After the dust settles, the dynamical system simplifies to, within our approximations and truncations (taking $\alpha = 1$, which is an assumption about the overall amplitude) 
\begin{widetext}
\begin{equation}
\label{DYNfin}
\begin{split}
    \frac{\dd}{\dd \bar t} C^{\rm low} =& \ 0 \\
    \frac{\dd}{\dd \bar t} c^{\rm high}_1 =& \ 
    \  \bar \ell_1 c^{\rm high}_1 \Re(C^{\rm low}) + 
    \bar\ell_1^{-\frac{3}{2}} \int\limits_c^{\varepsilon^{1/2-q}} \int\limits_c^{\varepsilon^{1/2-q}}
    \dd \bar \ell_2 \dd \bar \ell_3 
    \Big[ \Theta(\bar \ell_1 - |\bar \ell_2-\bar \ell_3|) -
    \Theta(\bar \ell_1 - |\bar \ell_2+\bar \ell_3|)
    \Big]
    \frac{\tan \left(\frac{\chi_{1}}{2}\right)}{(\bar\ell_1 + \bar\ell_2 + \bar\ell_3)^{\frac{3}{2}}} \times \\
    \times &\bigg\{ \frac{1}{2} \left[ -6\bar\ell_2^2\bar\ell_3^2
        \csc^2\left(\frac{\chi_1}{2}\right)+(\bar\ell_2+\bar\ell_3)^4 
        \right]
        \cos (\chi_{2}-\chi_{3}) c^{\rm high}_2
        c^{\rm high}_3 - 2\bar \ell_2^4 
        \cos (3\chi_{3}+\chi_{2}) c^{\rm high}_2
        (c^{\rm high}_3)^* \bigg\} .   \end{split}
\end{equation}
\end{widetext}
The homogeneous scaling behavior of the above integral kernel under rescalings of the $\bar \ell_i$ (noting that $\chi_i$ are the scale invariant angles of a triangle with side lengths $\bar \ell_i$) now suggests that we make the ansatz
\begin{equation}
    c^{\rm high,eq}_{\bar \ell} \propto \bar \ell^p
\end{equation}
for some $p$ to be determined. If the integration boundaries of the double integral \eqref{DYNfin}
were $(0,\infty)$ instead of $(c,\varepsilon^{1/2-q})$, the two terms on the right side of the second equation in \eqref{DYNfin} would cancel for a suitable proportionality factor of the order of $C^{\rm low,eq}$, the total QNM amplitude of the low $\ell$ QNMs, and for $p=-2$. For these choices, the double integral is actually divergent for both large and small $\bar \ell_{2,3}$, but we may regulate these divergences e.g., by a dimensional renormalization prescription. 
Such a prescription can be understood as an implicit assumption on the nature of the intermediate QNMs having $\ell \lesssim \varepsilon^{-1/2}$ and QNMs with very large $\ell \gtrsim \varepsilon^{-q}$, for which the approximations used to arrive at \eqref{DYNfin} do not necessarily hold. 

Thus, our equilibrium distribution for the high $\ell$ QNM amplitudes is
\begin{equation}
    c^{\rm high,eq}_{\bar \ell} \sim C^{\rm low,eq} \cdot \bar \ell^{-2}.
\end{equation}
By Eq. \eqref{Phisum}, the Hertz potential $\phi^{\rm eq}$ in equilibrium is given by a sum of the corresponding QNMs weighted with these amplitudes. Taking into account our renormalizations \eqref{renormc} and the definition \eqref{normalization:axisymmetric}, this translates into\footnote{The proportionality factor includes an $L^{-1} = \sqrt{\varepsilon}$.}:
\begin{equation}
\label{eqHertz}
\begin{split}
   & \phi^{\rm eq}(\bar x^\mu)
    \propto \\ 
    & \text{(low)} + 
     C^{\rm low, eq} \sum_{\varepsilon^{-q} \gtrsim \ell \gtrsim \varepsilon^{-1/2}} 
    2^{-\frac{\ell}{2}} 
    \ell^{-\frac{7}{2}} \cdot {}_{-2} \Upsilon_{0\ell 0}^{\rm near}(\bar t, \bar x, \bar \theta) ,
    \end{split}
\end{equation}
in the near horizon zone where ${}_{-2} \Upsilon^{\rm near}_{0\ell 0}(\bar x^\mu)$ are the near zone, zero-damped QNMs
as in Eq. \eqref{Upsaxi2}. ``Low'' stands for the contribution of the low $\ell$ QNMs. 
Since they are matched to the far zone modes (see Sec. \ref{sec:far sol}), we also have 
\begin{equation}
\label{eqHertz1}
\begin{split}
& \phi^{\rm eq}(x^\mu)
    \propto \\
    & \text{(low)} + C^{\rm low, eq} \sum_{\varepsilon^{-q} \gtrsim \ell \gtrsim \varepsilon^{-1/2}} 
    2^{-\frac{\ell}{2}}
    \ell^{-\frac{7}{2}}  \cdot {}_{-2} \Upsilon_{0\ell 0}^{\rm far}(t, x , \theta) ,
    \end{split}
\end{equation}
with $x^\mu = (t,x,\theta,\phi)$ the far zone coordinates, related to BL coordinates $(t,r,\theta,\phi)$ by 
$x=(r-r_+)/r_+$, and with far zone, zero-damped QNMs ${}_{-2} \Upsilon_{0\ell 0}^{\rm far}(t, x , \theta) \propto 
{}_{-2} Y_{\ell 0}(\theta) {}_{-2} R^{\rm far}_{0\ell 0}(x) e^{-i\omega_{0\ell 0}t}$ given in terms of the corresponding radial solutions \eqref{Rfar}.
These modes decay as $e^{-\varepsilon(\ell+1)t/(2M)}$ in BL time. Since, in the range of modes considered, we have $\ell \lesssim \varepsilon^{-q}$ and $1/2<q<1, \varepsilon \ll 1$, these decaying exponentials are 
in fact practically $=1$ for a parametrically long BL time of order $t/M = O(\varepsilon^{q-1})$.

Finally, the full metric is given by Eq. \eqref{GHZ}, which in equilibrium is 
\begin{equation}
\label{GHZeq}
g_{ab}^{\rm eq} = g_{ab}^{\rm kerr} + h_{ab}^{\rm IRG,eq} +   x_{ab}^{\rm eq},
\end{equation}
where $h_{ab}^{\rm IRG,eq}$ is the reconstructed metric \eqref{hIRG} corresponding to the equilibrium Hertz potential $\phi^{\rm eq}$, and where $x_{ab}^{\rm eq}$ the corresponding corrector, see Eq. \eqref{xrecon}.

The upshot is that, by Eqs. \eqref{eqHertz}, or \eqref{eqHertz1}, the equilibrium metric has dyadically small amplitudes for large $\ell$ QNMs, and thus the equilibrium metric, which is constant in time over a parametrically large BL time of order $t$,
has no large $\ell$-pieces. Assuming the equilibrium metric to be the endpoint of a dynamical evolution governed by our dynamical system in the weakly non-linear regime, 
it follows that high $\ell$ contributions must eventually die out, resembling an inverse cascade. This dying out is attributable solely to the weakly non-linear effects, as decaying exponentials in the linear QNMs are practically order one over the time scales that we consider, which are parametrically large (scaling as an inverse power of the extremality parameter) in BL time.

\section{Discussion}
\label{sec:discussion}

In this paper we have derived a non-linear dynamical system for the zero-damped QNM amplitudes of a near extremal Kerr black hole. Our derivation is based on the leading non-linear approximation of the EEs and the assumptions that (a) for a small extremality parameters $\varepsilon$ there exists a parametrically long epoch $t = O(1/\sqrt{\varepsilon})$ in which these QNMs are the dominant contributions to the non-linear metric perturbation, (b) the angular momentum of the QNMs can be considered either as ``small'' $\ell = O(1)$ or large $\ell = O(1/\sqrt{\varepsilon})$, and that intermediate and very large angular momentum QNMs can be neglected in an approximate description of the dynamical system, and (c) the non-linear interaction between these QNMs takes place predominantly in the near (nNHEK) zone. 

Under these assumptions, we then derived an approximate equilibrium solution to the dynamical system which is time independent for a parametrically long BL time. During this time, the exponentially decaying factor in the QNM functions is practically unity, so the form of the equilibrium solution is dictated entirely by non-linear effects. Our equilibrium solution has QNM amplitudes that are zero for non-axisymmetric $(m \neq 0)$ modes and that are dyadically exponentially small in $\ell$. Although we have not shown that our equilibrium solution is an attractor of the dynamical system, we view this as evidence that high angular momentum $(m,\ell)$ QNM contributions become exponentially suppressed, hence of a kind of inverse cascade. It is, of course, possible that the question of inverse versus direct cascade might be impacted by the restriction to $m=0$ in deriving the equilibrium solutions. This restriction effectively reduces the dimensionality of the system, which is known to affect cascade directions~\cite{Adams:2013vsa, Green:2013zba}.

Since the dynamical system has been explicitly computed, it would in principle be possible to perform a numerical simulation of a sufficiently large but finite subsystem of QNMs to check whether the attractor and cascade hypothesis are true. 

It might also be fruitful to introduce a statistical element into our description, following a standard procedure in the theory of weak wave turbulence, see \cite{zakharov2012kolmogorov} for a review. In such a description the phases of the QNM amplitudes governed by our dynamical system would be considered as random, and one would first derive a corresponding system for the absolute values of the amplitudes, as done in \cite{zakharov2012kolmogorov}. Our dynamical system does not appear to be Hamiltonian unlike the systems studied in \cite{zakharov2012kolmogorov}, but we do not think that this would be an essential obstacle. 
Finally, it would be worth exploring possible connections
to ``non-modal stability'' in  systems where the linear modes all decay (as is the case for QNMs) but due to their ``non-normality'' sum to cause a non-linear transition to turbulence, see e.g., \cite{Schmid2007}.

\begin{acknowledgments}
C.I. and S.H. would like to thank M. Casals for useful discussions. C.I. has been supported by the International Max Planck Research School
for Mathematics in the Sciences, Leipzig  and by the Nonlinear Algebra Group at MPI-MiS. S.R.G. is supported by a UKRI Future Leaders Fellowship (grant number MR/Y018060/1). S.R.G would like to acknowledge L. Sberna for helpful discussions.
\end{acknowledgments}

\appendix

\section{Spin coefficients in nNHEK}
\label{appA}

The complete list of optical scalars, spin coefficients, and Weyl components associated with the Kinnersley frame and BL coordinates in Kerr can be found e.g., in 
\cite{Price07}. In the body of the text, we mainly require those of the nNHEK geometry associated with the frame \eqref{nNHEK_NP} and the coordinates $\bar x^\mu = 
(\bt, \bx, \bar \theta, \bphi)$ of Eq. \eqref{scaledcts}. 
These spin coefficients and the non-vanishing optical scalars are, in GHP notations \cite{Compere2018}
\begin{subequations}
\begin{align}
\tau & =\frac{-i \sin \bar \theta}{M \sqrt{2}\left(1+\cos ^2 \bar \theta\right)}, \\ 
\tau' & =-\frac{i \sin \bar \theta}{M \sqrt{2}(1-i \cos \bar \theta)^2}, \\
\beta & =\frac{\cot \bar \theta}{2 \sqrt{2} M(1+i \cos \bar \theta)}, \\ 
\beta' & =-\frac{ 2 i\sin \bar \theta -(1-i\cos \bar \theta ) \cot \bar \theta }{2 \sqrt{2} M (1-i\cos \bar\theta )^2}, \\
\epsilon &= 0,\\
\epsilon' &=-\frac{\bx + 1}{2 M^2\left(1+\cos ^2 \bar \theta\right)}.
\end{align}
\end{subequations}
The only non-vanishing Weyl component is
\begin{equation}
\Psi_2= -\frac{1}{M^2(1-i \cos \bar \theta)^3}.
\end{equation}

\section{GHP relations in nNHEK}

In the nNHEK geometry, we have 
$\rho=\rho'=\kappa=\kappa'=\sigma=\sigma'=0$ and $\Psi_i = 0, i \neq 2$, i.e., the metric is type D and the principal null geodesics are non-shearing, non-expanding, and non-twisting. As a consequence, we have
the following GHP invariant identities involving the non-vanishing optical scalars $\tau,\tau'$, and non-vanishing Weyl component $\Psi_2$,
\begin{equation}
\edth \tau= \tau^2,\quad \thorn\tau= 0,\quad \thorn \tau'=0, \quad \edth'\tau= \Psi_2+\tau \bar\tau,
\end{equation}
together with their primed, complex conjugated and primed-complex conjugated versions. The GHP commutators read 
\begin{equation}
    \begin{split}
        [\th,\th']&=(\bar\tau-\tau')\edth+(\tau-\bar\tau')\edth'-p(\Psi_2-\tau \tau')\\&-q(\bar\Psi_2-\bar \tau \bar\tau'),\\
        [\edth,\edth']&=-p\Psi_2-q\bar \Psi_2,\\
        [\th,\edth]&=-\bar\tau'\th
    \end{split}
\end{equation}
when acting on a GHP scalar of weights $\circeq \GHPw{p}{q}$. The full set of GHP commutators is obtained by taking the GHP prime and complex conjugate of these relations.

\section{Discrete isometries and GHP intertwining relations}\label{discreteisometries}

Besides the continuous rotation and time-translation isometries, the Kerr metric has a group $(\mathbb{Z}_2)^2$ of isometries generated by $I,J$, where 
\begin{equation}
J: (t,\phi) \to (-t,-\phi), \quad I: \theta \to \pi-\theta,
\end{equation}
referring to BL coordinates. These isometries act on ordinary scalar or tensor fields by the usual tensor transformation rules (pull-back). 
However, defining their action on GHP scalars with non-trivial weights is somewhat more subtle. For the case of $J$, this was given in \cite{Green2022a}, as we now briefly recall.

Being an isometry, $J$ maps any 
NP frame aligned with the principal null directions to another such frame. In fact, it swaps the null directions $l^a$ and $n^a$ and changes the
orientation on the orthogonal complement of these null directions spanned by $m^a$, $\bar m^a$, i.e., there is  $\Lambda_J>0$, $\Gamma_J \in \mathbb{R}$ (depending on the chosen NP frame)
  such that $J_* l^a = -\Lambda_J n^a$, $J_* n^a = -\Lambda^{-1}_J l^a$, 
  $J_* m^a = e^{i\Gamma_J} \bar m^a$. Here we have defined $J$ to act on tensors by the push-forward, and we combine $\lambda^2_J := \Lambda_J e^{i\Gamma_J}$
  Following \cite{Green2022a}, we define the $\mathbb{C}$-linear action of $J$ on $\eta \circeq \GHPw{p}{q}$ by
  \begin{equation}
    \label{Jdef}
    \mathcal J \eta(x) := i^{p+q} \lambda_J(x)^{-p} \bar \lambda_J(x)^{-q} \eta(J(x))  
  \end{equation}
  in the given NP frame. It follows \cite{Green2022a} that $\mathcal J \eta \circeq \GHPw{-p}{-q}$ is an invariantly defined properly weighted GHP scalar. In the 
  Kinnersley frame \eqref{eq:Kintet} and BL coordinates, we have e.g. \cite{Green2022a}, 
\begin{equation}\label{eq:boostParams}
  \lambda_J = \frac{\sqrt{2} (r-ia \cos \theta)}{\sqrt{\Delta}}.
\end{equation}
In nNHEK in the frame \eqref{nNHEK_NP} and the coordinates \eqref{scaledcts}, we have
\begin{equation}
  \lambda_J = M \sqrt{2}\left(\frac{1-i \cos \bar \theta}{	\sqrt{f}}\right)
\end{equation}
instead, where $f=\bx(\bx+2)$.

For the case of $I$ one may give an analogous construction. In this case, we have $I_* l^a = \Lambda_I l^a$, $I_* n^a = \Lambda^{-1}_I l^a$, 
  $I_* m^a = e^{i\Gamma_I} \bar m^a$, where differently from the case of $J$, the principal null directions are not exchanged, and where
  $\Lambda_I>0$, $\Gamma_I \in \mathbb{R}$. As before, we combine $\lambda^2_I := \Lambda_I e^{i\Gamma_I}$, and then we define the $\mathbb{C}$-linear 
  action of $I$ on $\eta \circeq \GHPw{p}{q}$ by
  \begin{equation}
    \label{Idef}
    \mathcal{I} \eta(x) := \lambda_I(x)^{q} \bar \lambda_I(x)^{p} \eta(I(x))  
  \end{equation}
  in the given NP frame. It again follows that $\mathcal I \eta$ is an invariantly defined properly weighted GHP scalar, but differently from 
  $\mathcal J$, the weights are now $\mathcal{I} \eta \circeq \GHPw{q}{p}$. Both in the Kinnersley frame in Kerr, and in the frame \eqref{nNHEK_NP}
  in nNHEK, we have e.g., 
  \begin{equation}\label{eq:boostParamsI}
  \lambda_I = i.
\end{equation}

The Teukosky operators ${}_s \O, s\ge 0$ and their adjoints have useful intertwining relations with $\mathcal{I}, \mathcal{J}$. For the case of $\mathcal{J}$, this 
intertwining relation is \cite{Green2022a}
\begin{equation}
\label{intertwJ}
{}_{s} \O \Psi_2^{\frac{2s}{3}} \mathcal{J} = \Psi_2^{\frac{2s}{3}} \mathcal{J}  {}_{s}  \O^\dagger
\end{equation}
where $\Psi_2$ is the non-vanishing background Weyl scalar of Kerr. For the case of $\mathcal{I}$, this 
intertwining relation is 
\begin{equation}
\label{intertwI}
{}_s \O^\dagger \Psi_2^{-\frac{2s}{3}} \mathcal{I} =  \Psi_2^{-\frac{2s}{3}} \mathcal{I}  {}_{s}  \O^{\prime *}.
\end{equation}
The last formula may be checked using Eq. \eqref{Odef} and the actions $\mathcal{I}\rho = \bar \rho, \mathcal{I}\tau = \bar \tau, \mathcal{I}m^a = \bar m^a, 
\mathcal{I} n^a = n^a, \mathcal{I}l^a = l^a$, the formula for the adjoint,  
\begin{equation}
{}_s \O^\dagger \eta = \left[g^{ab}(\Theta_a - 2s B_a)(\Theta_b - 2s B_b) - 4s^2 \Psi_2 \right] \eta, 
\end{equation}
for $\eta \circeq \GHPw{-2s}{0}$, and the formula \cite{Green2022a}
\begin{equation}\label{eq:gradPsi}
  \Theta_a \Psi_2 = -3 (B_a + B'_a) \Psi_2,
\end{equation}
which yields $\Psi_2^{\frac{2s}{3}} {}_{s}\O^\dagger \Psi_2^{-\frac{2s}{3}} = {}_{s}\O'$ and which is in agreement with \eqref{intertwJ} after applying $\mathcal J$ both sides. The intertwining relations for ${}_s \O, s\le 0$ follow by applying a GHP priming operation.

Another well-known set of intertwining relations is closely related to the Teukolsky-Starobinsky (TS) identities
\cite{starobinskii1973amplification,teukolsky1974perturbations}  (see e.g., \cite[App. K]{hollands2024metric} for their GHP covariant forms used here), as we now recall. Let $\mathcal{W}$ be the linear differential operator that produces the perturbed Weyl scalar ${}_{+2} \psi$ from a metric perturbation. In GHP notation, 
\begin{subequations}
\begin{align}
    \label{eq:T}
    &\mathcal{W}^{ab} h_{ab} = \frac{1}{2}(\edth-\bar \tau')(\edth-\bar \tau') h_{ll} + \frac{1}{2} (\thorn - \bar \rho)(\thorn - \bar \rho) h_{mm} \nonumber \\
            &-\frac{1}{2}\Big[(\thorn- \bar \rho)(\edth-2\bar \tau') + (\edth -\bar \tau')(\thorn -2\bar \rho)\Big] h_{(lm)}\\
\label{eq:Tdag}
& \mathcal{W}^\dagger_{ab} \eta = \frac{1}{2} l_a l_b(\edth-\tau)(\edth-\tau) \eta + \frac{1}{2} m_a m_b (\thorn-\rho)(\thorn-\rho) \eta \nonumber\\
  & - \frac{1}{2} l_{(a} m_{b)}\Big[(\edth + \bar{\tau}' -\tau)(\thorn - \rho) + (\thorn - \rho + \bar \rho)(\edth-\tau) \Big] \eta,
\end{align}
\end{subequations}
where the second line gives the formal adjoint acting on $\eta \circeq \GHPw{-4}{0}$. 

Then the essence of the Teukolsky formalism may be succinctly summarized in the operator equation
$\S \E = \O \mathcal{W}$ \cite{wald1978construction}. 
The ``radial'' TS identities can be stated covariantly as 
\begin{subequations}
\begin{align}
\S \mathcal{W}^{\dagger*} =\ & -\frac{1}{2}
\left[\thorn^2 - 4 (\rho + \bar{\rho}) \thorn + 12 \rho \bar{\rho} \right] \thorn^2 \\
\label{P4}
\mathcal{W} \S^{\dagger *} =\ & \frac{1}{4} \thorn^4. 
\end{align}
\end{subequations}
Acting with $\O$ from the left, then using 
$\S \E = \O \mathcal{W}$, $\E = \E^*=\E^\dagger$, and 
$\E \S^{\dagger*} = \mathcal{W}^{\dagger *} \O^{\dagger *}$ gives
\begin{equation}
\label{intertw3}
\O \thorn^4 = -2
\left[\thorn^2 - 4 (\rho + \bar{\rho}) \thorn + 12 \rho \bar{\rho} \right]\thorn^2 \O^{\dagger *}.
\end{equation}

	\section{Normalization of QNMs}  
 \label{normalization}
	In this section, we provide some details on 
 the computation of the normalization constants ${}_{\pm 2} A_{N\ell m}$, see Eq.~\eqref{norm}, in the near extremal approximation $\varepsilon \ll 1$. Some peculiarities arise in the case of axisymmetric modes, $m=0$. That case is therefore treated separately in App.~\ref{axisym}.
 
 For $\Upsilon_i \circeq \GHPw{-4}{0}, i=1,2$, the definition of the scalar product \eqref{sprod} is explicitly 
	\begin{equation}
 \begin{split}
		\langle\langle\Upsilon_1,\Upsilon_2 \rangle \rangle_t=\int\limits_{\mathscr C}\ud S^a &\left[\left(\Psi_2^{\frac{4}{3}}\mathcal J \Upsilon_1\right) \left(\Theta_a-4B_a\right) \Upsilon_2\right.\\&\left.-\Upsilon_2 \left(\Theta_a+4B_a\right) \left(\Psi_2^{\frac{4}{3}}\mathcal J \Upsilon_1\right) \right],
\label{bilinearteukolsky}
\end{split}
	\end{equation}
 where $\mathscr C$ is any surface of constant $t$ in Kerr. The integration element on $\mathscr C$, $\dd S^a = u^a \dd S$, is defined in terms of the unit forward normal $u_a$ to $\mathscr C$ and the intrinsic integration element, $\dd S$.
 
 In case $\Upsilon_1, \Upsilon_2 \circeq \GHPw{-4}{0}$ are two QNMs, a complex $r$-integration contour as described in \cite{Green2022a} is understood in order to regulate the divergence of the integral as $r_* \to \pm \infty$.
 This integral is then split into a near-zone- and a far-zone part, as in Sec. \ref{nNHEK_QNM}. 
 Consequently, in the near-zone approximation, $\mathscr C$ is a constant $\bt$ slice
 with appropriately defined analytic extension in $\bx$, and we have
	\begin{equation}
 \ud S = 2M^3 \sin \bar\theta \sqrt{ \frac{1 + \cos^2 \bar \theta}{f}} \,\ud \bx \,\ud \bar \theta\, \ud \bar \phi,
 \end{equation}
and
\begin{equation}
 \label{inducedvolume}
		u^a =\frac{1}{2} \left(\frac{\sqrt{f}}{M \sqrt{\cos ^2\bar \theta+1}}l^a+\frac{2 M \sqrt{\cos ^2\bar \theta+1}}{\sqrt{f}}  n^a\right)
	\end{equation}
 which is valid in our frame \eqref{nNHEK_NP}. 
 We also have $u^aB_a=0$ since $\rho=0$ in nNHEK, and thereby
	\begin{equation}
		\begin{split}
		&\Psi_2^{\frac{4}{3}} \mathcal J \Upsilon_1 =\frac{4M^{\frac{4}{3}}}{f^2}\Upsilon_1\Bigg|_{
  \begin{smallmatrix}
  \bt \to -\bt\\
  \bphi \to -\bphi
  \end{smallmatrix}
  }\\
		&u^a \Theta_a \Upsilon_2= \frac{ \left[ f'\left(2 -\partial_{\bar \phi}\right) +2\partial_{\bar t}  \right]}{2M \sqrt{f^{}} \sqrt{1+\cos^2 \bar \theta}} \Upsilon_2\\
  &u^a \Theta_a \left(\Psi_2^{\frac{4}{3}} \mathcal J \Upsilon_1 \right)
  = \frac{ \left[ f'\left(2 +\partial_{\bar \phi}\right) -2\partial_{\bar t}  \right]}{2M \sqrt{f^5} \sqrt{1+\cos^2 \bar \theta}} \Upsilon_1 \Bigg|_{
  \begin{smallmatrix}
  \bt \to -\bt\\
  \bphi \to -\bphi
  \end{smallmatrix}
  }.
  \end{split}
  \end{equation}
  
Using these formulas, the near zone contribution to the scalar product between two QNMs [see Eq. \eqref{nearQNM}]
${}_{-2} \Upsilon_{N_1\ell_1 m_1}, {}_{-2} \Upsilon_{N_2\ell_2 m_2}$ is found to be
	\begin{equation}
	\begin{split}
			&\langle\langle
   {}_{-2} \Upsilon_{N_1\ell_1 m_1}, {}_{-2} \Upsilon_{N_2\ell_2 m_2}
   \rangle \rangle_{\rm near}=\\
&\delta_{\ell_1\ell_2}\delta_{m_1m_2}
\int\limits_{0}^{c/\sqrt{4\varepsilon}} \ud \bx \frac{4 \sqrt{2}M^\frac{10}{3}}{f^3} {}_{-2}R_{N_1\ell_1 m_1}\,{}_{-2}R_{N_2\ell_1 m_1}\times \\&\left[f'(2-im_1)-i(k_1+k_{2}-2m_1)\right].
			\end{split}
	\end{equation}
 In this formula, as in the text, we set the 
 point delimiting the near from the far-zone 
 to be $\bx = \frac{c}{\sqrt\varepsilon}$. Furthermore, we have dropped the superscript ``near'' from $R^{\rm near}_{N\ell m}$ in ${}_{-2} \Upsilon_{N\ell m}$ for easier readability, and we use the shorthand notation $k_1=\bar \omega_{N_1\ell_1 m_1}+m_1$ etc. already introduced above. The orthogonality of the spin-weighted spheroidal harmonics ${}_{-2} S_{\ell m}$
 in ${}_{-2} \Upsilon_{N\ell m}$ was already used to obtain the Kronecker deltas. 
 
 We now use the representation of radial wavefunction in terms of finite polynomials \eqref{radialwaves}, leading to
	\begin{equation}
	\begin{split}
	&\langle\langle
   {}_{-2} \Upsilon_{N_1\ell_1 m_1}, {}_{-2} \Upsilon_{N_2\ell_2 m_3}
   \rangle \rangle_{\rm near}=\\&-\delta_{\ell_1 \ell_2}\delta_{m_1m_2}2 \sqrt{2}M^{\frac{10}{3}} 2^{-\frac{i}{2} (k_1+k_{2})}	\sum_{j=0}^{N_1+N_2}{}_{-2}P_j^{(N_1,N_2)} \\& \times \int\limits_{0}^{c/\sqrt{4\varepsilon}} \ud y  ( \hat \gamma-\hat e y) \left(1+y\right)^{\hat \beta} y^{\hat \alpha} (-y)^{j-3} \label{intermediate}
	\end{split}
	\end{equation}
	where we substituted $\bx=2y$ and used the definitions \eqref{parameters}. The integral over $y$ has to be integrated, a priori, over a complex contour described in \cite{Green2022a} in order to make it convergent. However, as described in \cite{cannizzaro2024relativistic}, this procedure is equivalent to a ``minimal subtraction scheme'', where the lower boundary of the integral is replaced by a small positive regulator, and then all divergent parts of the integral are subtracted off. This, in turn, is equivalent, to evaluating the integral in Eq. \eqref{intermediate}, using the analytic extension of the Euler beta-function. We find that, as the extremality parameter $\varepsilon\to 0$,
 
  \begin{widetext}
 \begin{equation}
     \begin{split}
       \int\limits_{0}^{c/\sqrt{4\varepsilon}} \ud y (\hat \gamma-\hat e y) \left(1+y\right)^{\hat \beta} y^{\hat \alpha} (-y)^{j-3} =  &\frac{(-1)^j \Gamma (j+\hat \alpha -2 ) \Gamma (1-j-\hat \alpha -\hat \beta) [\hat \gamma  (\hat \alpha +\hat \beta +j-1)+\hat e  (\hat \alpha +j-2)]}{\Gamma (-\hat \beta )}\\&+\left(\frac{\sqrt{4\varepsilon}}{c}\right)^{1-\hat \alpha -\hat \beta -j} \left(\frac{(-1)^j \hat e  }{j+\hat \alpha +\hat \beta -1}+O\left(\varepsilon \right)\right)
     \end{split}
 \end{equation}
       \end{widetext}
 where we estimate the real part of the exponent of the correction as
\begin{equation}
\begin{split}
    \Re\left(1-\hat \alpha -\hat \beta -j\right)&=\Re(h_{1+}) +\Re(h_{2+})-j+N_1+N_2\\&\geq \Re(h_{1+})+\Re(h_{2+})>0
    \end{split}
\end{equation}
 since, we have $j\leq N_1+N_2$ in the $j$-sum.
 This gives Eq. \eqref{normalizationUpsilon}. For $N_1=N_2=N$, this final result may be simplified using identities for sums involving quotients of Gamma-functions, leading to the expression given above in Eq. \eqref{PZsimply},
 using the notation ${}_{-2}A^{\rm near}_{N\ell m}$ for the limit $\varepsilon \to 0$ of 
 $\langle\langle {}_{-2} \Upsilon_{N\ell m}, {}_{-2} \Upsilon_{N\ell m} \rangle \rangle_{\rm near}$.
 
Eq. \eqref{PZsimply} may be further simplified for overtone number $N=0$:
	\begin{equation}
 \label{normalization0overtone}
{}_{-2}A_{0\ell m}^{\rm near}=M^\frac{10}{3} 2^{\frac{7}{2}-(h_+ +im)}\frac{ \Gamma (2 h_+) \Gamma (-h_+-i m+3)}{\Gamma (h_+-i m+2)}.
	\end{equation}
For $s=+2$, we can use \eqref{Apm} and get 
\begin{equation}
 \label{normalization0overtone+}
\begin{split}
   {}_{+2}A_{0\ell m}^{\rm near}
   =&M^{\frac{2}{3}}2^{-\frac{1}{2}-(h_++i m)}\times\\&\frac{\Gamma (2 h_+) \Gamma (-h_+-i m+3) \Gamma (h_++i m+2)}{ |\Gamma (h_+-i m-2)|^2}.
   \end{split}
\end{equation}

These normalization factors  may be approximated for $\ell \gg 1$ by means of Stirling's formula \cite[5.11.7]{nist} and the Euler reflection formula \cite[5.5.3]{nist} for the Gamma function, leading to ($m \neq 0$):
	\begin{equation}
		{}_{-2} A_{0\ell m}\sim M^{\frac{10}{3}} 2^{\ell-i m+
        \frac{7}{2}} i \sqrt{\pi } (-1)^\ell \ell^{\frac{1}{2}} \text{csch}(\pi  m), \label{normdecay-}
	\end{equation}
 and
 \begin{equation}
		{}_{+2} A_{0\ell m}\sim  M^{\frac{2}{3}} 2^{\ell-i m-\frac{1}{2}} i \sqrt{\pi } (-1)^\ell \ell^{\frac{17}{2}} \text{csch}(\pi  m). \label{normdecay+}
	\end{equation} 

\subsection{Axisymmetric QNMs}
\label{axisym}
The treatment of QNMs in the nNHEK approximation $\varepsilon \ll 1$
is qualitatively different, and in fact more subtle, for axisymmetric ($m=0$) modes than 
for $m \neq 0$, as has been noted in several works before, see e.g., \cite{bredberg2010black,casals2016horizon}. The conceptual reason for this is that the branch cut starting at $\omega = 0$ in the complex frequency plane characterizing the contribution in the retarded Green's function
(see Sec. \ref{Greensf}) gets parametrically close to the isolated poles at the QNM frequencies $\omega \sim -i\varepsilon(N+\ell+1)/(2M)$. 

For the $m=0$ QNMs, one may first observe that our approximation of the angular Teukolsky equation \eqref{maxspheroidal} in the regime $\varepsilon \ll 1$ leads to the equation for axisymmetric ($m=0$) spin-weighted spherical harmonics \eqref{eq:Yph eq}, rather than that for the spin-weighted spheroidal harmonics \eqref{eq:SphNHEK}, as would be the case for $m \neq 0$. The angular eigenvalue 
of the spin-weighted spherical harmonic equation is 
${}_s E_{\ell 0} = {}_s \bar E_{\ell} = \ell(\ell+1)$, and substituting this value into the definition \eqref{eq:hpm} of ${}_sh_+$, we find that ${}_s h_+ = \ell+1$. In view of Eqs.
\eqref{normalization0overtone}, \eqref{normalization0overtone+}
we would thereby conclude for instance that the scalar product of an axisymmetric QNM would diverge for $s=\pm 2, m=0, N=0$. A similar conclusion can be drawn for all axisymmetric QNMs. 

In order to be self-consistent, a more precise analysis is necessary, which we now provide. Firstly, 
we must improve our approximation \eqref{maxspheroidal} of the solutions to the angular Teukolsy equation by including terms of $O(\varepsilon)$. For this, we self-consistently assume in the full angular Teukolsky equation \eqref{eq:Sph eq}  that $m=0$ and that $\omega_{N\ell 0} = -i\varepsilon(h_+ +N)/(2M) = -i\varepsilon(\ell+1+N)/(2M)$ up to terms of order $O(\varepsilon^2)$, cf. Eqs. \eqref{Hod}, \eqref{bomom}.
Neglecting $O(\varepsilon^3)$-terms, the approximation to the angular Teukolsky equation replacing Eq. \eqref{maxspheroidal} becomes
\begin{widetext}
\begin{align}\label{eq:Sph eq2}
  \left[\frac{1}{\sin \bar \theta} \frac{ \dd}{\dd \bar \theta}\left(\sin \bar \theta \frac{\dd \,}{\dd \bar \theta} \right) \right. \left. + \left( {}_s E_{\ell 0} -  \frac{s^2}{\sin^2 \bar \theta} - \frac{1}{4} \varepsilon^2(\ell+1+N)^2 \cos^2 \bar \theta+i\varepsilon(\ell+1+N) s \cos \theta \right) \right] {}_s S_{\ell 0}(\bar \theta) = 0.
\end{align}
\end{widetext}
The $\varepsilon$ dependent terms are next considered as perturbations to the $m=0$ spin-weighted spherical harmonic operator along similar lines described already described below Eq. \eqref{eq:Yph eq}. Using e.g., results in  \cite{Berti06}, we get
\begin{equation}
\label{Epert}
    {}_s E_{\ell 0}={}_s \bar E_{\ell 0}
    -\frac{\varepsilon^2}{4}(\ell+N+1)^2({}_s u_{(\ell+1)}-{}_s u_\ell-1) + O(\varepsilon^3).
\end{equation}
Here
\begin{equation}
    {}_s u_\ell=\frac{2(\ell^2-s^2)^2}{(4\ell^2-1)\ell},
\end{equation}
and ${}_s \bar E_{\ell 0} = \ell(\ell+1)$ is the unperturbed eigenvalue. Next, we recall that [Eq. \eqref{eq:hpm} for $m=0$]
\begin{equation}
    {}_s h_{+\ell 0} =\frac{1}{2}+\frac{1}{2}{}_s \eta_{\ell 0}, \ \ {}_s \eta_{\ell 0}=\sqrt{1+4{}_s E_{\ell0}}.
\end{equation}
For small $\varepsilon$ we obtain
\begin{equation}
\label{Hsmall}
    {}_s h_{+\ell 0}\sim \ell+1 +\frac{[2 \ell (\ell+1)-1] \varepsilon^2 (\ell+N+1)^2}{4 (2 \ell-1) (2 \ell+1) (2 \ell+3)}+ O(\varepsilon^3),
\end{equation}
which, for $\ell \gg 1, \varepsilon \ell \ll 1$, is approximated by
\begin{equation}
\label{haaprox}
    {}_s h_{+\ell 0}\sim \ell+1+\frac{1}{16} \varepsilon^2 ( \ell+2 N)+ O(\varepsilon^3).
\end{equation}  
In particular, the quantity ${}_s h_{+\ell 0}$ equation gets corrections only at second order in $\varepsilon$.

We must then also improve our approximation \eqref{maxspheroidal} of the solutions to the angular Teukolsy equation by including terms of order $O(\varepsilon)$ i.e., rather than setting $a\omega=0$ in the case $m=0$, 
we should set $a\omega = -i\varepsilon(\ell+1+N)/2$. Consequently, 
the first relevant terms in the series of the angular eigenfunction are obtained by setting $a\omega$ to this value in the small $a\omega$-expansion \cite{Kavanagh26}, i.e.
\begin{equation}
\label{Yexpand}
\begin{split}
    &{}_s S_{\ell 0} =   {}_s Y_{\ell 0}-\\ 
    &\frac{is\varepsilon(\ell+1+N)}{2} \left( 
    \frac{{}_s \alpha_{\ell 0}}{\ell} {}_s Y_{(\ell-1) 0}
    -
    \frac{{}_s \alpha_{(\ell+1)0}}{\ell+1} {}_s Y_{(\ell+1) 0}
    \right) + 
    \dots
    \end{split}
\end{equation}
Note that, to be self-consistent, we must have $\varepsilon (N+\ell) \ll 1$, so the correction term is very small, and ${}_s \alpha_{\ell 0}$
has been defined in Eq. \eqref{aldef}.

Next we need to analyze the radial Teukolsky equation \eqref{eq:radial} including the leading correction in $\varepsilon \ll 1$ for $m=0$. 
It turns out that the leading correction occurs already at $O(\varepsilon)$, so we can neglect the leading $O(\varepsilon^2)$-correction that we found for ${}_s E_{\ell 0}$ from the angular equation.
In fact, starting e.g., from Eq. \eqref{eq:radial}, we  find
\begin{widetext}
\begin{equation}
\left[ f^{-s} \frac{\mathrm{d}}{\mathrm{d} \bx}\left(f^{s+1} \frac{\dd}{\dd \bx} \right)-{}_s V_{k\ell 0}^{\rm near}(\bx) 
-2k\varepsilon \left( 
\frac{(\bar x+1)is + k}{\bar x+2}+is
\right) + O(\varepsilon^2)
\right] {}_s R^{\rm near}_{k\ell 0}=0, 
\label{radialeqnsigma}
\end{equation}
\end{widetext}
where the potential 
$_s V_{k\ell 0}^{\rm near}$ is given by Eq. \eqref{Vklm}.
By inspection, the total potential in the above equation is 
given by 
\begin{equation}
 \begin{split}  
 &{}_s V_{k\ell 0}^{\rm near}(\bx) 
+2k\varepsilon \left( 
 \frac{(\bar x+1)is+k}{\bar x+2}+is
\right) \\
=& {}_s V_{k\ell (-i\varepsilon(\ell+N+1))}^{\rm near} + O(\varepsilon^2),
  \end{split}
\end{equation}
which follows from Eq. \eqref{Vklm} by a straightforward calculation, keeping terms
up to and including of $O(\varepsilon)$. In other words, for $m=0$, up to and including $O(\varepsilon)$, we should simply make the replacement  
\begin{equation}
\label{subst}
 m \to -i\varepsilon(\ell+N+1)  
\end{equation}
in the potential \eqref{Vklm} for the near zone radial Teukolsky equation for the $m\neq 0$ case, keeping all $\varepsilon$-dependent terms up to and including $O(\varepsilon)$. 
A similar analysis with the same conclusion can be 
carried out for the far zone equation. The rest of the analysis then proceeds precisely as before for $m \neq 0$: To find the solutions $k_N$ of the matching condition \eqref{eq:QNM condition}, we should make the substitution \eqref{subst}, and use the approximation \eqref{eq:QNM condition}
for $h_+$. Since the $O(\varepsilon^2)$-term in Eq. \eqref{haaprox} is subleading, we obtain $k_N = m-i(h_++N) \to -i(1+\varepsilon)(\ell+N+1)$. In other words, we should 
simply make the change \eqref{subst} to $m$ while leaving $h_+$ as it is in all formulas in the $m\neq 0$ case.\footnote{
A similar conclusion has been reached before in the context of extremal black holes, see e.g., \cite{bredberg2010black,casals2016horizon}.}. 

For example, for the normalization constant ${}_{\pm 2} A_{N\ell 0}^{\rm near}$ (scalar product of QNM as $\varepsilon \ll 1$ for $m=0$) as given in Eqs. \eqref{PZsimply}, \eqref{PZsimply2} for $m\neq 0$, we find that, identifying $\ell+1=h_+$ to shorten the expressions:
\begin{equation}
\label{-2Aapp} 
\begin{split}
    &{}_{-2} A_{N\ell 0}^{\rm near} =(-1)^{N} M^\frac{10}{3} 2^{-(h_++N)+3}  N!  \times\\& \frac{\sqrt{2}\Gamma (2h_++N) \Gamma [-h_+-  \varepsilon(h_++N)-N+3]}{\Gamma [h_+-  \varepsilon(h_++N)+2] [h_++ \varepsilon(h_++N)-2]_N}.
    \end{split}
\end{equation}

For $s=2$, there is a similar formula which is obtained from Eqs. \eqref{Apm}, \eqref{Ncoeff}:
\begin{equation}
\label{+2Aapp}
\begin{split}
    {}_{+2} A_{N \ell 0}^{\rm near} = &\ (-1)^N M^{\frac{2}{3}} 2^{-(h_++N)} N! \times \\&\frac{  \Gamma\left(h_++\varepsilon(h_++N)+2\right) \Gamma [2 h_++N] }{\sqrt{2}\Gamma\left[h_+-\varepsilon(h_++N)-2\right]} \times \\
    &\frac{\Gamma[-h_+-\varepsilon(h_++N)-N+3]}
    {\Gamma [h_++\varepsilon(h_++N)+N-2]}.
    \end{split}
\end{equation}
Taking additionally $\ell \gg 1$ while $\varepsilon \ell \ll 1$, we find for $N=0$:
	\begin{equation} \label{normalization:axisymmetric}
 \begin{split}
{}_{-2} A_{0 \ell 0}^{\rm near}
\approx&-\frac{M^{\frac{10}{3}}(-1)^{\ell}  2^{\ell+\frac{7}{2}}\ell^{-\frac{1}{2}}}{\sqrt{\pi} \varepsilon},
\\{}_{+2} A_{0 \ell 0}^{\rm near}
\approx &-\frac{M^{\frac{2}{3}}(-1)^{\ell}  2^{\ell-\frac{1}{2}}\ell^{\frac{15}{2}}}{\sqrt{\pi}  \varepsilon}.
\end{split}
	\end{equation}
 In particular, we capture the precise leading form of the divergence of this normalization constant as $\varepsilon \to 0$.
 \vspace{2.5cm}

\section{Computation of \texorpdfstring{$\S\T$}{}  in nNHEK}
\label{app:ST}
Here we give the expression for 
$\S\T$ in nNHEK, acting on two symmetric tensors 
$\hat h_{1\, ab}, \hat h_{2\, ab}$
in ingoing radiation gauge. Recall that $\S$, $\T$ were given
by Eqs. \eqref{eq:S}, \eqref{G2E}. 
We first give the expression as a quadratic form, i.e. when $\hat h_{1\, ab}= \hat h_{2\, ab} = \hat h_{ab}$:

\hspace{1cm}

\begin{widetext}
    \begin{equation}
\begin{split}
\hspace{-1.5cm}	\S\T[\hat h,\hat h] =	&\frac{1}{4\pi}  \left\{2 \tau'^2\hat h_{mm} \thorn^2\hat h_{mm} +4 \tau' (\thorn \hat h_{nm}) \thorn^2\hat h_{mm} +4 \tau' \hat h_{mm} \thorn^3 \hat h_{nm} -\tau' \hat h_{mm} \thorn^2\edth'\hat h_{mm}-(\thorn^2\hat h_{mm}) (\edth'\hat h_{mm}) \tau' \right.\\&\left.+2 \left(\tau'^2+\bar \tau^2\right) (\thorn \hat h_{mm})^2+2 (\thorn^2\hat h_{nm})^2+2 \hat h_{mm} \bar \tau^2 \thorn^2\hat h_{mm}+2 \hat h_{\bar m \bar m} \tau \bar\tau' \thorn^2\hat h_{mm}+2 \bar \tau (\thorn \hat h_{nm}) (\thorn^2\hat h_{mm})\right.\\&\left.-4 \tau (\thorn \hat h_{n \bar m}) (\thorn^2 \hat h_{mm})+2 \bar\tau' (\thorn \hat h_{n \bar m}) (\thorn^2\hat h_{mm})-2 (\thorn^2\hat h_{n n}) (\thorn^2 \hat h_{mm})+2 \hat h_{mm} \tau \bar\tau' \thorn^2 \hat h_{\bar m \bar m}+2 (\thorn \hat h_{nm})( \thorn^3\hat h_{nm})\right.\\&\left.+2 \hat h_{mm} \tau (\thorn^3\hat h_{n \bar m})+2 \hat h_{mm} \bar\tau' (\thorn^3\hat h_{n \bar m})-3 \hat h_{n \bar m} \tau (\thorn^3\hat h_{mm})+\hat h_{nm} \bar \tau (\thorn^3\hat h_{mm})-3 (\thorn \hat h_{n n}) (\thorn^3\hat h_{mm})\right.\\&\left.+\hat h_{mm} (\thorn^4\hat h_{n n})-\hat h_{n n} (\thorn^4\hat h_{mm})-2 \hat h_{mm} (\thorn^3\edth \hat h_{n \bar m})+2 \hat h_{n \bar m} (\thorn^3\edth \hat h_{mm})-2 \hat h_{mm} (\thorn^3\edth'\hat h_{nm})+2 \hat h_{nm} (\thorn^3\edth' \hat h_{mm})\right.\\&\left.+3 \hat h_{\bar m \bar m} \tau (\thorn^2 \edth \hat h_{mm})-\hat h_{\bar m \bar m} \bar\tau' (\thorn^2 \edth \hat h_{mm})+3 (\thorn \hat h_{n \bar m}) (\thorn^2 \edth \hat h_{mm})-2 \hat h_{mm} \tau (\thorn^2 \edth \hat h_{\bar m \bar m})-2 \hat h_{mm} \bar\tau' (\thorn^2 \edth \hat h_{\bar m \bar m})\right.\\&\left.-\hat h_{\bar m \bar m} (\thorn^2\edth^2\hat h_{mm})+\hat h_{mm} (\thorn^2\edth^2 \hat h_{\bar m \bar m})-\hat h_{mm} \bar \tau (\thorn^2 \edth'\hat h_{mm})+5 (\thorn \hat h_{nm}) (\thorn^2\edth'\hat h_{mm})+4 (\thorn^2\hat h_{mm}) (\thorn \edth \hat h_{n \bar m})\right.\\&\left.-4 (\thorn^2 \hat h_{mm}) (\thorn \edth'\hat h_{nm})+4 (\thorn^2\hat h_{nm}) (\thorn\edth'\hat h_{mm})+(\thorn \hat h_{mm}) \left[2 \bar\tau'^2 \thorn \hat h_{\bar m \bar m}+4 \thorn^2 \hat h_{n \bar m} \bar\tau'-2 \thorn \edth \hat h_{\bar m \bar m} \bar\tau' +\tau \thorn^2\hat h_{n \bar m}\right.\right.\\&\left.\left.+(8 \tau'+\bar \tau) \thorn^2 \hat h_{nm}+\thorn^3 \hat h_{n n}-\thorn^2\edth \hat h_{n \bar m}-5 \thorn^2 \edth' \hat h_{nm}-2 \tau \thorn\edth \hat h_{\bar m \bar m}-2 \tau' \thorn\edth'\hat h_{mm}-2 \bar \tau \thorn\edth'\hat h_{mm}\right]\right.\\&\left.+3 (\thorn^3\hat h_{mm}) (\edth \hat h_{n \bar m})+\tau (\thorn^2\hat h_{\bar m \bar m}) (\edth \hat h_{mm})-\bar\tau' (\thorn^2\hat h_{\bar m \bar m}) (\edth \hat h_{mm})-(\thorn^3\hat h_{n \bar m})( \edth \hat h_{mm})+(\thorn^2\edth \hat h_{\bar m \bar m})( \edth \hat h_{mm})\right.\\&\left.+4 \tau (\thorn^2\hat h_{mm}) (\edth \hat h_{\bar m \bar m})-2 \bar\tau' (\thorn^2\hat h_{mm}) (\edth \hat h_{\bar m \bar m})-3 (\thorn^2\edth(\hat h_{mm}) (\edth \hat h_{\bar m \bar m})-2 (\thorn^2\hat h_{mm})( \edth^2\hat h_{\bar m \bar m})-(\thorn^3\hat h_{mm}) (\edth'\hat h_{nm})\right.\\&\left.-\bar \tau (\thorn^2\hat h_{mm}) (\edth'\hat h_{mm})+(\thorn^3 \hat h_{nm}) (\edth'\hat h_{mm})\right\}
\end{split}
\end{equation}
\end{widetext}
For the general case, we simply apply the polarization formula relating bilinear and quadratic forms, 
\begin{equation}
    \S\T[\hat h_1, \hat h_2] = 
    \frac{1}{4}\{\S\T[\hat h_1+\hat h_2, \hat h_1 +\hat h_2] 
    - 
    \S\T[\hat h_1-\hat h_2, \hat h_1 -\hat h_2]\}.
\end{equation}

	\section{ \texorpdfstring{$\thorn$,$\thorn'$}{} as ladder operators in nNHEK} 
 \label{radwave}
In this section we show that in nNHEK, the GHP operators $\th$ and $\th'$ act as ladder operators on a suitably defined set of modes closely related to the mode solutions of the Teukolsky equation. 

We begin by defining 
\begin{widetext}
\begin{equation}
\begin{split}
{}_{s,\nu} R_{k \ell m}(\bar x) &= {}_s C_{k\ell m} {\bx}^{-(s+\nu)-\frac{i k}{2}} \left(\frac{\bx}{2}+1\right)^{-(s+\nu)+i \left(\frac{k}{2}-m\right)}\times \\& \hspace{1cm}\, _2F_1\left({}_s h_+-i m-(s+\nu),{}_s h_--i m-(s+\nu);1-i k-(s+\nu);-\frac{\bx}{2}\right),
\end{split} \label{tildeR}
\end{equation}
\end{widetext}
where ${}_s h_\pm \equiv {}_s h_{\pm \ell m}$ are defined by Eq.
\eqref{eq:hpm} in terms of $s$ ({\it not} $s+\nu$) via the eigenvalue ${}_s E_{\ell m}$
of the angular equation in nNHEK. 
$\nu = 0,\pm 1, \pm 2,\dots$ is a new index. An alternative representation valid at a QNM frequency $k_N \equiv {}_s k_{N\ell m}$ is
\begin{equation}
\label{tildeR1}
\begin{split}
{}_{s,\nu} R_{N\ell m}(\bar x) =& {}_{s}  C_{N\ell m} \bar x^{-(s+\nu)-\frac{i k_N}{2}} \left(1+\frac{\bar x}{2}\right)^{-i \left(\frac{k_N}{2}-m\right)}  \\
& \times \sum_{j=0}^N {}_{s,\nu}P^{(N)}_j \left(-\frac{\bar x}{2}\right)^j
\end{split}
\end{equation}
where
\begin{equation}
\label{PN1}
{}_{s,\nu} P^{(N)}_{j}=\frac{(-N)_j (1-2{}_sh_+-N)_j}{\left(1-{}_sh_+-N-im-(s+\nu)\right)_j j!}.
\end{equation}
E.g., for $s=-2$, $\nu=0$ 
we recover the QNMs in nNHEK, 
${}_{-2,0} R_{N \ell m} = {}_{-2}R^{\rm near}_{N\ell m}$. 
Likewise, since ${}_s E_{\ell m}$
and ${}_s h_\pm$ depend on $s$
only through $|s|$, we also have 
${}_{-2,4} R_{N \ell m} \propto {}_{+2}R^{\rm near}_{N\ell m}$. 
In this sense, the new radial functions ${}_{s,\nu} R_{k \ell m}$
interpolate the QNMs for opposite spins $\pm s$. As we will see, the 
operators $\thorn, \thorn'$ induce ladder operators raising and lowering $\nu$. 

It can be shown that the 
functions ${}_{s,\nu} R_{k \ell m}$
satisfy the following differential equation:
\begin{equation}
\left[f^{-(s+\nu)} \frac{\mathrm{d}}{\mathrm{d} \bx}\left(f^{s+\nu+1} \frac{\mathrm{d}}{\mathrm{d} \bx}\right)-{}_{s,\nu }V(\bx)\right] {}_{s,\nu} R_{k\ell m}=0\label{spinraiseeqn}
\end{equation}
with the potential
\begin{equation}
\begin{split}
{}_{s,\nu}V&=-\frac{3}{4}m^2-(s+\nu)(s+\nu+1)+{}_{s}E_{\ell m}-2i(s+1)m\\&+\frac{(m\bx+k)[2i(s+\nu)-k+2\bx i(s+\nu)-m\bx]}{f},
\end{split}
\end{equation}
where we stress again that ${}_{s}E_{\ell m}$ is the spin $s$ eigenvalue ({\it not} $s+\nu$) of the angular equation in nNHEK.
For the boundary conditions, we compute
\begin{equation}
\label{BCtilde}
\begin{split}
{}_{s,\nu}R_{k\ell m}&\sim{}_{s}C {\bx}^{-(s+\nu)-\frac{i k}{2}}, \ \ \bx\to 0\\
{}_{s,\nu} R_{k\ell m} &\sim{}_{s}C\left({}_{s,\nu} a_-\, \bar x^{-{}_{s}h_--(s+1)}+ {}_{s,\nu} a_+\, \bar x^{-{}_{s}h_+-(s+1)}\right), \\&\hspace{4.5cm} \bx\to \infty\,,
\end{split}
\end{equation}
with
\begin{equation}
\label{BCtildeCoeff}
    \begin{split}
        {}_{s,\nu} a_+&= \frac{2^{{}_{s}h_+-\frac{i k}{2}} \Gamma (1-2 {}_{s}h_+)  \Gamma (1-i k-(s+\nu))}{\Gamma (1-{}_{s}h_+ +i m-i k) \Gamma (1-{}_{s}h_+-i m-(s+\nu))}\\
        {}_{s,\nu} a_-&={}_{s,\nu}
        a_+\vert_{h_+\to h_-}.
    \end{split}
\end{equation}
Consider now a mode 
${}_{s}\Upsilon_{k \ell m}^{\rm near} \circeq \GHPw{2s}{0}$
\begin{equation}
\label{Upsnear}
{}_{s}\Upsilon_{k \ell m}^{\rm near}(\bar x^\mu)=e^{-i \bar \omega \bar t+im\bar \phi} {}_{s}S_{\ell m}(\bar \theta){}_{s}R_{k\ell m}^{\rm near}(\bar x),
\end{equation}
with the usual identification 
of $k=\bar \omega +m$ in the near zone. To this mode, we apply $\thorn$ using the values of the spin coefficients in nNHEK \cite{Compere2018} recalled in App. \ref{appA}. We find
\begin{equation}
\begin{split}
& \th [{}_{s}\Upsilon_{k\ell m}^{\rm near}(\bar x^\mu)] 
= e^{-i  \bar \omega \bar t+im\bar\phi} {}_{s}S_{\ell m}(\bar \theta) \\
&\times \left(
\frac{\mathrm{d}}{\mathrm{d} \bx}
-i \frac{k +m  \bx }{f}\right) {}_{s}R_{k\ell m}^{\rm near}(\bar x).
\end{split}
\end{equation}
By inspection, the last term
involving the differential operator acting on ${}_{s}R_{k\ell m}^{\rm near}$ is a solution to 
Eq. \eqref{spinraiseeqn} for $\nu=1$, satisfying the boundary condition 
\eqref{BCtilde} up to the constant 
$-ik-s$. Consequently, 
we have 
\begin{equation}
\begin{split}
& \th [{}_{s}\Upsilon_{k\ell m}^{\rm near}(\bar x^\mu)] 
= e^{-i  \bar \omega \bar t+im\bar\phi} {}_{s}S_{\ell m}(\bar \theta) \\
&\times (-ik-s){}_{s,1}R_{k\ell m}(\bar x).
\end{split}
\label{thact}
\end{equation}
Iterating this argument $\nu$ times, we find that 
\begin{widetext}
\begin{equation}
\label{highorderthorn}
\thorn^\nu
[{}_{s}\Upsilon_{k\ell m}^{\rm near}(\bar x^\mu)]
=
e^{-i\bar \omega \bar t+im\bar\phi}
{}_s S_{\ell m}(\bar \theta)
 \left\{
\prod_{j=0}^{\nu-1}\left[-ik-(s+j)\right] \right\} {}_{s,\nu}R_{k\ell m}(\bar x).
\end{equation}
\end{widetext}
Thus, we see that $\thorn$
is a kind of raising operator for the index $\nu$ in the system of functions ${}_{s,\nu}R_{k\ell m}(\bar x)$. The relation \eqref{Ncoeff} follows as the special case $\nu=4,s=-2$ of Eq. \eqref{highorderthorn}.\footnote{Observe that
\begin{equation}
\label{rtilder}
    {}_{+2}R^{\rm near}_{N\ell m}={}_{+2}C_{N\ell m}\,{}_{-2,4} R_{N\ell m}
\end{equation}
}\footnote{
Using the QNM approximation, for $m=0$ and at order $O(\varepsilon)$, the coefficient
\begin{equation}
\prod_{j=0}^{\nu-1}(-ik_N-(s+j))=\prod_{j=0}^{\nu-1}(-N-{}_{s}h_+-(s+j))
\end{equation}
vanishes whenever
\begin{equation}
N=-1-\ell-s-j.
\end{equation}
where we used ${}_{s}h_+=\ell+1+O(\varepsilon)$.}

The operator $\thorn'$ can be 
analyzed in an analogous manner. As in the main text, we use the shorthand $\zeta = (-\Psi_2)^{-\frac{1}{3}} = M^{\frac{2}{3}} (1-i\cos \bar \theta)$, noting that $|\zeta|^2$ and $\thorn'$ commute.

Analogously to Eq. \eqref{thact}, we compute
\begin{equation}
\begin{split}
& \thorn' [{}_{s} \Upsilon_{k\ell m}^{\rm near}(\bar x^\mu)] =
-\frac{1}{2M^{\frac{2}{3}}|\zeta|^2} e^{-i\bar \omega\bar t+im\bar\phi}{}_{s}S_{\ell m}(\bar \theta) \\&\times 
f\left(  
\frac{\mathrm{d}}{\mathrm{d} \bx}
+ \frac{sf'+i (k +m  \bx  )}{f} \right) {}_{s}R_{k\ell m}^{\rm near} (\bar x).
\end{split}
\end{equation}
By inspection, the term in the last line
involving the differential operator acting on ${}_{s}R_{k\ell m}^{\rm near}$ is a solution to 
Eq. \eqref{spinraiseeqn} for $\nu=-1$, satisfying the boundary condition 
\eqref{BCtilde} up to a constant 
that is easily computed. Consequently, this function must in fact be equal to 
${}_{s,-1}R_{k\ell m}$ times that constant. More precisely,
\begin{equation}
\begin{split}
&  \thorn' [{}_{s} \Upsilon_{k\ell m}^{\rm near}(\bar x^\mu)] = e^{-i\bar \omega \bar t+im\bar\phi}
{}_s S_{\ell m}(\bar \theta) \  \\
& \times 
\frac{({}_{s}h_+-i m-s) ({}_{s}h_++i m+s-1) }{2M^{\frac{2}{3}}|\zeta|^2(ik+s-1)} \ {}_{s,-1}R_{k\ell m}(\bar x).
\end{split}
\end{equation}
Iterating this formula $\nu$
times, we similarly find\footnote{
Mind that, unlike $\thorn$, the form of $\thorn'$ in the given tetrad explicitly depends on the GHP weights of the quantity that it acts on. This dependence must be properly taken into account in order to obtain the following expression.
} 
\begin{widetext}
\begin{equation}
\label{lower}
\thorn^{\prime \nu} [{}_{s} \Upsilon_{k\ell m}^{\rm near}(\bar x^\mu)] = e^{-i\bar \omega \bar t+im\bar\phi}
{}_s S_{\ell m}(\bar \theta)   \left\{ \prod_{j=0}^{\nu-1}
\frac{({}_{s}h_+-i m-s+j) ({}_{s}h_++i m+s-j-1) }{2M^{\frac{2}{3}}|\zeta|^2[ik+( s-j-1)]}\right\}
{}_{s,-\nu}R_{k\ell m}(\bar x).
\end{equation}
\end{widetext}
Thus, we see that $\thorn'$
is a kind of lowering operator for the index $\nu$ in the system of functions ${}_{s,\nu}R_{k\ell m}(\bar x)$. The formula \eqref{Dnear}
can be obtained from the above raising and lowering relations
taking $s=-2$ and $\nu = 4$ first in Eq. \eqref{highorderthorn} and then $s=2$ and $\nu=4$ in Eq. \eqref{lower}, using ${}_{2,-4}R_{N\ell m}={}_{+2}C_{N\ell m}\, {}_{-2,0}R_{N\ell m}$.

\section{\texorpdfstring{$\edth,\edth'$ }{} and ladder operators in nNHEK}
\label{edthladder}
The GHP operators $\edth,\edth'$ are related to spin-lowering and spin-raising operators in nNHEK. In the NP tetrad \eqref{nNHEK_NP}, and when acting on a GHP scalar with GHP weights $\circeq \GHPw{p}{q}$, spin $s=(p-q)/2$, and harmonic $\bar \phi$-dependence $e^{im\bar \phi}$, we may effectively substitute 
\begin{equation}
\begin{split}
\edth \to \frac{1}{\sqrt{2}M(1+i\cos \bar\theta)}\bigg( & \frac{\mathrm{d}}{\mathrm{d} \bar \theta}-m \csc \bar \theta-
s\cot\bar\theta\\&+\frac{m\sin\bar\theta}{2}+\frac{qi\sin\bar\theta}{1+i\cos\bar\theta}\bigg)\\
\edth' \to \frac{1}{\sqrt{2}M(1-i \cos\bar \theta)}\bigg(&  \frac{\mathrm{d}}{\mathrm{d} \bar \theta}+m \csc \bar \theta +s\cot\bar\theta\\& -\frac{m\sin\bar\theta}{2}
-\frac{pi\sin\bar \theta}{1-i\cos\bar \theta} \bigg)  
\end{split}
\end{equation}
In these expressions, we can recognize the operators 
\begin{subequations}
\label{eq:Chandop}
\begin{align}
&{}_s \mathcal{L}_{m}^\dagger 
 =-\frac{1}{\sqrt{2}} \left( \frac{\mathrm{d}}{\mathrm{d} \bar \theta}
- m \csc \bar \theta  - s \cot \bar \theta \right), \\ 
&{}_s \mathcal{L}_{m}   = -\frac{1}{\sqrt{2}} \left( \frac{\mathrm{d}}{\mathrm{d} \bar \theta} +m \csc \bar \theta + s \cot \bar \theta \right).
\end{align}
\end{subequations}
The operators ${}_s \mathcal{L}_{m}$ and ${}_s \mathcal{L}_{m}^\dagger$ are known (see e.g., \cite{chandrasekhar1998mathematical}) to be spin-lowering respectively spin-raising operators for the spin-weighted spherical harmonics ${}_s Y_{\ell m}$ \cite{Goldberg1967}, 
in the sense that
\begin{subequations}\label{eq:SpheroidalHarmonicsIdentities}
\begin{align}
{}_{-s} \mathcal{L}^\dagger_{m} \, {}_s Y_{\ell m} = \sqrt{\frac{(\ell - s)(\ell + s + 1)}{2}} {}_{s+1} Y_{\ell m},\\
{}_{s} \mathcal{L}_{m} \, {}_s Y_{\ell m} = -\sqrt{\frac{(\ell + s)(\ell - s + 1)}{2}} {}_{s-1} Y_{\ell m}
.\label{eq:bar sYlm}
\end{align}
\end{subequations}
These relations, and the initial condition ${}_0 Y_{\ell m} = Y_{\ell m}$, where the latter denote the ordinary unweighted spherical harmonics without the 
harmonic factor $e^{im\bar \phi}$, may be seen as a possible definition of the spin-weighted spherical harmonics. 

\section{Large \texorpdfstring{$\ell_i$}{} analysis of angular integrals}\label{statphase}

%
%
%
%

In this appendix we show how to evaluate the angular overlap integrals, $[123]$ [see Eq. \eqref{a123}] for $\ell_i \gg 1$, either for all $i=1,2,3$ or for a subset if $i$'s. To simplify the discussion, 
we shall assume that $m_i$ are arbitrary but fixed, but a 
variation of the argument would lead to the same conclusion for 
possibly large $|m_i|$ so long as $|m_i| \ll \ell_i$.

We first need to clarify the nature of our large $\ell_i$ limit. Generally, we will distinguish different cases.  
First, we consider the case when \underline{all $\ell_i, i=1,2,3$ go to infinity.} To have a single large semiclassical parameter, $L$, we set, as in the main text $\ell_i = L\bar \ell_i$, where $\bar \ell_i \ge c > 0, i=1,2,3$. 
We then view the angular overlap integral  
\eqref{a123} as a function $[123](\bar \ell_1,\bar \ell_2,\bar \ell_3)$ that is parameterized by $L$. We will generally say that such a function, $g_L(\bar \ell_1,\bar \ell_2,\bar \ell_3)$ converges to a function 
$g(\bar \ell_1,\bar \ell_2,\bar \ell_3)$, if, for any smooth compactly supported testfunction 
$\varphi$ with support restricted to $\bar \ell_i \ge c$, we have
\begin{equation}
\begin{split}
    &\lim_{L \to \infty} \int \dd^3 \bar \ell (g_L \varphi) (\bar \ell_1,\bar \ell_2,\bar \ell_3)\\
    =& 
    \int \dd^3 \bar \ell (g \varphi) (\bar \ell_1,\bar \ell_2,\bar \ell_3).
\end{split}
\end{equation}
Thus, our notion of convergence is in the weak (distributional) sense. We will write this as 
\begin{equation}
    g_L \sim_w g.
\end{equation}
This notion of convergence is useful because we would like to average over any local oscillation of the overlap integral $[123]$ in the $\ell_i$'s, and because we will need to sum the overlap coefficients over the $\ell_i$ in the formulation of our dynamical system. 

We first recall asymptotic expansions for the spin-weighted spherical harmonics. 
For fixed $m\leq0$, $s=0$, an asymptotic formula (uniform in $0\leq\theta\leq\pi/2$) for large $\ell$ has been given by \cite{Bakaleinikov2020}. Combining their 
result with well-known reflection formulas for the ordinary spherical harmonic 
under $\bar \theta \to \pi-\bar \theta$, the asymptotic formula may be presented 
for $s=0$ as\footnote{As stated, the formula is valid for $m \le 0$. For $m>0$, one can use the mentioned reflection formula.}
\begin{widetext}
	\begin{equation}
	Y_{\ell m}(\bar \theta) = \sqrt{\frac{\ell}{2\pi}}  \times	\begin{cases}
	\sqrt{\frac{\bar \theta}{\sin\bar \theta}} J_{-m}\left[\left(\ell+\frac{1}{2}\right) \bar \theta\right]\ \ &0\leq\theta\leq\frac{\pi}{2}\\ (-1)^{\ell-m}\sqrt{\frac{\pi-\bar\theta}{\sin\bar \theta} } J_{-m}\left[\left(\ell+\frac{1}{2}\right) (\pi-\bar \theta)\right] \ \ &\frac{\pi}{2}<\bar \theta\leq\pi
	\end{cases}
 + O\left(\frac{m}{\ell}\right)
 , \label{sphericalharmonicsbehaviour}
	\end{equation}
\end{widetext}
 where $J_\nu$ is a Bessel function, and where $O(m/\ell)$ roughly indicates a function of this order, uniformly in $\theta \in [0,\pi]$.
 The precise error bound is actually more subtle 
 since it involves the ``envelope'' of the Bessel function given in terms of its zeros. 
 Since the distribution of zeros is itself a difficult issue, we will proceed heuristically here and leave a more precise discussion aside.

In order to transfer this estimate for the unweighted spherical harmonics $Y_{lm}$ to the spin-weighted harmonics ${}_{s}Y_{lm}$, we use the asymptotic relationship
\begin{widetext}
	\begin{equation}
	{}_{s}Y_{\ell m}(\bar \theta) = 
 \begin{cases}
	(-1)^s Y_{\ell(m+s)} (\bar \theta)
 \ \ & 0 \le \bar\theta\leq\frac{\pi}{2}\\ 
 (-1)^{s+\ell +m} Y_{\ell(m-s)}(\pi-\bar \theta) 
 \ \ &\frac{\pi}{2}<\theta\leq\pi,
	\end{cases} 
+O\left(\frac{m}{\ell}\right)	,
 \label{expansionYLM}
	\end{equation}
 \end{widetext}
 where the convergence is uniform in $\theta \in [0,\pi]$, and the error is understood in the same sense. This relation can be obtained from \eqref{eq:SpheroidalHarmonicsIdentities} and 
 \eqref{sphericalharmonicsbehaviour}, noting that the asymptotic relation  \eqref{sphericalharmonicsbehaviour} 
 can be bootstraped to arbitrarily high derivatives when combined with \eqref{eq:SpheroidalHarmonicsIdentities} and a standard recurrence formula for the Legendre functions \cite[14.10.4]{nist}, and for the Bessel functions $J_n$ \cite[10.6.2]{nist}.

We have thereby reduced the asymptotic analysis of the angular integrals $[123]$ to the integrals
    \vspace{1cm}
	\begin{equation}
 \label{brac0}
 \begin{split}
	[123]_0 :=  L^{\frac{3}{2}} \sqrt{\frac{\bar \ell_1 \bar \ell_2 \bar \ell_3}{8\pi^3}}
 \int\limits_{0}^{\pi/2}  & \dd \bar \theta   \frac{1 - i\cos\bar \theta}{1 + i \cos \bar \theta} (\sin \bar \theta)^{-\frac{1}{2}} \bar \theta^{\frac{3}{2}} \\
 &\times
  \prod_{j=1}^3 J_{\mu_j}[(\ell_j+\tfrac{1}{2})\bar \theta],
\end{split}
	\end{equation}
 where $\mu_i = m_i+s_i$. 
 The relationship to the original 
 overlap integral \eqref{a123} is
 \begin{equation}
 \label{brac3}
     L^{\frac{1}{2}} \cdot [123] \sim_w L^{\frac{1}{2}} \cdot 
     \left\{[123]_0 + (-1)^{\mu_S+\ell_S} [123]_0^*\big\vert_{s_i\to -s_i} \right\}
 \end{equation}
 where here and in the following, we use ``$S$''
for the sum as in e.g., $\ell_S = \ell_1+\ell_2+\ell_3$.

Due to the somewhat subtle notion of error in our asymptotic estimate for the spherical harmonics, the above arguments constitute nor rigorous proof of the asymptotic relation $\sim_w$ in Eq. \eqref{brac3}. But we have tested it numerically and found very good agreement already for $\ell_i$'s
of the order of $10^2$ and $m_i$ of order unity.

We have not found a way to carry out the integral \eqref{brac0} in closed form, so we now proceed with further analysis reducing it, in the sense of $\sim_w$, to a manageable integral. 

 Into Eq. \eqref{brac0}, 
 we substitute a standard integral representation \cite[10.9.2]{nist} of the Bessel function $J_n$, which leads to
 \begin{widetext}
	\begin{equation}
	[123]_0 = 
	 L^{\frac{3}{2}} i^{\mu_S }\sqrt{
  \frac{\bar \ell_1 \bar \ell_2 \bar \ell_3 }{8 \pi^9}} \int\limits_0^{\pi/2} \dd\bar \theta \int\limits_{[0,\pi]^3} \dd^3 \boldsymbol{z} \, 
	e^{i L f(\bar \theta,\boldsymbol{z})} g(\bar \theta,\boldsymbol{z})  
	\end{equation}
 \end{widetext}
	where $\boldsymbol{z}=(z_1,z_2,z_3)$ and where 
	\begin{equation}
	\begin{split}
	f(\bar \theta,\boldsymbol{z})&=\bar \theta \sum_i \bar \ell_i \cos z_i\\
	g(\bar \theta,\boldsymbol{z})&=	\frac{e^{\frac{i\bar\theta}{2} \sum_i \cos z_i}}{(\sin\bar\theta)^{1/2}} \left[ \prod_i \cos (\mu_i z_i) \right] \frac{1- i\cos\bar\theta}{1+i \cos \bar\theta}\bar\theta^{\frac{3}{2}}.
	\end{split}
	\end{equation}
The form of the $\boldsymbol{z}$-integral suggests using the 
stationary phase method for large semiclassical parameter $L$ for $\bar \theta$ values such that $L\bar \theta \gg 1$. With this in mind, we split the $\bar \theta$ integration into the interval from $0$
to $1/\sqrt{L}$, and the interval from $1/\sqrt{L}$ to $\pi/2$. 
For the latter integral, denoted by $[123]_0^{\ge 1/\sqrt{L}}$ we may then confidently apply the stationary phase method to the $\boldsymbol{z}$ integral.

According to this method, the dominant contribution to $[123]_0^{\ge 1/\sqrt{L}}$ is determined by the points of stationary phase where $\nabla_{\boldsymbol{z}}f=0$, which are at $z_i = 0,\pi$. Evaluating that contribution leads to  
\begin{widetext}
	\begin{equation}\label{basicA123}
	\begin{split}
	L^{\frac{1}{2}} \cdot [123]_0^{\ge 1/\sqrt{L}} 
  &\sim_w \frac{\sqrt{L} e^{\frac{i\pi}{4}} i^{\mu_S }}{8 \pi ^3}\sum_{\{\pm\}}\left\{\prod_i p_i[(\pm 1)_i] \right\} \int\limits_{1/\sqrt{L}}^{\pi/2} \dd\bar \theta\frac{ (\cos \bar \theta+i) \sqrt{\csc \bar \theta} }{ \cos \bar \theta-i} e^{i L\bar \theta \sum_i (\pm 1)_i\bar \ell_i},
	\end{split}
	\end{equation}
\end{widetext}
where an error term of order $O(L^{-1/2} \log L)$
from the corresponding error in the stationary phase approximation has been discarded in Eq. \eqref{basicA123}.
The first sum is over a distinct sign $(\pm 1)_i$ for each $i=1,2,3$, corresponding to the different saddles in the stationary phase approximation and
\begin{equation}
    p_i(x)=\begin{cases}
        1, &x=+1\\
        i (-1)^{\mu_i}, & x=-1.
    \end{cases}
\end{equation}

If we now integrate the expression \eqref{basicA123} against a 
testfunction $\varphi(\bar \ell_1, \bar \ell_2, \bar \ell_3)$
compactly supported in $\bar \ell_i \ge c>0$
and use a standard integration by parts trick, then it is easy to see that the resulting expression decays faster than any inverse power of $L$. 

Hence, we see that $L^{1/2} \cdot [123]_0^{\ge 1/\sqrt{L}} \sim_w 0$, and we can 
concentrate on $L^{1/2} \cdot [123]_0^{\le 1/\sqrt{L}}$. In that integral, we may safely replace $\frac{(1- i\cos\bar\theta)\bar \theta^{1/2}}{(1+i \cos \bar\theta)(\sin \bar \theta)^{1/2}}$ by its value at $\bar \theta = 0$, i.e. $-i$. To see this more clearly, we note that the error incurred by this replacement is estimated as follows after smearing with a testfunction $\varphi(\bar \ell_1, \bar \ell_2, \bar \ell_3)$, and using again the integral representation \cite[10.9.2]{nist} of the Bessel function:
\begin{widetext}
\begin{equation}
\begin{split}
    |{\rm error}| 
    = &  \  \bigg| L^2
    \int\limits_{[0,\pi]^3} \dd^3 \boldsymbol{z} \int\limits_0^{1/\sqrt{L}} \dd \bar \theta \, O(\bar \theta^2) \ \hat \varphi(L\bar \theta \cos z_1,L\bar \theta \cos z_2,L\bar \theta \cos z_3) \bigg| \\
    \lesssim & \ L^{2-\frac{p}{2}}
    \int\limits_{[0,\pi]^3} \dd^3 \boldsymbol{z} \int\limits_0^{1/\sqrt{L}} \dd \bar \theta \, \bar \theta^{2-p} \left(1+ L^2\bar \theta^2 \sum_i \cos^2 z_i\right)^{-M} \\
    \lesssim & \ L^{-1+\frac{p}{2}}
    \int\limits_{[0,\pi]^3} \frac{\dd^3 \boldsymbol{z}}{(\sum_i \cos^2 z_i)^{(3-p)/2}} \int\limits_0^{\sqrt{L}} \dd \bar \theta \, \bar \theta^{2-p} \left(1+ \bar \theta^2 \right)^{-M}
    \lesssim  \ L^{-1+\frac{p}{2}},
\end{split}
\end{equation}
\end{widetext}
where the $\boldsymbol{z}$ integral converges provided that $p>0$, and where we used that the Fourier transformed testfunction, $\hat \varphi$, is rapidly decaying, i.e., $M$ is as large as we like, say $M>(3-p)/2$ in order to make the $\bar \theta$ integral converge for arbitrary $L$. Therefore, taking e.g., $p=1$, we see that the error can be neglected in the sense of $\sim_w$.

Thus we conclude that 
\begin{equation}
 \label{brac1}
 \begin{split}
	L^{\frac{1}{2}} \cdot [123]_0 \sim_w  & L^{\frac{1}{2}} \cdot [123]^{\le 1/\sqrt{L}}_0 \\
 \sim_w  & -iL^2 \sqrt{\frac{\bar \ell_1 \bar \ell_2 \bar \ell_3}{8\pi^3}}  \int\limits_{0}^{1/\sqrt{L}}   \dd \bar \theta  \ \bar \theta 
  \prod_{j=1}^3 J_{\mu_j}[(\ell_j+\tfrac{1}{2})\bar \theta]\\
  \sim_w  & -i\sqrt{\frac{\bar \ell_1 \bar \ell_2 \bar \ell_3}{8\pi^3}}  \int\limits_{0}^{\sqrt{L}}   \dd \bar \theta  \ \bar \theta 
  \prod_{j=1}^3 J_{\mu_j}(\bar \ell_j\bar \theta),
\end{split}
	\end{equation}
The large $L$ limit of the last integral has to be understood in the sense of distributions in the $\bar \ell_j$. It has been evaluated in the case, relevant for us, that $\mu_3=\mu_1+\mu_2, \bar \ell_j>0$ \cite[Table I]{Gervois1984}, 
\begin{widetext}
	\begin{equation}
 \begin{split}\int\limits_{0}^{\infty}   \dd \bar \theta  \ \bar \theta 
  \prod_{j=1}^3 J_{\mu_j}(\bar \ell_j\bar \theta) 
= \begin{cases}
\frac{1}{\pi \bar \ell_1 \bar \ell_2} \frac{ \cos(\mu_2 \chi_{1}- \mu_1 \chi_2)}{\sin\chi_3} & \text{if
$|\bar \ell_1-\bar \ell_2|< \bar \ell_3 < \bar \ell_1+\bar \ell_2$},
\\
0 & \text{$\bar \ell_3 < |\bar \ell_1-\bar \ell_2|$ or $\bar \ell_3>\bar \ell_1+\bar \ell_2$.}
\end{cases}
\end{split}
	\end{equation}
 \end{widetext}
 In the first case, $\bar \ell_j$
 are the side lengths of a triangle, and the angle opposite $\bar \ell_i$ is called $\chi_i$.
Inserting this back into relation \eqref{brac3} and remembering that $\mu_i=m_i+s_i$ finally leads to 
\begin{widetext}
\begin{equation}
\label{brac2}
\begin{split}
L^{\frac{1}{2}} \cdot [123] \sim_w &\frac{i}{  \sin (\chi_3)}  \sqrt{\frac{\bar \ell_3}{2\pi^5 \bar \ell_1 \bar \ell_2}} \Big[
\Theta(\bar \ell_3-|\bar \ell_1-\bar \ell_2|)
-\Theta(\bar \ell_3-|\bar \ell_1+\bar \ell_2|) \Big] \\
& \times
\begin{cases}
    \sin (m_2 \chi_1-m_1\chi_2) \sin (s_2 \chi_1-s_1 \chi_2), & \ell_S \text{ even,}\\
     - \cos (m_2 \chi_1-m_1\chi_2) \cos (s_2 \chi_1-s_1 \chi_2), &\ell_S \text{ odd.}
\end{cases}
\end{split}
\end{equation}
\end{widetext}

A similar analysis can be carried out for integrals of the form

\begin{equation}
    [123] :=\int\limits_{0}^{\pi} \dd \bar \theta
    (\sin \bar \theta) S_1Y_2Y_3(\bar \theta), 
\end{equation}
where $S_1$ is some well behaved function of $\bar \theta$ e.g., smooth on $[0,\pi]$ up to and including the boundary of the interval, which is independent of $\ell_2, \ell_3$. $Y_2, Y_3$ are spin weighted spherical harmonics, as before. In our computation of the overlap coefficients in the body of the paper, $S_1$ is a spin-raising or lowering operator applied to a spheroidal harmonic for a fixed $s_1,\ell_1,m_1$ times certain trigonometric factors. But the analysis is just the same for any $S_1$ with the above properties, so we will not specify it.

The limit that we need to consider is that \underline{only $\ell_i, i=2,3$ go to infinity.} To have a single large semiclassical parameter, $L$, we again set, as in the main text $\ell_i = L\bar \ell_i$, where $\bar \ell_i \ge c > 0, i=2,3$. 

As in the previous case, we first observe that the symmetries of the spin-weighted spherical harmonics allow us to consider separately and analogously the integral from 
$0$ to $\pi/2$, and from $\pi/2$ to $\pi$. 
In each of these intervals, we approximate the spin-weighted spherical harmonics in the large $L$ limit by Bessel functions, as in 
Eqs. \eqref{expansionYLM},\eqref{sphericalharmonicsbehaviour}. Since both resulting integrals have the same structure, we may restrict attention to only one of them, say
\begin{equation}
    [123]_0 = \frac{L\sqrt{\bar \ell_2 \bar \ell_3}}{2\pi} \int\limits_0^{\pi/2}
    \dd \bar \theta \, \bar \theta S_1(\bar \theta) J_{\mu_2}[(\ell_2+\tfrac{1}{2})\bar \theta] J_{\mu_3}[(\ell_3+\tfrac{1}{2})\bar \theta].
\end{equation}
As before, we now split the 
$\bar \theta$ integration into the interval from $0$ to $1/\sqrt{L}$ and the interval from
$1/\sqrt{L}$ to $\pi/2$. As before, the latter integral may be safely treated with the method of stationary phase, and one sees in that way that $L [123]_0^{\ge 1/\sqrt{L}} \sim_w 0$. The first integration is first transformed to the integration variable $x=L \bar \theta$. Using the smoothness of $F$ up to and including $\bar \theta = 0$, one shows by an integration by parts argument that $S_1(\bar \theta)$ may safely replaced by $S_1(0)$ up to subleading terms in inverse powers of $L$, and be pulled out of the integral. Thus, we  
can say that
\begin{equation}
    L \cdot [123]_0 \sim_w \frac{\sqrt{\bar \ell_2 \bar \ell_3} S_1(0)}{2\pi} \int\limits_0^{\sqrt{L}}
    \dd x \, x J_{\mu_2}(\bar \ell_2 x) J_{\mu_3}(\bar \ell_3 x),
\end{equation}
In the large $L$ limit, the resulting integral must be understood as a distribution in the continuous variables $\bar \ell_i$ \cite[Eq. 3.2a]{Gervois1984}, yielding altogether (recall $\mu_i=m_i+s_i$)
\begin{widetext}
\begin{equation}
    \begin{split}
     &L \cdot [123] \sim_w\Theta(\mu_3+\mu_2)\bigg\{S_1(0)[(m_2,s_2,\ell_2),(m_3,s_2,\ell_3)] 
     +(-1)^{\ell_2+\ell_3+\mu_2+\mu_3} S_1(\pi)[(m_2,-s_2,\ell_2),(m_3,-s_3,\ell_3)]  
     \bigg\} +\\
     &(-1)^{\mu_3+\mu_2}\Theta(-\mu_3-\mu_2-1)\bigg\{S_1(0) 
     [(-m_2,-s_2,\ell_2),(-m_3,-s_3,\ell_3)] 
     +(-1)^{\ell_2+\ell_3+\mu_2+\mu_3} S_1(\pi)
     [(-m_2,s_2,\ell_2),(-m_3,s_2,\ell_3)] 
     \bigg\}
    \end{split}
\end{equation}
with
\begin{equation}
\label{angdef}
\begin{split}
&[(m_2,s_2,\ell_2),(m_3,s_2,\ell_3)]
    =\frac{\sqrt{\bar \ell_2 \bar \ell_3} }{2\pi} \Bigg[\frac{\cos[\tfrac{\pi}{2}(\mu_3-\mu_2)]}{\bar \ell_2} \delta(\bar \ell_2-\bar \ell_3)+  {\rm P.P.} \left( \frac{\mu_3^2-\mu_2^2}{\bar \ell_3^2 - \bar \ell_2^2} \right) \times\\
      & \left\{ \Theta(\bar \ell_2-\bar \ell_3)
     \frac{\sin[\tfrac{\pi}{2}(\mu_3-\mu_2)]
     \Gamma[\tfrac{1}{2}(\mu_2+\mu_3)]
     \Gamma[\tfrac{1}{2}(\mu_3-\mu_2)]
     }{2\pi \Gamma(\mu_3+1)} 
     \left( \frac{\bar \ell_3}{\bar \ell_2} \right)^{\mu_3}
     {}_2 F_1 \left( 
     \frac{1}{2}(\mu_2+\mu_3),
     \frac{1}{2}(\mu_3-\mu_2),
     \mu_3+1
     ; \frac{\bar \ell_3^2}{\bar \ell_2^2}\right)
     + (2\leftrightarrow 3)
     \right\}
     \Bigg]
    \end{split}
\end{equation}
\end{widetext}
Here, $\mathrm{P.P.}$ denotes the principal part prescription for the distribution $1/x$.\footnote{Using a transformation formula for $_2 F_1$ the expression in curly brackets $\{ \dots \}$ is seen to be continuous at $\bar \ell_2 = \bar \ell_3$, with a derivative diverging at most logarithmically in $\bar \ell_3-\bar \ell_2$. 
Hence, the distributional product with the $P.P.$ term is well-defined.}

Lastly, we require in the main part of the paper integrals of the form 
\begin{equation}
    [123] :=\int\limits_{0}^{\pi} \dd \bar \theta
    (\sin \bar \theta) S_1S_2Y_3(\bar \theta), 
\end{equation}
where $S_1, S_2$ are well behaved functions of $\bar \theta$ e.g., smooth on $[0,\pi]$ up to and including the boundary of the interval. $Y_3$ is spin weighted spherical harmonic with parameters $(s_3, \ell_3, m_3)$. In our computation of the overlap coefficients in the body of the paper, $S_1, S_2$ are spheroidal harmonic for a fixed $(s_1,\ell_1,m_1), (s_2,\ell_2,m_2)$ acted upon by spin-raising or lowering operators, times certain trigonometric factors. But the analysis is just the same for any $S_1, S_2$ so we will not specify them.

The limit that we need to consider is that \underline{only $\ell_3$ go to infinity.} As before, we write $\ell_3 = L\bar \ell_3$, where $\bar \ell_3 \ge c > 0$. Similar to the previous cases, we may reduce consideration to the integral 
\begin{equation}
    [123]_0 = L^{\frac{1}{2}}\sqrt{\frac{\bar \ell_3}{2\pi}} \int\limits_0^{\pi/2}
    \dd \bar \theta \, \bar \theta 
    \left( \frac{\bar \theta}{\sin \bar \theta}\right)^{\frac{1}{2}}S_1S_2(\bar \theta) J_{\mu_3}\left[(\ell_3+\tfrac{1}{2})\bar \theta\right].
\end{equation}
Furthermore, again by arguments analogous to those in the previous cases, one sees that 
\begin{equation}
    L^{\frac{3}{2}} \cdot [123]_0 \sim_w \sqrt{\frac{\bar \ell_3}{2\pi}} S_1 S_2(0) \int\limits_0^{\sqrt{L}}
    \dd \bar \theta \, \bar \theta 
\, J_{\mu_3}(\bar \ell_3\bar \theta).
\end{equation}
The large $L$ limit of this integral must again be understood in the sense of distributions in $\bar \ell_3$. This integral may be evaluated using formulas of \cite[Sec. 1.2]{Rosenheinrich} or\footnote{With a Gaussian cutoff instead of the sharp cutoff at the upper integration boundary $\sqrt{L}$, which one expects to give equivalent limits.} \cite[10.22.54, 10.2.23]{nist}. Using also that $\bar \ell_3\ge c>0$, 
one finds 
\begin{equation}
    L^{\frac{3}{2}} \cdot [123]_0 \sim_w \frac{\mu_3}{\sqrt{2\pi} \bar\ell_3^{\frac{3}{2}}}  S_1S_2(0).
\end{equation}
which holds for $\mu_3>-2$. The other cases can be dealt with using the symmetries of the Bessel function. This leads to 
\begin{widetext}
\begin{equation}
    L^{\frac{3}{2}} \cdot [123] \sim_w \frac{1}{\sqrt{2\pi}
    \bar \ell_3^{\frac{3}{2}}}
    \left[
    (m_3+s_3)S_1S_2(0) + (-1)^{m_3+\ell_3}(m_3-s_3)S_1S_2(\pi) \right][\Theta(m_3+s_3)-(-1)^{m_3+s_3}\Theta(-m_3-s_3-1)].
\end{equation}
\end{widetext}

\section{Large \texorpdfstring{$\ell_i$}{} analysis of radial integrals}
\label{radoverlap}

In this section, we give some details concerning the evaluation of the radial overlap integrals $\{123\}$ [see Eq. \eqref{r123}]. These involve the generalized radial functions \eqref{tildeR1} $R_i \equiv {}_{\nu_i,s_i}R_{N_i \ell_i m_i}(\bar x),$ where the indices $s_i,\,\nu_i,\, \ell_i, \mu_i, N_i$ for $i=1,2,3$ are understood below but are often suppressed to lighten the notations. 

In accordance with the discussion in Sec. \ref{axisym}, see Eqs. \eqref{Upsaxi1}, \eqref{Upsaxi2}, of the axisymmetric $(m=0)$
modes, the index $\mu_i$ means
\begin{equation}
    \mu_i = \begin{cases}
        m_i & \text{if $m_i = \pm 1, \pm 2, \dots$,}\\
        -i\varepsilon(N_i+\ell_i+1) & \text{if $m_i=0$.}
    \end{cases}
\end{equation}
To simplify our formulas, we will also omit the normalization constant ${}_s C_{N\ell m}$ in Eq. \eqref{tildeR1}, and we will correspondingly denote the overlap integral \eqref{r123} $\{123\}$ by $\{123\}_0$, to emphasize this distinction.

Similarly to Eq. \eqref{product}, we define the generalized symbols $P_j^{(N_1,N_2,N_3)}$ by the triple product in the integrand of Eq. \eqref{r123},

\begin{widetext}
\begin{equation}
    f^2 R_1 R_2 R_3 = 
    \sum_{j=0}^{N_1+N_2+N_3}(-1)^j P^{(N_1,N_2,N_3)}_{j}  \left(1+\frac{\bar x}{2}\right)^{{\hat\beta}}\left(\frac{\bar x}{2} \right)^{j+\hat \alpha},
\end{equation}
\end{widetext}
where
\begin{equation}
\begin{split}
    {\hat\alpha}&:=2-
    \left(s_S + \nu_S +\frac{ik_S}{2}\right)\\
    {\hat \beta}&:=2-i\left(\frac{k_S}{2}-\mu_S\right),
    \end{split}
\end{equation}
where again, a subscript ``$S$'' means the sum e.g., 
\begin{equation}
s_S := \sum_{i=1}^3 s_i, 
\quad 
\text{or}
\quad
k_S := \sum_{i=1}^3 {}_{s_i} k_{N_i \ell_i m_i}.
\end{equation}

The integral \eqref{r123} can now be computed in the limit $\varepsilon\to0$ similarly as we did in Sec. \ref{sec:BilinearNnHEK}, which leads to
\begin{equation}
\label{radialoverlapN}
\begin{split}
    \{123\}_0=&2^{{\hat \alpha}+1} \sum_{j=0}^{N_1+N_2+N_3}(-1)^jP_{j}^{(N_1,N_2,N_3)} \\&\times\frac{\Gamma (j+{\hat\alpha} +1) \Gamma (-j-{\hat\alpha} -{\hat\beta} -1)}{\Gamma (-{\hat\beta} )}.
    \end{split}
\end{equation}
Similarly as in the case of the normalization of the QNMs treated in Sec. \ref{sec:BilinearNnHEK}, the analysis of the radial overlap coefficients \eqref{radialoverlapN} appears to be difficult due to the complicated sums appearing in the expression \eqref{radialoverlapN}, and implicitly in $P_{j}^{(N_1,N_2,N_3)}$. However, we have made progress when the angular momenta \underline{$\ell_i$ are large compared to both $m_i, N_i$ for all $i=1,2,3$}. To have a single large parameter, $L$, in formulas with multiple $\ell_i$'s, we set $\ell_i = L\bar \ell_i$ as in the main text, where $\bar \ell_i \ge c, i=1,2,3$, where $c$ is some strictly positive constant.

We begin by considering the symbol ${}_{s,\nu} P^{(N)}_j$ \eqref{PN1} for large $\ell$. Using Eqs. \eqref{eq:hpm} and \eqref{Eexpand} for the large $\ell$ expansion of $h_+$, and using Stirling's formula \cite[5.11.7]{nist}, one has the asymptotic expansion to the needed order
\begin{widetext}
\begin{equation}
\label{PNsing}
\begin{split}
    {}_{s,\nu}P^{(N)}_j = &\ \frac{2^j (-N)_j}{j!}\Bigg\{1  +\frac{j}{ 4 \ell}   \left[j-4 i m-2 N-4(s+\nu)+1\right] \\&
    + \frac{j}{32\ell^2  }  \Bigg[j^3+j^2 [-8 i m-4 N-8 (s+\nu)+6]+ 4j \left[-4 m^2+2 i m (2 N+4 s+4\nu-3)+(N+2 s+2\nu)^2\right]-16j N \\&+2 \left[-8 m^2+8 i m (2 N+2 s+2\nu+1)+6 N^2+2 N (8 s+8\nu+1)+8 (s+\nu) (s+\nu+1)-3\right]-24 (s+\nu) j-j\Bigg] \Bigg\}\\
    &+O(\ell^{-3}).
  \end{split}
\end{equation}
\end{widetext}
Note that the leading order term is independent of $s,\nu$ and $m$, though the subleading terms are not.
From Eq. \eqref{PNsing}, one can obtain a corresponding asymptotic formula for 
the symbols $P^{(N_1,N_2)}_j$ which appear in Eq. \eqref{product}, and which are sums of the $P^{(N)}_j$, the leading term of which is e.g.,
\begin{equation}
P^{(N_1,N_2)}_j = \frac{2^j \Gamma (-N_1-N_2+j)}{\Gamma (j+1) \Gamma (-N_1-N_2)} +O\left(L^{-1}\right).
\end{equation}
Iterating this sum, we obtain the asymptotic behavior of the 
symbols $P_{j}^{(N_1,N_2,N_3)}$ which are triple sums of the $P^{(N)}_j$, the leading term of which is e.g.,
\begin{equation}
   P_j^{(N_1,N_2,N_3)} = \frac{2^{j} \Gamma (-N_1-N_2-N_3+j)}{\Gamma (j+1) \Gamma (-N_S)} + O(L^{-1}).
\end{equation}
From the asymptotic expansion, carried out at the needed order,  
\begin{widetext}
\begin{equation}
\label{Gammaasympt}
\begin{split}
&\frac{\Gamma (j+{\hat\alpha} +1) \Gamma (-j-{\hat\alpha} -{\hat\beta} -1)}{\Gamma (-{\hat\beta} )}
=\frac{\sqrt{\pi }}{\sqrt{\ell_S}}(-1)^{1+j} 2^{-j+\ell_S+N_S+(s_S+\nu_S)-\frac{5}{2}} \sec \left[\frac{1}{2} \pi  (\ell_S+i \mu_S+N_S+2(s_S+\nu_S))\right]\\&\times\Bigg\{1-\frac{15 \log (2) \left(\ell_1 \ell_2 m_3^2+\ell_1 \ell_3 m_2^2+\ell_2 \ell_3 m_1^2\right)}{16 \ell_1 \ell_2 \ell_3}+\frac{2 [m_S-i (s_S+\nu_S-j)]^2-2 N_S-2 (s_S+\nu_S)+2 j+5}{4 \ell_S}\\& +\frac{B_0}{32 \ell_S^2}+\frac{1}{\ell_S}\sum_{i=1}^3\frac{B_{i}}{\ell_i}+\log(2)\sum_{i=1}^3\frac{15m_i^2}{32\ell_i^2}\left(\frac{15 m_i^2 \log (2)}{6}+1\right)+\frac{225 \log^2(2)}{256}\left(\frac{m_1^2 m_2^2}{\ell_1 \ell_2}+\frac{ m_1^2 m_3^2 }{ \ell_1 \ell_3}+\frac{ m_2^2 m_3^2 }{ \ell_2 \ell_3}\right)\\&+\frac{15}{32\ell_S^2}\left(\frac{ \ell_2 m_1^2}{ \ell_1 }+\frac{ \ell_1 m_2^2}{ \ell_2 }+\frac{ \ell_3 m_1^2}{ \ell_1 }+\frac{ \ell_1 m_3^2}{ \ell_3 }+\frac{ \ell_3 m_2^2}{ \ell_2 }+\frac{ \ell_2 m_3^2}{ \ell_3 }\right)\Bigg\}+ O(2^{\ell_S} L^{-\frac{7}{2}})
    \end{split}
\end{equation}
where 
\begin{equation}
    \begin{split}
        B_0&=m_S^2 \Big[-24 \Big(j^2-2 j (s_S+\nu_S)+(s_S+\nu_S)^2+s_S+\nu_S\Big)+24 j-24 N_S+67\Big]\\&+4 \Big[(s_S+\nu_S-j)^2 \Big(j^2-2 j (s_S+\nu_S+3)+(s_S+\nu_S)^2+6 (s_S+\nu_S)-10\Big) \\
       & \ \ \ \ \ \ \ +6 N_S (s_S+\nu_S-j) (-j+s_S+\nu_S+1)+3 N_S^2\Big]\\&-16 i m_S^3 (s_S+\nu_S-j)+8 i m_S (s_S+\nu_S-j) \Big[(6-4 j) (s_S+\nu_S)+2 (j-3) j+6 N_S+2 (s_S+\nu_S)^2-13\Big]\\&+60 j-30 (m_1 m_2+m_1 m_3+m_2 m_3)+4 m_S^4-60 N_S-60 s+73\\
        B_i&=-\frac{15}{64}  m_i^2 \log (2) \Big[2 (m_S-i (s_S+\nu_S)+i j)^2-2 N_S-2 (s_S+\nu_S)+2 j+5\Big],
    \end{split}
\end{equation}
\end{widetext}
obtained using Eq. \eqref{omegadoublescal} and $k=\bar \omega+m$, Stirling's formula \cite[5.11.7]{nist}, the Euler reflection formula \cite[5.5.3]{nist}, and the functional 
identity \cite[5.5.1]{nist} of the Gamma function.
Using the leading large $L$ terms in Eqs. \eqref{PNsing}, \eqref{Gammaasympt}, \eqref{eq:hpm} and \eqref{Eexpand}, we obtain the leading large $L$ asymptotic 
behavior of the sum in \eqref{radialoverlapN} as:
\begin{widetext}
\begin{equation}
\begin{split}
&\sum_{j=0}^{N_1+N_2+N_3}(-1)^jP_{j}^{(N_1,N_2,N_3)} \frac{\Gamma (j+{\hat\alpha} +1) \Gamma (-j-{\hat\alpha} -{\hat\beta} -1)}{\Gamma (-{\hat\beta} )} \\
=& -\sec \left[\frac{1}{2} \pi  (\ell_S+i \mu_S+N_S+2 (s_S+\nu_S))\right] \frac{ \sin (\pi  N_S) 2^{\ell_S+N_S+s_S+\nu_S-\frac{5}{2}} }{N_S\sqrt{\ell_S\pi }} +O\left(2^{\ell_S}L^{-\frac{3}{2}}\right).
   \end{split}
\end{equation}
\end{widetext}
We note that the explicit term on the right side vanishes unless $N_i=0$ for all $i=1,2,3$, or equivalently unless $N_S=0$. In the latter case, we obtain the leading large $L$ behaviour as
\begin{widetext}
   \begin{equation}
   \label{I12}
    \begin{split}
    \{123\}_0^{N_S=0} =& 
     -\sqrt{\pi}2^{\frac{1}{2}(\ell_S-im_S-2)} (L\bar \ell_S)^{-\frac{1}{2}}  
     \sec \left[\frac{1}{2} \pi  (\ell_S+i \mu_S+2(s_S+\nu_S))\right] 
     \\&+O\left(2^{\frac{\ell_S}{2}}L^{-\frac{3}{2}}\right).
    \end{split}
\end{equation} 
\end{widetext}
However e.g., for $N_S=1,2$, to find the leading large $L$ behaviour it is necessary that we repeat the calculations including next to leading terms in the large $L$ expansions
\eqref{PNsing}, \eqref{Gammaasympt}, \eqref{eq:hpm} and \eqref{Eexpand}. This gives the following leading large $L$ behavior for $N_S=2$ at order $2^{\ell_S/2}L^{-3/2}$:
\begin{widetext}
   \begin{equation}
   \label{radialoverlap:N2}
    \begin{split}
    \{123\}_0^{N_S=2} =& 
     -\sqrt{\pi}2^{\frac{1}{2}(\ell_S-i m_S)} (L\bar \ell_S)^{-\frac{3}{2}} 
     \sec \left[\frac{1}{2} \pi  \big(\ell_S+i \mu_S+2(s_S+\nu_S)\big)\right] F^{N_S=2}
     \\&+O\left(2^{\frac{\ell_S}{2}}L^{-\frac{5}{2}}\right),
    \end{split}
\end{equation} 
\end{widetext}
where 
\begin{equation}
\label{F2}
    F^{N_S=2} := 
     \begin{cases}
     -2 & N_i = N_j = 1, N_k=0\\
     \frac{\bar\ell_j+\bar\ell_k-\bar\ell_i}{\bar \ell_i} & N_i=2, N_j=N_k=0,
     \end{cases}
\end{equation}
Increasing $N_S$ further, the previously leading term in $\{123\}_0$ at order $2^{\ell_S/2}L^{-3/2}$ is now seen to vanish for $N_S>2$. Thus, to find the leading term, next to next to leading terms in the large $L$ expansions \eqref{PNsing}, \eqref{Gammaasympt}, \eqref{eq:hpm} and \eqref{Eexpand} must be used in the intermediate steps. The leading behavior of the result is now at order $2^{\ell_S/2}L^{-5/2}$ for $N_S=3,4$, and the expression for $N_S=4$ is, in fact, 
\begin{widetext}
\label{radialoverlap:N4}
    \begin{equation}
    \begin{split}
        \{123\}_0^{N_S=4}=&-\sqrt{\pi} 2^{\frac{1}{2}(\ell_S-im_S +2)} (L\bar\ell_S)^{-\frac{5}{2}} \sec \left[\frac{1}{2} \pi  \big(\ell_S+i \mu_S+2 (s_S+\nu_S)\big)\right] F^{N_S=4} \\
        &+O\left(2^{\frac{\ell_S}{2}}L^{-\frac{7}{2}}\right)
        \end{split}
    \end{equation}
\end{widetext}
where 
\begin{equation}
\label{F4}
            F^{N_S=4}:=
            \begin{cases}
            \frac{3 (\bar \ell_i+\bar\ell_j-\bar\ell_k)^2 }{\bar \ell_k^2 } &\qquad N_i=N_j=0,\, N_k=4
           \\ -\frac{6 (\bar\ell_i+\bar\ell_j-\bar\ell_k)  }{\bar\ell_k}  &\qquad N_i=0,\,N_j=1,\, N_k=3\\
           \frac{ \bar\ell_i^2-\bar\ell_j^2+10 \bar\ell_j\bar\ell_k-\bar\ell_k^2  }{\bar\ell_j\bar\ell_k } &\qquad N_i=0,\, N_j=N_k=2\\
           -\frac{2 (\bar\ell_i+\bar\ell_j-5 \bar\ell_k) }{\bar\ell_k } &\qquad N_i=N_j=1,\, N_k=2.
        \end{cases}
\end{equation}
The expressions for the radial overlap integrals $\{123\}$ for $N_S=0,2,4$ as given have the following similarities: 
\begin{enumerate}
\item[(i)] In all cases the dependence on $s_i,\nu_i$ and $m_i$ is only through the sum $s_S+\nu_S$ and $m_S$, 
\item[(ii)] the leading large $L$
behavior is $(L\bar \ell_S)^{-\lfloor (N_S+1)/2 \rfloor - 1/2}$, 
\item[(iii)] the remaining terms are homogeneous Laurent polynomials in the $\bar \ell_i$'s, where the maximum power is $\lfloor N_S/2 \rfloor$. We also note that terms with odd $N_S$ have the analogous scaling in $L$ [see (ii)]; however the $U^{\rm near}_{123}$ and $V^{\rm near}_{123}$ are subleading in those cases, due to the factors $L^{+N_i/2}$ in our normalization for the $c_q$. 
\end{enumerate}
This evidence, and further numerical experimentation that we do not describe in detail here, suggests that, for general even $N_S$ 
\begin{widetext}
\begin{equation}
    \begin{split}
    \{123\}_0^{N_S}=&
    - \sqrt{\pi} 2^{\frac{1}{2}(\ell_S-im_S-2+N_S)} (L\bar \ell_S)^{-\lfloor \frac{N_S+1}{2} \rfloor - \frac{1}{2}} \sec \left[\frac{1}{2} \pi  \big(\ell_S+i \mu_S+2(s_S+\nu_S)\big)\right] \\
     &\times F^{N_S}(\{ N_i, \bar \ell_i\}) 
     +O\left(2^{\frac{\ell_S}{2}}L^{-\lfloor \frac{N_S+1}{2} \rfloor-\frac{3}{2}}\right),
    \end{split}
\end{equation}
\end{widetext}
where $F^{N_S}$ is a homogeneous rational function of the $\bar \ell_i$'s (the scaled angular momenta).

A similar analysis can be carried out, in principle, when
e.g., \underline{only $\ell_i, i=2,3$ go to infinity,} but $\ell_1$ remains finite. To have a single large semiclassical parameter, $L$, we set, as in the main text $\ell_i = L\bar \ell_i$, but now only for $\bar \ell_i \ge c > 0, i=2,3$. Then we find e.g., 
\begin{widetext}
   \begin{equation}
    \begin{split}
    \{123\}_0^{N_S=0} =&-\sqrt{\pi } 2^{\frac{1}{2} (i k_1+\ell_2+\ell_3- im_1-im_S-3 )}(
    \ell_2+\ell_3)^{-\frac{1}{2}}   \csc \left[\frac{1}{2} \pi  \big(ik_1-im_1+\ell_2+\ell_3+i m_S+2 (s_S+\nu_S) \big)\right]
     \\&+O\left(2^{\frac{\ell_2+\ell_3}{2}}L^{-\frac{3}{2}}\right).
    \end{split}
\end{equation} 
\end{widetext}
The formula is consistent with \eqref{I12}. Indeed, for large $\ell_1$ we have $k_1=-i h_1+m_1\sim-i(\ell_1+1)+m_1$, whereas for $\ell_1\ll \ell_i, \, (i=2,3)$ we can say that $\ell_S\sim\ell_2+\ell_3$.

Finally, when only \underline{only $\ell_3=L \bar \ell_3$ goes to infinity,} but $\ell_1,\,\ell_2$ remain finite, we can reproduce the same analysis as the previous cases and get e.g.,
\begin{widetext}
   \begin{equation}
    \begin{split}
    \{123\}_0^{N_S=0} =&\sqrt{\pi } 2^{\frac{1}{2} (i k_1+ik_2+\ell_3- im_1-im_2-im_S-2 )}
    \ell_3^{-\frac{1}{2}}   \sec \left[\frac{1}{2} \pi  \big(ik_1+ik_2-im_1-im_2+\ell_3+i m_S+2 (s_S+\nu_S) \big)\right]
     \\&+O\left(2^{\frac{\ell_2+\ell_3}{2}}L^{-\frac{3}{2}}\right).
    \end{split}
\end{equation} 
\end{widetext}

\section{Overlap coefficients for \texorpdfstring{$N_S>0$}{}}
\label{app:higher N}
For completeness, we discuss here the overlap coefficients for non-zero overtone numbers $N_i$. 
We use the convention $N_S=N_1+N_2+N_3$. \underline{$N_S=2$}:

\vspace{0.5cm}

\begin{widetext}
\begin{equation}
\label{UVdefn2}
\begin{split}
V_{123}^{N_S=2} &= \bar \ell_S^{-1}\frac{\bar \ell_1^{\frac{N_1}{2}}\bar \ell_2^{\frac{N_2}{2}}\bar \ell_3^{\frac{N_3}{2}}}{\sqrt{N_1! N_2! N_3!}} F^{N_S=2}_{123} V^{N_S=0}_{123},\\
    U^{N_S=2}_{123} &= 
    \bar \ell_S^{-1}\frac{\bar \ell_1^{\frac{N_1}{2}}\bar \ell_2^{\frac{N_2}{2}}\bar \ell_3^{\frac{N_3}{2}}}{\sqrt{N_1! N_2! N_3!}} F^{N_S=2}_{123}  U^{N_S=0}_{123},
\end{split}
\end{equation}
\end{widetext}
where $F^{N_S=2}_{123}$ is the homogeneous Laurent polynomial of the $\bar \ell_i$'s defined by Eq. \eqref{F2}.
Again, we are discarding consistently terms of order $ O(L^{-1/2})$.

\underline{$N_S=4$}:
\begin{widetext}
\begin{equation}
\label{UVdefn3}
\begin{split}
V_{123}^{N_S=4} &=  \bar \ell_S^{-2}\frac{\bar \ell_1^{\frac{N_1}{2}}\bar \ell_2^{\frac{N_2}{2}}\bar \ell_3^{\frac{N_3}{2}}}{\sqrt{N_1! N_2! N_3!}} F^{N_S=4}_{123}   V_{123}^{N_S=0},\\
    U^{N_S=4}_{123} &= \
    \bar \ell_S^{-2}\frac{\bar \ell_1^{\frac{N_1}{2}}\bar \ell_2^{\frac{N_2}{2}}\bar \ell_3^{\frac{N_3}{2}}}{\sqrt{N_1! N_2! N_3!}} F^{N_S=4}_{123}  U_{123}^{N_S=0} ,
    \end{split}
\end{equation}
\end{widetext}
where $F^{N_S=2}_{123} $ is the homogeneous Laurent polynomial of the $\bar \ell_i$'s defined by Eq. \eqref{F4}.
Again, we are discarding consistently terms of order $ O(L^{-1/2})$.

The general structure of the overlap coefficients evident in the above expressions for $N_S \le 4$ depends on certain non-trivial structural properties of the radial overlap integrals that we have found in App. \ref{radoverlap}, (i)--(iii), for $N_S \le 4$, and that we conjecture to be the case generally.  In particular, as seen from the above expressions, the $U_{123}^{N_S}, V_{123}^{N_S}$ are homogeneous Laurent polynomials in $\bar \ell_i$, with no explicit dependence upon $L$, and this will be the case for all $N_S$, if the features of the radial overlap integrals in App. \ref{radoverlap}, (i)--(iii), hold true for all $N_S$, as numerical experiments suggest is the case. These structural properties would also imply that the overlap coefficients $U_{123}^{N_S}, V_{123}^{N_S}$ when $N_S$ is odd are  subleading in the large $L$ limit, 
and so could be put to zero consistently in our regime.

\section{Overlap coefficients}
\label{sec:exploverlap}

\subsection{(high),(low) \texorpdfstring{  $\to$}{} (high)}
Here we give the explicit form of the overlap coefficients for one low $\ell$ and two high $\ell$ for vanishing overtone numbers, i.e., $N_i=0$. 

In the following formulas, $\ell_3$ corresponds to the low $\ell$ QNM. 
Since $U^{\rm near}_{123}$ is symmetric in $(23)$, the case when $\ell_2$
is the low $\ell$ QNM is obtained by a relabelling. On the other hand, 
$V^{\rm near}_{123}$ is not symmetric in $(23)$, but the case when $\ell_2$
is the low $\ell$ QNM gives a subleading result.

$h_3 \equiv {}_2h_{\ell_3 m_3}$ is as given in Eq. \eqref{eq:hpm} with a ``$+$'' and is assumed to be real, otherwise $h_3$ should be replaced by its 
complex conjugate in the formula for $V^{\rm near}_{123}$. The symbols $\left[\left(m_1, s_1, \ell_1\right),\left(m_2,s_2,\ell_2\right)\right]$ are defined in Eq. \eqref{angdef}. ${}_sS_{\ell m}$
are the spin-weighted spheroidals
as defined in Eq.~\eqref{maxspheroidal} when $m\neq 0$, and in Eq.~\eqref{Yexpand} when $m=0$. The spin raising and lowering operators ${}_s \mathcal{L}_m^{\dagger}$ are defined in Eqs. \eqref{eq:Chandop}.
As discussed more fully in App. \ref{axisym}, in the case $m_3=0$, $h_3$
should be replaced by Eq. \eqref{Hsmall}, and in other places, every occurrence stemming from the $m_i=0$ radial functions should be replaced by $m_i\to -i\varepsilon(\ell_i+1)$.

\begin{widetext}
\begin{equation}
    \begin{split}\label{Uhlh}
         &U^{\rm near}_{123}  =2^{ \frac{3}{2}} \pi^2  (-1)^{m_3} \delta_{m_1,m_2+m_3} \\&\times \frac{\bar \ell_2 {}^{\frac{9}{2}}\bar \ell_1{}^{-3}}{\sqrt{\bar \ell_2+\bar \ell_1}}  \left[\frac{\mathrm{ev}(\ell_1+\ell_2)}{\sin \left(\frac{\pi  h_3}{2}\right) i \coth (-\pi  m_1)+\cos\left(\frac{\pi  h_{3}}{2}\right) }-\frac{i \mathrm{odd}(\ell_1+\ell_2)}{\cos\left(\frac{\pi h_3}{2}\right) i \coth (-\pi  m_1)-\sin\left(\frac{\pi h_3}{2}\right)}\right] \\&\times\Bigg\{\frac{1}{2}\Big[- \Theta\left(m_1+m_2+2\right)\Big(i{}_{-2}S_{\ell_3 -m_3}(0)\left[\left(m_1, 2, \ell_1\right),\left(m_2,0,\ell_2\right)\right]\\&\hspace{3cm}-(-1)^{\ell_1+\ell_2+m_1+m_2}i{}_{-2}S_{\ell_3 -m_3}(\pi)\left[\left(m_1,-2, \ell_1\right),\left(m_2,0, \ell_2\right)\right]\Big) \\
& -(-1)^{m_1+m_2} \Theta\left(-m_1-m_2-3\right)\Big(i{}_{-2}S_{ \ell_3 -m_3}(0)\left[\left(-m_1,-2, \ell_1\right),\left(-m_2,0, \ell_2\right)\right]\\&\hspace{3cm}-(-1)^{\ell_1+\ell_2+m_1+m_2}i{}_{-2}S_{\ell_3 -m_3}(\pi)\left[\left(-m_1, 2, \ell_1\right),\left(-m_2,0,\ell_2\right)\right]\Big)\Big]\prod_{j=0}^1(- h_3-im_3 +2-j)  \\&\hspace{1cm}+ \Big[\Theta\left(m_1+m_2+4\right)\left\{ -i\left({}_{-1}\mathcal L_{-m_3}^{\dagger}\, {}_{-2}\mathcal L_{-m_3}^{\dagger}\, {}_{-2}S_{ \ell_3-m_3}\right)(0) \left[ \left(m_1, 2, \ell_1 \right),\left(m_2, 2, \ell_2\right) \right]\right.\\&\left.\hspace{3cm}+(-1)^{\ell_1+\ell_2+m_1+m_2}i \left({}_{-1}\mathcal L_{-m_3}^{\dagger}\, {}_{-2}\mathcal L_{-m_3}^{\dagger}\, {}_{-2}S_{\ell_3-m_3}\right)(\pi)\left[ \left(m_1,-2, \ell_1\right),\left(m_2,-2, \ell_2 \right)\right]\right\} \\
& +(-1)^{m_1+m_2} \Theta\left(-m_1-m_2-5\right)\left\{-i\left({}_{-1}\mathcal L_{-m_3}^{\dagger}\, {}_{-2}\mathcal L_{-m_3}^{\dagger}\, {}_{-2}S_{\ell_3-m_3}\right)(0)\left[\left(-m_1,-2, \ell_1\right),\left(-m_2,-2, \ell_2\right)\right]\right.\\&\left.\hspace{3cm}+(-1)^{\ell_1+\ell_2+m_1+m_2} i \left({}_{-1}\mathcal L_{-m_3}^{\dagger}\, {}_{-2}\mathcal L_{-m_3}^{\dagger}\, {}_{-2}S_{ \ell_3 -m_3}\right)(\pi)\left[\left(-m_1, 2, \ell_1\right),\left(-m_2, 2, \ell_2\right)\right]\right\}\Big]\\&\hspace{1cm} +\sqrt{2} \Big[ \Theta\left(m_1+m_2+3\right)\left\{-i\left({}_{-2}\mathcal L^{\dagger}_{-m_3}\,_{-2}S_{\ell_3 -m_3}\right)(0)\left[\left(m_1, 2, \ell_1\right),\left(m_2, 1, \ell_2\right)\right]\right.\\&\left.\hspace{3cm}-(-1)^{\ell_1+\ell_2+m_1+m_2}i \left({}_{-2}\mathcal L^{\dagger}_{-m_3}\,_{-2}S_{\ell_3 -m_3}\right)(\pi)\left[\left(m_1,-2, \ell_1\right),\left(m_2,-1, \ell_2\right)\right]\right\} \\
& -(-1)^{m_1+m_2} \Theta\left(-m_1-m_3-4\right)\left\{-i\left({}_{-2}\mathcal L^{\dagger}_{-m_3}\,_{-2}S_{\ell_3 -m_3}\right)(0)\left[\left(-m_1,-2, \ell_1\right),\left(-m_2,-1, \ell_2\right)\right]\right.\\&\left.\hspace{1cm}-(-1)^{\ell_1+\ell_2+m_1+m_2}i\left({}_{-2}\mathcal L^{\dagger}_{-m_3}\,_{-2}S_{\ell_3-m_3}\right)(\pi)\left[\left(-m_1, 2, \ell_1\right),\left(-m_2, 1, \ell_2\right)\right]\right\}\Big](- h_3-i m_3+2) \Bigg\}
    \end{split}
\end{equation}
and 
\begin{equation}
    \begin{split} \label{Vhlh}
       & V^{\rm near}_{123}= 2^{ \frac{3}{2}} \pi^2 (-1)^{m_3} \delta_{m_1,m_2-m_3}\\
    &\times\frac{\bar \ell_2{}^{\frac{9}{2}}\bar \ell_1{}^{-3}}{\sqrt{\bar\ell_1+\bar\ell_2}} \left[\frac{\mathrm{ev}(\ell_1+\ell_2)}{\sin \left(\frac{\pi h_3}{2}\right) i \coth (-\pi  m_1)+\cos\left(\frac{\pi h_3}{2}\right) }-\frac{i \mathrm{odd}(\ell_1+\ell_2)}{\cos\left(\frac{\pi h_3}{2}\right) i \coth (-\pi  m_1)-\sin\left(\frac{\pi h_3}{2}\right)}\right]\\&
    \times\Bigg\{\frac{1}{2}\Big[ - \Theta\left(m_1+m_2+6\right)\Big(-i {}_{-2}S_{\ell_3 -m_3}(0)\left[\left(m_1,2 , \ell_1\right),\left(m_2, 4, \ell_2\right)\right]\\&\hspace{3cm}+(-1)^{\ell_1+\ell_2+m_1+m_2} i{}_{-2}S_{\ell_3 -m_3}(\pi)\left[\left(m_1,-2, \ell_1\right),\left(m_2,-4, \ell_2\right)\right]\Big) \\
& -(-1)^{m_1+m_2} \Theta\left(-m_1-m_2-7\right)\Big( -i{}_{-2}S_{\ell_3 -m_3}(0)\left[\left(-m_1,-2, \ell_1\right),\left(-m_2,-4, \ell_2\right)\right]\\&\hspace{3cm}+(-1)^{\ell_1+\ell_2+m_1+m_2} i{}_{-2}S_{\ell_3 - m_3}(\pi)\left[\left(-m_1, 2, \ell_1\right),\left(-m_2, 4, \ell_2\right)\right]\Big)\Big] \prod_{j=0}^1(-h_3+im_3 +2-j) \\&\hspace{1cm} - \Big[\Theta\left(m_1+m_2+4\right)\left\{-i\left({}_{-1}\mathcal L_{-m_3}^{\dagger}\, {}_{-2}\mathcal L_{-m_3}^{\dagger}\, {}_{-2}S_{\ell_3 -m_3}\right)(0)\left[\left(m_1, 2, \ell_1\right),\left(m_2, 2, \ell_2\right)\right]\right.\\&\left.\hspace{3cm}+(-1)^{\ell_1+\ell_2+m_1+m_2}i \left({}_{-1}\mathcal L_{-m_3}^{\dagger}\, {}_{-2}\mathcal L_{-m_3}^{\dagger}\, {}_{-2}S_{\ell_3-m_3}\right)(\pi)\left[\left(m_1,-2,\ell_1\right),\left(m_2,-2, \ell_2\right)\right]\right\} \\
& +(-1)^{m_1+m_2} \Theta\left(-m_1-m_2-5\right)\left\{-i\left({}_{-1}\mathcal L_{-m_3}^{\dagger}\, {}_{-2}\mathcal L_{-m_3}^{\dagger}\, {}_{-2}S_{\ell_3-m_3}\right)(0)\left[\left(-m_1,-2, \ell_1\right),\left(-m_2,-2, \ell_2\right)\right]\right.\\&\left.\hspace{3cm}+(-1)^{\ell_1+\ell_2+m_1+m_2} i \left({}_{-1}\mathcal L_{-m_3}^{\dagger}\, {}_{-2}\mathcal L_{-m_3}^{\dagger}\, {}_{-2}S_{\ell_3-m_3}\right)(\pi)\left[\left(-m_1,2, \ell_1\right),\left(-m_2, 2, \ell_2\right)\right]\right\} \Big]\\&\hspace{1cm}+\sqrt{2} \Big[\Theta\left(m_1+m_2+5\right)\left\{-i\left({}_{-2}\mathcal L^{\dagger}_{-m_3}\,_{-2}S_{\ell_3 -m_3}\right)(0)\left[\left(m_1,2, \ell_1\right),\left(m_2, 3,\ell_2\right)\right]]\right.\\&\left.\hspace{5cm}-(-1)^{\ell_1+\ell_2+m_1+m_2} i \left({}_{-2}\mathcal L^{\dagger}_{-m_3}\,_{-2}S_{\ell_3 -m_3}\right)(\pi)\left[\left(m_1,-2, \ell_1\right),\left(m_2,-3, \ell_2\right)\right]\right\} \\
& -(-1)^{m_1+m_2} \Theta\left(-m_1-m_2-6\right)\left\{-i\left({}_{-2}\mathcal L^{\dagger}_{-m_3}\,_{-2}S_{\ell_3-m_3}\right)(0)\left[\left(-m_1,-2, \ell_1\right),\left(-m_2,-3, \ell_2\right)\right]]\right.\\&\left.\hspace{1cm}-(-1)^{\ell_1+\ell_2+m_1+m_2} i\left({}_{-2}\mathcal L^{\dagger}_{-m_3}\,_{-2}S_{\ell_3-m_3}\right)(\pi)\left[\left(-m_1, 2, \ell_1\right),\left(-m_2, 3, \ell_2\right)\right]\right\} \Big](-h_3+i m_3+2)\Bigg\} .
    \end{split}
\end{equation}
\end{widetext}

We note the following simplifications in these formulae when all $m_i=0$.
According to Eq. \eqref{Yexpand}, 
every  ${}_s S_{\ell 0}$ becomes a 
${}_s Y_{\ell 0}$, i.e. a spin weighted spherical harmonic. These vanish at $\bar \theta = 0,\pi$ unless $s=0$. Furthermore, by Eq. \eqref{eq:bar sYlm}, ${}_s \mathcal{L}^\dagger_m$ acts as a spin raising operator. Consequently, unless we have a term where precisely two such operators act on a ${}_{-2}Y_{\ell 0}$, such a term vanishes, giving rise to significant simplifications. A further simplification arises when we look at the symbols $\left[\left(m_1, s_1, \ell_1\right),\left(m_2,s_2,\ell_2\right)\right]$, defined in Eq. \eqref{angdef}, when $m_i=0$ and $s_1=s_2$: In this case, only the $\delta$-function term in Eq. \eqref{angdef} survives. The resulting formulas when all the $m_i$'s vanish are thereby found to be:
\begin{widetext}
\begin{equation}
         U^{\rm near}_{123}= 
         -\frac{\pi}{2} i^{\ell_3}
         [(\ell_3-1)\ell_3(\ell_3+1)(\ell_3+2)]^{\frac{1}{2}} \bar \ell_1
         \mathrm{odd}(\ell_S) \Big[-
         {}_{0}Y_{\ell_3 0}(0) +(-1)^{\ell_1+\ell_2} 
         {}_{0}Y_{\ell_3 0}(\pi) \Big] \delta(\bar \ell_1-\bar \ell_2) =-V^{\rm near}_{123}.
\end{equation}
\end{widetext}

\subsection{(high),(high) \texorpdfstring{  $\to$}{} (low)}

In the following formulas, $\ell_1$ corresponds to the low $\ell$, whereas $\ell_2, \ell_3$ are large e.g., $\bar \ell_2, \bar \ell_3 \ge c>0$, in terms of the rescaled angular momenta $\bar \ell = \ell/L$. We assume vanishing overtone numbers, i.e., $N_i=0$.
\begin{widetext}
 \begin{equation}
        \begin{split}
         &U^{\mathrm{near}}_{123}\\
            &= \delta_{m_1,m_2+m_3}\frac{ i^{-\ell_2-\ell_3}(-1)^{m_1}\pi^{\frac{5}{2}}\Gamma (-1+h_1-i m_1 ) \bar \ell_2{}^{\frac{1}{2}} \bar \ell_3{}^{\frac{1}{2}}  \csc\left[\frac{\pi}{2}(h_1+\ell_2+\ell_3+2 i m_1)\right]}{4\sqrt{2} {}_{+2}C_{0\ell_1 m_1}M^{\frac{4}{3}}  \Gamma (-2-h_1-im_1 )   \Gamma(h_1) \Gamma\left(h_1+\frac{1}{2}\right) [ h_1{} ^2+ (-m_1+2 i)^2]  \sqrt{\bar\ell_2+\bar \ell_3}}\\&
            \times \Bigg\{- (\bar \ell_2 +\bar \ell_3{})^4 \Big[ \Theta\left(-m_2-m_3+2\right)\Big(i {}_{2} S_{\ell_1 m_1}(0)\left[\left(m_2, 2, \ell_2\right),\left(m_3, 0, \ell_3\right)\right]\\&\hspace{5cm}-(-1)^{\ell_2+\ell_3+m_2+m_3} i {}_{2} S_{\ell_1 m_1}(\pi)\left[\left(m_2,-2, \ell_2\right),\left(m_3,0, \ell_3\right)\right]\Big) \\
& +(-1)^{m_2+m_3} \Theta\left(-m_2-m_3-3\right)\Big(i {}_{2} S_{\ell_1 m_1}(0)\left[\left(-m_2,-2, \ell_2\right),\left(-m_3,0,\ell_3\right)\right]\\&\hspace{5cm}-(-1)^{\ell_2+\ell_3+m_2+m_3} i {}_{2} S_{\ell_1 m_1}(\pi)\left[\left(-m_2, 2, \ell_2\right),\left(m_3,0,\ell_3\right)\right]\Big)\Big]\\&- \left(\bar\ell_2 ^4+4 \bar\ell_2 ^3 \bar\ell_3{}-6 \bar\ell_2 ^2 \bar\ell_3{}^2+4 \bar\ell_2  \bar\ell_3{}^3+\bar\ell_3{}^4\right) \\ &\times \Big[\Theta\left(m_2+m_3+2\right)\Big(i {}_{2} S_{\ell_1 m_1}(0)\left[\left(m_3,1,\ell_3\right),\left(m_2,1,\ell_2\right)\right]\\&\hspace{5cm}-(-1)^{\ell_2+\ell_3+m_2+m_3} i {}_{2} S_{\ell_1 m_1}(\pi)\left[\left(m_2,-1, \ell_2\right),\left(m_3,-1, \ell_3\right)\right]\Big) \\
& +(-1)^{m_2+m_3} \Theta\left(-m_2-m_3-3\right)\Big(i {}_{2} S_{\ell_1 m_1}(0)\left[\left(-m_3,-1, \ell_3\right),\left(-m_2,-1,\ell_2\right)\right]\\&\hspace{5cm}-(-1)^{\ell_2+\ell_3+m_2+m_3} i {}_{2} S_{\ell_1 m_1}(\pi)\left[\left(-m_3, 1, \ell_3\right),\left(-m_2,1, \ell_2\right)\right]\Big)\Big]\Bigg\}
            \end{split}
            \end{equation}
            and 
             \begin{equation}
                \begin{split}
                     & V^{\rm near}_{123}\\
             &=\delta_{m_1,m_2-m_3}\frac{ i^{\ell_3-\ell_2} (-1)^{m_1+m_3}\pi^{\frac{5}{2}} \Gamma (-1+h_1-i m_1 ) \csc\left[\frac{\pi}{2}(h_1+\ell_2+\ell_3+2 i m_1)\right] \bar\ell_2{}^{\frac{9}{2}}\bar\ell_3^{\frac{1}{2}}}{4\sqrt{2} {}_{+2}C_{0\ell_1 m_1}M^{\frac{4}{3}}  \Gamma (-2-h_1-im_1 ) \Gamma(h_1) \Gamma\left(h_1+\frac{1}{2}\right) [ h_1{} ^2+ (-m_1+2 i)^2]  \sqrt{\bar\ell_2+\bar\ell_3}}   \\&\times\Bigg\{ \Theta\left(-m_3+m_2+2\right)\\&\hspace{0.5cm}\times\Big[i {}_{2} S_{\bar\ell_1 m_1}(0)\Big(\left[\left(-m_3,-2,\ell_3\right),\left(m_2, 4, \ell_2\right)\right] +\left[\left(m_2,2, \ell_2\right),\left(-m_3, 0, \ell_3\right)\right]-2\left[\left(-m_3, -1,\ell_3\right),\left(m_2, 3,\ell_2\right)\right]\Big)\\&-(-1)^{\ell_2+\ell_3+m_3+m_2} i {}_{2} S_{\ell_1 m_2}(\pi)\Big(\left[\left(-m_3,2, \ell_3\right),\left(m_2,-4, \ell_2\right)\right] +\left[\left(m_2,-2, \ell_2\right),\left(-m_3,0,\ell_3\right)\right]\\&\hspace{5cm}-2\left[\left(-m_3,1, \ell_3\right),\left(m_2,-3, \ell_2\right)\right]\Big)\Big] \\
& -(-1)^{m_2+m_3} \Theta\left(m_3-m_2-3\right)\\&\times\left[-i {}_{2} S_{\ell_1 m_1}(0)\Big(\left[\left(m_3,2, \ell_3\right),\left(-m_2,-4,\ell_2\right)\right]+\left[\left(-m_2,-2, \ell_2\right),\left(m_3,0, \ell_3\right)\right]-2\left[\left(m_3,1, \ell_3\right),\left(-m_2,-3, \ell_2\right)\right]\Big)\right.\\&\left.+(-1)^{\ell_2+\ell_3+m_2+m_3} i {}_{2} S_{\ell_1 m_1}(\pi)\Big(\left[\left(m_3, -2, \ell_3\right),\left(-m_2,4,\ell_2\right)\right]+\left[\left(-m_2, 2, \ell_2\right),\left(m_3, 0, \ell_3\right)\right]\right.\\&\left.\hspace{5cm}-2\left[\left(m_3, -1, \ell_3\right),\left(-m_2,3,\ell_2\right)\right]\Big)\right]\Bigg\}.
                \end{split}
            \end{equation}
\end{widetext}
Similar simplifications as before apply to the case when 
all $m_i=0$. In fact, because the spin-weighted spheroidal harmonics vanish in that case and it is thereby found that $U^{\rm near}_{123}, V^{\rm near}_{123}=0$
to leading order in $L$.

\subsection{(high),(low)\texorpdfstring{  $\to$}{} (low)}

In the following formulas, $\ell_1$ and $\ell_3$ correspond to the low $\ell$'s, whereas $\ell_2$ is large e.g., $\bar \ell_2 \ge c>0$, in terms of the rescaled angular momenta $\bar \ell = \ell/L$. We assume vanishing overtone numbers, i.e., $N_i=0$.
\begin{widetext}
      \begin{equation}
            \begin{split}
        &U^{\rm near}_{123}\\
        &=- \delta_{m_1,m_2+m_3}\frac{i^{-\ell_2} (-1)^{m_2} \pi^{\frac{7}{4}} \Gamma (-1+ h_1-i m_1 )   \sec \left[\frac{1}{2} \pi  (h_1+h_3+\ell_2+2 i m_1)\right] }{ 2^{\frac{3}{4} }M^{\frac{5}{3}}\bar \ell_2^{2} {}_{+2}C_{0\ell_1 m_1} \Gamma (-2- h_1-im_1)  \Gamma (2 h_1 ) [ h_1 ^2+ (-m_1+2 i)^2]}  \\& \times\Big\{\frac{i}{4}\prod_{j=0}^1(-h_3+im_3 +2-j)
    [\Theta(m_2)-(-1)^{m_2}\Theta(-m_2-1)]\\&\hspace{2cm}\times\left[-
    m_2{}_{-2}S_{\ell_1 m_1}{}_{-2}S_{\ell_3-m_3}(0) +(-1)^{m_2+\ell_2}m_2 S_{\ell_1 m_1}{}_{-2}S_{\ell_3-m_3}(\pi) \right]\\&-\frac{i}{\sqrt{2}}(-h_3+im_3 +2)[\Theta(m_2+1)+(-1)^{m_2}\Theta(-m_2-2)]\\&\times\left[
    (m_2+1){}_{-2}S_{\ell_1 m_1}\left({}_{-2}\mathcal L^{\dagger}_{-m_3}\,_{-2}S_{\ell_3-m_3}\right)(0) - (-1)^{m_2+\ell_2}(m_2-1){}_{-2}S_{\ell_1 m_1}\left({}_{-2}\mathcal L^{\dagger}_{-m_3}\,_{-2}S_{\ell_3-m_3}\right)(\pi) \right] \\&-\frac{i}{2}[\Theta(m_2+2)-(-1)^{m_2}\Theta(-m_2-3)]\left[
    (m_2+2){}_{-2}S_{\ell_1 m_1}\left({}_{-1}\mathcal L_{-m_3}^{\dagger}\, {}_{-2}\mathcal L_{-m_3}^{\dagger}\, {}_{-2}S_{\ell_3-m_3}\right)(0) \right.\\&\left.\hspace{4cm}- (-1)^{m_2+\ell_2}(m_2-2){}_{-2}S_{\ell_1 m_1}\left({}_{-1}\mathcal L_{-m_3}^{\dagger}\, {}_{-2}\mathcal L_{-2}^{\dagger}\, {}_{-m_3}S_{\ell_3-m_3}\right)(\pi) \right]\Big\}
    \end{split}
\end{equation}
and
\begin{equation}
    \begin{split}
        &V^{\rm near}_{123}\\
        &=-  \delta_{m_1,m_2-m_3}\frac{i^{-\ell_2 } (-1)^{m_2}  \pi^{\frac{7}{4}} \Gamma (-1+h_1-i m_1 )   \sec \left[\frac{1}{2} \pi  (h_1+h_3+\ell_2+2 i m_1)\right] }{2^{\frac{3}{4}}M^{\frac{5}{3}}\bar\ell_2^{2}  {}_{+2}C_{0\ell_1 m_1} \Gamma (-2-h_1-im_1)  \Gamma (2 h_1 ) [  h_1{} ^2+ (-m_1+2 i)^2]}     \\&\times \Big\{
    \frac{i}{4} \prod_{j=0}^1(-h_3+im_3 +2-j)[\Theta(m_2+4)-(-1)^{m_2}\Theta(-m_2-5)]\\&\hspace{2cm}\times\left[
    (m_2+4){}_{2}S_{\ell_1m_1}{}_{-2}S_{ l_3-m_3}(0) - (-1)^{m_2+\ell_2}(m_2-4){}_{2}S_{\ell_1m_1}{}_{-2}S_{\ell_3 -m_3}(\pi) \right]\\&+  \frac{i}{2}[\Theta(m_2+2)-(-1)^{m_2}\Theta(-m_2-3)]\Big[
    (m_2+2){}_{2}S_{\ell_1m_1}\left({}_{-1}\mathcal L_{-m_3}^{\dagger}\, {}_{-2}\mathcal L_{-m_3}^{\dagger}\, {}_{-2}S_{\ell_3 -m_3}\right)(0) \\&\hspace{4cm}- (-1)^{m_2+\ell_2}(m_2-2){}_{2}S_{\ell_1m_1}\left({}_{-1}\mathcal L_{-m_3}^{\dagger}\, {}_{-1}\mathcal L_{-m_3}^{\dagger}\, {}_{-2}S_{\ell_3-m_3}\right)(\pi) \Big]\\&+\frac{i}{\sqrt{2}} (-h_3+im_3 +2)[\Theta(m_2+3)+(-1)^{m_2}\Theta(-m_2-4)]\\\times&  \left[
    -(m_2+3){}_{2}S_{\ell_1m_1}\left({}_{-2}\mathcal L^{\dagger}_{-m_3}\,_{-2}S_{\ell_3-m_3}\right)(0) + (-1)^{m_2+\ell_2}(m_2-3){}_{2}S_{\ell_3m_3}\left({}_{-2}\mathcal L^{\dagger}_{-m_3}\,_{-2}S_{\ell_3-m_3}\right)(\pi) \right]\Big\}
        \end{split}
        \end{equation}
\end{widetext}

Finally, we consider the case when 
all $m_i=0$. As in the previous channel, because the spheroidal harmonics with non-trivial spin weight vanish at $0, \pi$ in that case, we now find $U^{\rm near}_{123}, V^{\rm near}_{123}=0$ to leading order in $L$.

\section{nNHEK QNMs as \texorpdfstring{$\SLg$}{} modules} 
\label{SL2R}

The spacetime symmetry generators of nNHEK corresponding to infinitesimal $\SLg$ actions are given by
\begin{equation}
\begin{split}\label{eq:Killing NHEK}
 	H_{0}^a &= -(\partial_{\bar{t}})^a , \\
    H_{\pm}^a &= \frac{ e^{\pm \bar{t}} }{\sqrt{f}} 
				\Big[ (1+\bar{x})\partial_{\bar t} \mp f\partial_{\bar x} -\partial_{\bar \phi}\Big]^a.
\end{split}
\end{equation}
They satisfy the $\mathfrak{sl}_2(\mathbb R)$ commutation relations
\begin{equation}
[H_-, H_+]^a = 2 H_0^a ,\qquad [H_0, H_{\pm}]^a = H_{\pm}^a.
\end{equation}
Associated with each such Killing vector field of nNHEK 
we have a corresponding GHP covariant Lie-derivative \cite{edgar2000integration} ${\mathcal L}_X, X^a \in 
\{H_0^a, H_\pm^a\}$. These operators satisfy the same commutation relations and furthermore commute with the spin $s$ Teukolsky operators, $[{}_s \O, {\mathcal L}_X]=0$. 
Hence the complex linear span of the near zone QNM solutions [see Eq. \eqref{nearQNM}] to the spin $s$ Teukolsky equations is a module of $\mathfrak{sl}_2(\mathbb R)$ under the action of the GHP covariant Lie derivative. 

We now characterize the decomposition of this module into irreducible submodules. We first define the Casimir
\begin{equation}\label{eq:Cas}
\hat \Omega := \L_{H_0} (\L_{H_0}- 1) - \L_{H_+}\L_{H_-},
\end{equation}
which is useful to classify irreducible representations. 
Next, we observe that, 
\begin{equation}
    {\mathcal J} \L_{H_0} {\mathcal J} = -\L_{H_0}, \quad
    {\mathcal J} \L_{H_+} {\mathcal J} = -\L_{H_-},
\end{equation}
where ${\mathcal J}$ is the $t$-$\phi$ reflection operator on GHP scalars \eqref{Jdef}. By the same arguments as given in \cite{Green2022a} in the case of Kerr, the latter relations imply that 
\begin{equation}
\label{Hsa}
\begin{split}
    &\llangle \L_{H_0} \Upsilon_1, \Upsilon_2 \rrangle_{\rm near} = \llangle  \Upsilon_1, \L_{H_0} \Upsilon_2 \rrangle_{\rm near}, \\
    &\llangle \L_{H_+} \Upsilon_1, \Upsilon_2 \rrangle_{\rm near} = \llangle  \Upsilon_1, \L_{H_-} \Upsilon_2 \rrangle_{\rm near},
\end{split}
\end{equation}
for each pair of near zone modes. The near zone mode solutions 
\eqref{nearQNM} have $(\bar t, \bar \phi)$-dependence ${}_s \Upsilon^{\rm near}_{N\ell m} \propto e^{-i\bar \omega \bar t+im\bar \phi}$ with 
$\bar \omega = -i(N+{}_s h_{+\ell m})$ [see Eq. \eqref{eq:hpm} for the definition of $h_+ \equiv {}_s h_{+\ell m}$], which in view of the structure \eqref{eq:Killing NHEK} of $H_0^a, H_\pm^a$ at once implies the following. For any given $(s,N,\ell,m)$, ${}_s \Upsilon^{\rm near}_{N\ell m}$ must be an eigenfunction of $\L_{H_0}$, in fact with eigenvalue $N+h$. Due to the factors of $e^{\pm \bar t}$ in $H^a_\pm$, $\L_{H_\pm}$ map ${}_s \Upsilon^{\rm near}_{N\ell m} \to {}_s \Upsilon^{\rm near}_{(N\pm 1)\ell m}$. 

Therefore, we have 
\begin{equation}
\label{sl2rep}
    \L_{H_0} \Upsilon_N = (h_++N) \Upsilon_N, \quad
    \L_{H_\pm} \Upsilon_N = \alpha_N^\pm \Upsilon_{N\pm1}, 
\end{equation}
where we omitted the reference to $(s,\ell,m)$ and used e.g., the shorthand $\Upsilon_N \equiv \Upsilon^{\rm near}_{N\ell m}$ \eqref{nearQNM}, and where $\alpha_N^\pm$ are constants that depend on the choice 
of normalization for $\Upsilon_N$. E.g. in the normalization \eqref{tildeR1} of the 
radial functions $(\nu=0)$ without the factor $C_N$ in that equation, we would have
\begin{equation}
\label{alpm}
\begin{split}
    \alpha^+_N &= -\sqrt{2}(h_+ + N +im+s), \\ \alpha_N^- &= -\frac{N(N+2h_+-1)}{\sqrt{2}(h_+ + N -1 +im +s)}.
    \end{split}
\end{equation}
In particular, $\alpha_0^- = 0$ as must be the case since there are no QNMs with overtone numbers $N<0$. In this sense, the complex linear span ${}_s V_{\ell m}$ of the near zone QNMs with a fixed $(s,\ell,m)$ forms a module with raising/lowering operators $\L_{H_\pm}$ and lowest weight state ${}_s \Upsilon_{0\ell m}^{\rm near}$ annihilated by
$\L_{H_-}$. We also find from Eqs. \eqref{alpm}
\begin{equation}
    \hat \Omega = h_+(h_+-1)
\end{equation}
for the value of the Casimir in each irreducible module ${}_s V_{\ell m}$, so $h_+ \equiv {}_s h_{+\ell m}$ is indeed a conformal weight in the usual terminology. Note that the modules ${}_s V_{\ell m}$ are {\it not} unitary because the bilinear form 
\eqref{Hsa} is complex linear in each entry rather than sesquilinear positive definite, as would be required for a unitary representation.

In the normalization $\hat \Upsilon_N := 
\Upsilon_N/\sqrt{A_N}$ [see Eqs. \eqref{PZsimply}, \eqref{PZsimply2} for $|s|=2$], 
such that $\llangle \hat \Upsilon_N, \hat \Upsilon_N\rrangle_{\rm near}=1$, we would instead have
\begin{equation}
    \hat \alpha_N^\pm = \alpha_N^\pm \sqrt{ \frac{A_{N\pm 1}}{A_N} },  
\end{equation}
whereas by Eq. \eqref{Hsa}, we must also have $\hat \alpha^+_N = \hat \alpha^-_{N+1}$. 
This leads to the following recursion for the $A_N$.
\begin{equation}
    \frac{A_N}{A_{N-1}} = \frac{\alpha^-_{N}}{\alpha^+_{N-1}} = \frac{N(N+2h_+-1)}{2(h_+ + N -1 +im +s)^2}.
\end{equation}
This could be used e.g., in order to find $A_N, N>0$ from $A_0$ (see \eqref{normalization0overtone} and \eqref{normalization0overtone+}), and thereby obtain an analytical proof of Eqs. \eqref{PZsimply}, \eqref{PZsimply2} for $|s|=2$. QNM mode orthogonality in the near zone follows abstractly from the first relation in Eqs. \eqref{Hsa} by the same argument as in \cite{Green2022a}.

\section{Coefficients in Eqs. \texorpdfstring{\eqref{overlaptdep}}{}}
\label{tables}
In this section we give the explicit form for each term in the overlap integrals  $ V^{\rm near}_{123}$ and $U^{\rm near}_{123}$ \eqref{G41} at the leading order in $L$, referring to the ``(high),(high) $\to$ (high)'' $\ell$ channel. Each term factorizes into a radial and angular overlap integral as in Eq. \eqref{overlaptdep}. These in turn are evaluated in our limit in Apps. \ref{radoverlap} and \ref{statphase}.

We use the definition of
 \begin{equation}
     {}_{-2}\Upsilon_{-q}:=\zeta^4\mathcal J {}_{+2}\Upsilon_q=\frac{f^2}{4M^{\frac{4}{3}}}e^{i\omega_N \bar t-im\bar \phi}{}_{+2}R_{N\ell m}{}_{+2} S_{\ell m}.
 \end{equation}
and the relations 
\begin{equation}
\label{Rreality}
\begin{split}
    &{}_{s}Y_{\ell m}=(-1)^{s+m}{}_{-s} Y_{\ell (-m)}\\
    & {}_{s,\nu}R_{N\ell m}^*={}_{s,\nu}R_{N\ell (-m)}
    \end{split}
\end{equation}
where the first symmetry is standard \cite{Goldberg1967}, and second one can be directly read off from the definition \eqref{tildeR} and the fact that $h_\pm\in \mathbb R$ in our approximation regime. Again, when an $m_i$ vanishes, we have to make the replacement $m_i\to-i\varepsilon(\ell_i+N_i+1)$ in the solution to the radial Teukolsky equation (see App. \ref{axisym}) and in the quantities affected by this change, i.e., column 2 of the following table. In this regard, note that, for these modes, we have the reality condition
\begin{equation}
    \left({}_{s,\nu}R_{N\ell (-i\varepsilon(N+\ell+1))}\right)^*={}_{s,\nu}R_{N\ell(-i\varepsilon(N+\ell+1))}
\end{equation}
instead of Eq. \eqref{Rreality}. Therefore, in $\{123\}$
if $m_i=0$, we should take $\pm m_i\to-i\varepsilon(\ell_i+N_i+1)$
in those quantities for both signs.

The GHP operators appearing in the terms are related to ladder operators as described in Apps. \ref{radwave}, \ref{edthladder}, which 
allows us to remain in the class of integrals in Apps. \ref{radoverlap} and \ref{statphase}.


\begin{turnpage}
\begin{table}
  \centering
\begin{tabular}{ m{8.0cm}|m{2.5cm}| m{7.8cm}| m{5.1cm} }
\hline
\multicolumn{4}{c}{$v_{123}\{123\}[123]=v_{123}\big\{(s_1,\nu_1,N_1,\ell_1,m_1)(s_2,\nu_2,N_2,\ell_2,m_2)(s_3,\nu_3,N_3,\ell_3,m_3)\big\}\big[(s_1,\ell_1,m_1)(s_2,\ell_2,m_2)(s_3,\ell_3,m_3)\big]$}\\
\hline
\vspace{0.4cm} GHP form of the overlap coefficient & $v_{123}/M^{\frac{5}{3}}$ & Parameters of $\{123\}$ & Parameters of $[123]$ \\[0.4cm]
\hline
\vspace{0.4cm} $\frac{1}{A_2^* A_3}\int\limits_{ \mathscr{C} }  
 \bar T^a \dd \bar S_a 
({}_{-2}^{} \Upsilon_{-q_1}) 
\thorn^2 ({}_{+2}\Upsilon_{q_2}) \edth^2({}_{-2}\Upsilon_{-q_3^*})$ & $\frac{1}{4  A_2^* A_3}\ell_2^2 \ell_3^2$ & $\big\{(2,0,N_1,\ell_1,m_1)(2,2,N_2,\ell_2,m_2)(-2,0,N_3,\ell_3,-m_3)\big\}$  & $\big[(2,\ell_1,m_1)(2,\ell_2,m_2)(0,\ell_3,-m_3) \big]$ \\[0.4cm] 
\hline
\vspace{0.4cm} $\frac{-2}{A_2^* A_3}\int\limits_{ \mathscr{C} } 
 \bar T^a \dd \bar S_a 
({}_{-2}^{} \Upsilon_{-q_1})  \edth \thorn({}_{+2}\Upsilon_{q_2})\edth \thorn ({}_{-2}\Upsilon_{-q_3^*})$ & $-\frac{1}{2  A_2^* A_3}\ell_2^2 \ell_3^2$ & $\big\{(2,0,N_1,\ell_1,m_1)(2,1,N_2,\ell_2,m_2)(-2,1,N_3,\ell_3,-m_3)\big\}$  & $\big[(2,\ell_1,m_1)(3,\ell_2,m_2)(-1,\ell_3,-m_3)\big]$ \\[0.4cm] 
 \hline
\vspace{0.4cm} $\frac{1}{A_2^* A_3}\int\limits_{ \mathscr{C} }  
 \bar T^a \dd \bar S_a
({}_{-2}^{} \Upsilon_{-q_1})\edth^2 ({}_{+2}\Upsilon_{q_2}) \thorn^2 ({}_{-2}\Upsilon_{-q_3^*})  
$ & $\frac{1}{4  A_2^* A_3}\ell_2^2 \ell_3^2$ & $\big\{(2,0,N_1,\ell_1,m_1)(2,0,N_2,\ell_2,m_2)(-2,2,N_3,\ell_3,-m_3)\big\}$ & $\big[(2,\ell_1,m_1)(4,\ell_2,m_2)(-2,\ell_3,-m_3)\big]$  \\ [0.4cm] 
\end{tabular}
\vspace{0.4cm}

  \centering
\begin{tabular}{ m{8.0cm}|m{2.5cm}| m{7.8cm}| m{5.1cm} }
\hline
\multicolumn{4}{c}{$u_{123}\{123\}[123]=u_{123}\big\{(s_1,\nu_1,N_1,\ell_1,m_1)(s_2,\nu_2,N_2,\ell_2,m_2)(s_3,\nu_3,N_3,\ell_3,m_3)\big\}\big[(s_1,\ell_1,m_1)(s_2,\ell_2,m_2)(s_3,\ell_3,m_3)\big]^*$}\\
\hline
\vspace{0.4cm} GHP form of the overlap coefficient & $u_{123}/M^{\frac{5}{3}}$ & Parameters of $\{123\}$ & Parameters of $[123]^*$ \\[0.4cm]
\hline
\vspace{0.4cm} $   \frac{1}{A_{2}  A_{3}} \int\limits_{ \mathscr{C} }  
 \bar T^a \dd \bar S_a
({}_{-2} \Upsilon_{-q_1}) 
\thorn^2 ({}_{+2} \Upsilon_{q_2}) \edth'{}^2  ({}_{-2}^{} \Upsilon_{-q_3^*}^*)$ & $ \frac{(-1)^{m_3}}{4A_{2}  A_{3}} \ell_2^2 \ell_3^2$ & $\big\{(2,0,N_1,\ell_1,m_1)(2,2,N_2,\ell_2,m_2)(-2,0,N_3,\ell_3,m_3)\big\}$  & $[(2,\ell_1,m_1)(2,\ell_2,m_2)(0,\ell_3,m_3)]^*$ \\[0.4cm] 
\hline
\vspace{0.4cm} $   \frac{1}{A_{2}  A_{3}} \int\limits_{ \mathscr{C} }  
 \bar T^a \dd \bar S_a
({}_{-2} \Upsilon_{-q_1}) \edth'{}^2  ({}_{+2} \Upsilon_{q_2})
\thorn^2 ({}_{-2} \Upsilon^*_{-q_3^*}) $ & $\frac{(-1)^{m_3}}{4A_2 A_3}\ell_2^2\ell_3^2$ & $\big\{(2,0,N_1,\ell_1,m_1)(2,0,N_2,\ell_2,m_2)(-2,2,N_3,\ell_3,m_3)\big\}$   &$[(2,\ell_1,m_1)(0,\ell_2,m_2)(2,\ell_3,m_3)]^*$ \\[0.4cm] 
 \hline
\vspace{0.4cm} $   \frac{-2}{A_{2}  A_{3}} \int\limits_{ \mathscr{C} }  
 \bar T^a \dd \bar S_a
({}_{-2} \Upsilon_{-q_1}) 
\edth'\thorn ({}_{+2} \Upsilon_{q_2}) \edth'\thorn  ({}_{-2} \Upsilon_{-q_3^*}^*)$ & $\frac{(-1)^{m_3}}{2A_2 A_3}\ell_2^2 \ell_3^2$ & $\big\{(2,0,N_1,\ell_1,m_1)(2,1,N_2,\ell_2,m_2)(-2,1,N_3,\ell_3,m_3)\big\}$ & $[(2,\ell_1,m_1)(1,\ell_2,m_2)(1,\ell_3,m_3)]^*$ \\[0.4cm] 
 \hline
 \vspace{0.4cm} $   \frac{4}{A_{2}  A_{3}} \int\limits_{ \mathscr{C} }  
 \bar T^a \dd \bar S_a
({}_{-2} \Upsilon_{-q_1}) \thorn({}_{+2} \Upsilon_{q_2})
\edth'{}^2\thorn ({}_{-2} \Upsilon^*_{-q_3^*}) $ & $\frac{(-1)^{m_3}}{A_2 A_3} \ell_2 \ell_3^3$ & $\big\{(2,0,N_1,\ell_1,m_1)(2,1,N_2,\ell_2,m_2)(-2,1,N_3,\ell_3,m_3)\big\}$  & $[(2,\ell_1,m_1)(2,\ell_2,m_2)(0,\ell_3,m_3)]^*$ \\[0.4cm] 
 \hline
 \vspace{0.4cm} $   \frac{-8}{A_{2}  A_{3}} \int\limits_{ \mathscr{C} }  
 \bar T^a \dd \bar S_a
({}_{-2} \Upsilon_{-q_1})  \edth'({}_{+2} \Upsilon_{q_2})
\edth'\thorn^2  ({}_{-2} \Upsilon^*_{-q_3^*}) $ & $\frac{2(-1)^{m_3}}{A_{2}  A_{3}}\ell_2 \ell_3^3$ & $\big\{(2,0,N_1,\ell_1,m_1)(2,0,N_2,\ell_2,m_2)(-2,2,N_3,\ell_3,m_3)\big\}$  & $[(2,\ell_1,m_1)(1,\ell_2,m_2)(1,\ell_3,m_3)]^*$ \\[0.4cm] 
 \hline
 \vspace{0.4cm} $   \frac{1}{A_{2}  A_{3}} \int\limits_{ \mathscr{C} }  
 \bar T^a \dd \bar S_a
({}_{-2} \Upsilon_{-q_1})  \edth'{}^2\thorn^3({}_{-2}\Upsilon^*_{-q_2^*})
\thorn^3  ({}_{-2} \Upsilon^*_{-q_3^*}) $ & $\frac{(-1)^{m_2+m_3}}{4 A_{2}  A_{3}}\ell_2^5\ell_3^3$ & $\big\{(2,0,N_1,\ell_1,m_1)(-2,3,\ell_2,m_2)(-2,3,\ell_3,m_3)\big\}$  & $[(2,\ell_1,m_1)(0,\ell_2,m_2)(2,\ell_3,m_3)]^*$ \\[0.4cm] 
 \hline
 \vspace{0.4cm} $   \frac{6}{A_{2}  A_{3}} \int\limits_{ \mathscr{C} }  
 \bar T^a \dd \bar S_a
({}_{-2} \Upsilon_{-q_1})  ({}_{+2} \Upsilon_{q_2})
\edth'{}^2\thorn^2  ({}_{-2} \Upsilon^*_{-q_3^*}) $ & $ \frac{3(-1)^{m_3}}{2A_{2}  A_{3}}\ell_3^4$ & $\big\{(2,0,N_1,\ell_1,m_1)(2,0,\ell_2,m_2)(-2,2,\ell_3,m_3)\big\}$  & $[(2,\ell_1,m_1)(2,\ell_2,m_2)(0,\ell_3,m_3)]^*$ \\[0.4cm] 
 \hline
 \vspace{0.4cm} $   \frac{-3}{2A_{2}  A_{3}} \int\limits_{ \mathscr{C} }  
 \bar T^a \dd \bar S_a
({}_{-2} \Upsilon_{-q_1}) \edth'\thorn^3({}_{-2}\Upsilon^*_{-q_2^*}) \edth'\thorn^3({}_{-2}\Upsilon^*_{-q_3^*})
 $ & $ \frac{-3(-1)^{m_2+m_3}}{8A_{2}  A_{3}}\ell_2^4 \ell_3^4$ & $\big\{\big(2,0,N_1,\ell_1,m_1)(-2,3,N_2,\ell_2,m_2)(-2,3,N_3,\ell_3,m_3)\big\}$  & $[(2,\ell_1,m_1)(1,\ell_2,m_2)(1,\ell_3,m_3)]^*$ \\[0.4cm]
\end{tabular}

\end{table}

\end{turnpage}

\makeatletter\onecolumngrid@push\makeatother
\clearpage
\makeatletter\onecolumngrid@pop\makeatother

	\bibliographystyle{apsrev4-2}
\bibliography{NHEK}
\end{document}